\documentclass[aps,prd,twocolumn,amsmath,groupedaddress,nofootinbib]{revtex4}

\usepackage{graphicx}
\usepackage{amssymb}
\usepackage{bm}
\usepackage{dcolumn}

\usepackage{color}   

\newcommand{\be}{\begin{eqnarray}}     	\newcommand{\ee}{\end{eqnarray}}   
\newcommand{\ba}{\begin{array}}         \newcommand{\ea}{\end{array}} 
\newcommand{\bs}{\begin{subequations}}  \newcommand{\es}{\end{subequations}} 
\newcommand{\rf}[1]{~(\ref{#1})}  \newcommand{\rfd}[2]{~(\ref{#1},\,\ref{#2})}
\newcommand{\rfs}[2]{~(\ref{#1}--\ref{#2})}
\newcommand{\rft}[3]{~(\ref{#1},\,\ref{#2},\,\ref{#3})}
\newcommand{\rfp}[1]{~\ref{#1} on p.~\pageref{#1}} 
\newcommand{\ct}[1]{\ Ref.~\cite{#1}}  \newcommand{\cts}[1]{\ Refs.~\cite{#1}}
\newcommand{\lb}[1]{\label{#1}}  
\newcommand{\fig}[1]{Fig.~\ref{#1}}
\newcommand{\nn}{\nonumber}
\newcommand{\lf}{\left}     \newcommand{\rt}{\right}
\newcommand{\fr}{\frac}

\newcommand{\dst}{\displaystyle}
\newcommand{\Vsp}{\vphantom{\displaystyle{\hat I \over \hat I}}}
\newcommand{\vsp}{\vphantom{\displaystyle{I \over I}}}

\def\d{\delta}    \def\pd{\partial}  \def\D{\Delta}
\def\dd{\d_{\rm D}}
\def\t{\tau}           
\def\al{\alpha}    \def\b{\beta}     \def\g{\gamma}
\def\l{\lambda} 
\def\Nb{\nabla}	  \def\Nbi{\nabla_{\!i}}    \def\Nbj{\nabla_{\!j}}
\newcommand{\Nbs}[1]{\nabla_{\!#1}}
\def\L{\Lambda}

\def\fs{\varphi_{\!s}}      \def\f{\varphi}
         \def\chf{\chi}
\def\Om{\Omega}    \def\om{\omega}    
\def\eps{\epsilon} 
\def\H{{\cal H}}
\def\R{{\cal R}}
\def\Teff{\Theta^{\rm eff}}   \def\dTeff{{\dot\Theta}^{\rm eff}}   
\def\Neff{N_{\nu}}
\def\r{{\bf r}} 	\def\k{{\bf k}} 
\def\^{\hat}	        \def\~{\tilde}	
\def\n{\hat{\bf n}}     
\def\odot{^{\!\displaystyle\cdot}}
\def\eq{_{\rm eq}}         \def\dec{_{\g\,\rm dec}}
\def\Rnu{R_{\nu}}          \def\alnu{\alpha_{\nu}} 
\def\snu{_{\nu}}           \def\sg{_{\gamma}}  
\def\stot{}
\def\sd{_{\rm de}}
\def\sHe{_{\rm He}}
\def\rtt{\sqrt{3}}
\def\eg{{\it e.g.\/}}   \def\cf{{\it c.f.\/}\ }   \def\ie{{\it i.e.\/}}
\def\eV{\,{\rm eV}}

\def\i{_{\rm in}}

\newcommand{\const}{{\rm const}}
\newcommand{\sign}{\mathop{\rm sign}}
\newcommand{\ci}{\mathop{\rm ci}}     
\newcommand{\fp}{}  

\begin{document}

\title{\Large Signatures of Relativistic Neutrinos 
           \\ in CMB Anisotropy and Matter Clustering}

\author{Sergei Bashinsky}
\email{sergei@princeton.edu}
\author{Uro\v s Seljak}
\email{useljak@princeton.edu}

\affiliation{Department of Physics, 
              Princeton University, Princeton, NJ 08544}

\date{January 15, 2004}

\begin{abstract}  
  We present a detailed analytical study of ultra-relativistic 
neutrinos in cosmological perturbation theory and of the observable 
signatures of inhomogeneities in the cosmic neutrino background.  
We note that a modification of perturbation variables that removes
all the time derivatives of scalar gravitational potentials
from the dynamical equations simplifies their solution notably.
The used perturbations of particle number per
coordinate, not proper, volume are generally 
constant on superhorizon scales. 
In real space an analytical analysis can be extended
beyond fluids to neutrinos. 

The faster cosmological expansion due to the 
neutrino background changes the acoustic and damping 
angular scales of the cosmic microwave background (CMB). 
But we find that equivalent changes can be produced by varying
other standard parameters, including the primordial helium abundance.  
The low-$l$ integrated Sachs-Wolfe
effect is also not sensitive to neutrinos. 
However, the gravity of neutrino perturbations suppresses 
the CMB acoustic peaks for the multipoles with $l\gtrsim 200$
while it enhances the amplitude of matter fluctuations on these scales.  
In addition, the perturbations of relativistic neutrinos
generate a {\it unique phase shift\/} of the CMB
acoustic oscillations that for adiabatic initial conditions
cannot be caused by any other standard physics. 
The origin of the shift is traced to neutrino free-streaming
velocity exceeding the sound speed of the photon-baryon plasma.
We find that from a high
resolution, low noise instrument such as CMBPOL the effective number of light
neutrino species can be determined with an accuracy of 
$\sigma(N_{\nu})\simeq 0.05$ to $0.09$, depending on the constraints 
on the helium abundance.
\end{abstract}

\maketitle

\section{Introduction}

Neutrinos play a significant role in the evolution of the early universe.
They are expected to provide around $40\%$,
{\it e.g.\/}\ct{KTbook}, of the total energy density during 
the radiation era. 
The gravitational potentials (metric perturbations) induced by
inhomogeneities in the photon and neutrino backgrounds are
comparable.  Due to the internal photon and neutrino dynamics 
the potentials decay when the growing acoustic,
for photons, or particle, for neutrinos, horizon of the universe
becomes of the order of the perturbation scale, \ie\ 
as the perturbation modes ``enter the horizon''.
This decay, in contrast to the constancy of the 
potential generated during matter domination by freely collapsing cold 
dark matter~(CDM), leads to substantial difference between
the amplitude of the acoustic oscillations in 
the cosmic microwave background~(CMB) on the
scales that enter the horizon before and after the 
matter-radiation equality.  
For example, in the model without neutrinos
the amplitude generated by equal primordial power
on the smaller scales is 5~times larger,\ct{HuSugSmall}.
In addition, the gravity of both the photon and neutrino perturbations
at the horizon entry boosts CDM peculiar velocities, contributing
to matter clustering.

Neutrino contribution to the radiation energy density
reduces the redshift of the transition from
radiation to matter domination, bringing the transition
closer to CMB decoupling. 
This too leads to important consequences 
to both CMB anisotropies and matter clustering. 
The reasons are the larger amplitude of the acoustic 
oscillations entering the horizon in the radiation
universe, larger early-ISW effect from 
the transition proximity, 
and suppressed growth of matter 
fluctuations in the radiation epoch.
But these and other discussed later 
effects caused by neutrino background speeding up the 
cosmological expansion can generally be 
mimicked by variations of other standard cosmological parameters.
For example, the redshift of the radiation-matter equality
could be reduced not by neutrino background but by
CDM density being smaller than derived from the fits 
assuming the standard neutrino content.

The internal dynamics of neutrino perturbations bears almost 
no resemblance to the more familiar acoustic physics of the photon-baryon fluid. 
All the three main distinctions below arise from neutrinos being fully 
decoupled and free-streaming since a very early redshift~$z_{\nu\,\rm dec}\sim
10^{10}$, long before the hydrogen recombination and CMB photon 
decoupling at $z\dec\simeq 1090$.

First, the tightly coupled photon-electron-baryon fluid 
supports compressional acoustic waves.
These waves are little attenuated until the recombination.  
Neutrino perturbations propagate differently, by means of free 
streaming.  Neutrinos escape overdense regions in every direction;
the projection of their velocity on the density gradient
spans the entire interval~$[-1,\,1]$  (in units $c=1$.)
The dispersion of the perturbation transfer velocity along 
the density gradient, called ``directional 
dispersion''~\cite{BondSzalay}, damps subhorizon neutrino 
perturbations inversely proportional to time.  This damping was 
noted three decades ago~\cite{Stewart72}. But it was quickly 
realized~\cite{Peebles73} that, regardless of their evolution,
the subhorizon neutrino perturbations exert negligible
gravitational effects on other species.

Second, neutrino stress is locally anisotropic.
According to the Einstein's equations, the stress sources
the perturbations of space-time metric.
The anisotropic stress leads to richer structure of the 
metric perturbations than locally isotropic fluids can provide.

Third, neutrino perturbations propagate with the speed of light,
exceeding the sound speed of acoustic perturbations in the 
photon-baryon fluid.  As a result, the gravitational effect
of neutrino perturbations on CMB, viewed in real space, 
extends beyond the acoustic horizon of primordial inhomogeneities.  
We find that this leads to a unique phase 
shift of the CMB mode oscillations 
in the presence of neutrino gravity.

What new physics can be revealed by the imprint
of neutrino gravity on the more easily observable species,
such as CMB photons or non-relativistic matter?  
The considered neutrino signatures probe
the ratio of neutrino and photon energy densities,
evaluated when the observed scales enter the horizon.  
Complimentary constraints on the universe composition in the radiation era
are set by the predictions of Big Bang nucleosynthesis (BBN)
for the primordial abundances of light elements.
The baryon to photon ratios inferred
from BBN and CMB are in good agreement with each other,
but the presently low observational estimates of the primordial ${}^4$He 
abundance~\cite{Y_OS_95,Y_OS_97,Y_ITL_97,Y_ITL_98,Y_PPL_02} 
favor the effective number of neutrinos, 
$\Neff$, at BBN below its standard value~$3.04$.
Joint analyses~\cite{BargPLB03,HannestadConstr} of the current data
on the primordial ${}^4$He and $D/H$ abundances and of 
the cosmological constraints considered by
WMAP team~\cite{WMAPParams}
(CMB + LSS + Lyman~$\alpha$, fit by $\Lambda$CDM model)
give the $2\sigma$ limits $1.7<\Neff<3.0$.   
If a neutrino chemical potential, characterizing $\nu/\bar\nu$
asymmetries\footnote{
  Any initial differences among the individual $\nu/\bar\nu$ asymmetries 
  for the 3~generations of active neutrinos are equilibrated 
  by neutrino oscillations by the time BBN begins,\cts{LunardSmir01,
  Dolgov02,Wong02,Abazajian02}.
}, 
is treated as a free parameter to be marginalized over, 
the limits relax~\cite{Barger03} to $-1.7<\Neff<4.1$.
However, the constraints from BBN and CMB should be combined 
with caution.

One of the reasons is that the probed redshifts are separated
by many orders of magnitude.
The processes that determine the BBN yield of light elements 
extend from the freeze-out of $\nu_{\mu}$, $\nu_{\t}$,
and (shortly after) $\nu_e$ interactions at $z_{\nu\,\rm dec}\sim 10^{10}$
to the fusion of light nuclei at $z_{\rm ns}\sim 4\times 10^8$.
On the other hand, the CMB multipoles up to $l\sim 3000$ 
probe the neutrino density in the redshift range from
$z_{l\,\rm entry}\sim 6\times10^4$ to
$z\eq\simeq 3.2\times10^3$  
(assuming the ``standard'' cosmological 
parameters~\cite{WMAPParams,Tegmark2003_constr}.)
Either the photon entropy or the number of uncoupled 
relativistic species per comoving volume may 
change\footnote{
   The change of the photon entropy density
   is tightly constrained below 
   $z_{\g\, \rm chem\,eq}\simeq (2-5)\times 10^6$,\cts{Wright_etal_CMBdistort,
   HuSilk_CMBdistort,DaneseZotti_CMBdistort},
   by the Planckian shape of the CMB spectrum from COBE~\cite{COBE_mu_gamma}.
   Although energy released into the photon gas at smaller
   redshifts can still be redistributed among the photons by Compton 
   scattering, the photon production rates, from double Compton scattering
   ($e\g\to e\g\g$) and bremsstrahlung ($eN\to eN\g$), 
   become insufficient to change the photon {\it number\/} 
   to its new equilibrium value.
   This would lead to a Bose-Einstein CMB frequency spectrum
   with a non-zero chemical potential; see reviews 
   in\cts{DolgovReview,Ellis_etal92}.
   The present agreement between the BBN- and CMB-derived ratio of baryons to photons 
   is an additional evidence against a large change of the comoving photon entropy density.  
   Of course, the above considerations
   do not limit the change of the energy of 
   the uncoupled relativistic species; 
   see \eg\ct{Dicus_unstable78} for specific scenarios. 
} 
in the considerable span of the universe history from 
BBN to the redshifts probed by CMB. 
The responsible physical mechanisms could be, though are not limited to,
heating from decays of massive particles or fields, 
\eg~\cite{KapTur01,Kaw_decays_PRL,Kaw_decays_PRD,ACH03},
such as expected in thermal inflation~\cite{Randall_therm_inf,Lyth_therm_inf},
or cooling by thermal contact with other species,\cts{entropy_reduct,KapTur01}.

Another reason is that both BBN and CMB constraints depend
on certain properties of the uncoupled relativistic species,
in addition to their total energy density.  For BBN, 
the relevant characteristics include the asymmetry between 
the active neutrinos and their antiparticles, 
their interaction and mixing with other species beyond the standard 
model, and the cosmological expansion rate, which may be affected 
by the density of other uncoupled relativistic particles -- ``dark radiation''
(right-handed neutrinos, Goldstone bosons, moduli, etc.)\
and by more exotic phenomena such as early quintessence, 
non-minimally coupled fields, or extra-dimensional physics.
On the other hand,
the CMB anisotropies and matter clustering do not
discriminate between active neutrinos, their antiparticles, 
and other relativistic degrees of freedom.  
But the dynamics of cosmological perturbations
in the unseen relativistic background becomes important.   
In this paper we focus on the signatures caused by
ultra-relativistic decoupled particles.
Their energy spectrum need not be thermal.
Given initial conditions, their gravitational 
impact on the ``visible'' species indeed depends only on 
their combined energy density, parameterized by the effective 
number of neutrino species~$\Neff$, eq.\rf{Neff_def}.

While the impact of neutrinos on the light element production
at BBN has been studied in detail, 
the neutrino features in the CMB spectra 
are less well established.
Their comprehensive analysis and the investigation of their potential 
for probing the primordial radiation of non-electromagnetic origin
are presented in this paper.
The motivations for exploring these features
and the related constraints independently from the BBN physics
include: verification of the
``standard BBN'' model (SBBN); 
guidance in resolving the tensions between SBBN
predictions and observational estimates of the light 
element abundances (the tension presently exists for ${}^4$He 
but in future is also conceivable for other elements as potentially more 
sensitive experiments with less studied systematics appear);
probing the parameter space of extended BBN models in the directions of 
degeneracies (\eg\ the degeneracy in 
the $\nu$ chemical potential -- $\Neff$ plane~\cite{Barger03});
constraining the models of high energy physics, frequently leading to
decoupled relics, non-standard BBN, particle decays during or after BBN,
or modified cosmological expansion; finally,
clarifying robustness of 
the constraints derived from CMB anisotropies and matter clustering.

The possibility of identifying the background of decoupled 
ultra-relativistic species with CMB, 
sometimes complemented by other cosmological probes,
has been analyzed extensively~\cite{Lopez_et_al_nuCMB,
Bowen,HannestadConstr,HansenConstr,
PierpaoliConstr,PastorConstr,OritoConstr}  
in the past using numerical calculations with Boltzmann integrator codes,
such as CMBFAST or CAMB/CosmoMC~\cite{CMBFAST,CAMB,CosmoMC},
or with simpler codes in that Boltzmann hierarchy
is truncated at the quadrupole order
in a way that mimics free steaming~\cite{HETW99,GDM}.
Some of this work forecasted future constraints 
on the density of relativistic species using likelihood 
(Fisher matrix) analysis of specific experiments.

The authors of\cts{HuSugSmall,HuSugAnalyt}
noted that the CMB modes entering the horizon
in the radiation era are perturbed less and the CDM 
modes more in a model 
with a larger neutrino to photon ratio.  
Later work~\cite{HETW99} stressed the essential role
of neutrino {\it perturbations\/} in 
breaking the degeneracies between $\Neff$ and 
the density of non-relativistic matter, 
set by $\om_m\equiv\Om_m h^2$, either of which affects the redshift 
of radiation-matter equality 
$z\eq+1=\rho_{m,0}/(\rho_{\g,0}+\rho_{\nu,0})$.
Degeneracies between the variation of neutrino 
density and of other cosmological parameters
were studied numerically in\ct{Bowen}.  
This work pointed out that, with a fixed $z\eq$ and fixed 
angular size of the acoustic CMB horizon,
the remaining CMB spectrum variation with $\Neff$
to the third acoustic peak can be practically removed  
by a same sign change in the scalar spectral index~$n_s$,
and that the matter power spectrum breaks this degeneracy.
However, because of the normalization of the calculational 
results by the height of the first acoustic peak, the 
neutrino-induced suppression of CMB anisotropy on small scales 
was explained as increased ISW contribution 
on large scales.   This interpretation propagated into
several later papers. We argue at the end of Sec.~\ref{sec_radmat}
and Sec.~\ref{sbs_Cl} that this interpretation is incorrect.

The Fisher matrix likelihood analysis of\ct{Bowen} 
showed that, prior to WMAP, 
$\Neff$ could not be constrained by CMB alone.
With the WMAP data~\cite{WMAPGen} new
analyses~\cite{PastorConstr,PierpaoliConstr,HannestadConstr} 
set the upper limits  $\Neff\lesssim 7$ or somewhat better
if matter clustering or HST data are included.  
Ref.~\cite{PierpaoliConstr} reported a lower limit~$\Neff>1.6$
with $95\%$ confidence level from WMAP only and $\Neff>1.9$ with
HST data added. 
We find that these constraints can be improved dramatically with 
the future experiments and become comparable and tighter than
those presently derived using the standard BBN model 
from the primordial element abundances.

Recently,\ct{WeinbTensor} considered
the interaction of neutrino perturbations with tensor 
gravitational waves. The problem was reduced 
to an integro-differential equation using the so-called line-of-sight 
solution for free streaming particles, 
derived previously 
in a context of photons~\cite{AbbottSch86,HuSelWhtZal_ComplCMB,ZaldSelj_pol_los}.  
Numerical integration of this equation 
showed that neutrinos suppress the amplitude of 
the gravitational waves entering the horizon in the radiation era 
and of the related $B$-mode of CMB polarization by about $20\%$.
Even on the largest angular scales the neutrino damping of the tensor correlation 
functions is predicted to be close to~$10\%$.

In this work we focus on the more significant and, as of now,
the only accessible to observations scalar\footnote{
  As customary in cosmology, the term ``scalar perturbation''
  denotes the invariance of the perturbation Fourier
  modes with respect to the {\it little\/} 
  rotational group: the axial rotations that do not change a mode 
  wave vector~$\k$.
}
perturbations.  We use an analytic approach. 
It provides the physical insight into the cosmological role of neutrinos
and helps find a quantitatively small but unique signature
of neutrino perturbations, the phase shift, which turns out
to play the primary role in measuring the neutrino background density
with CMB experiments.
The analytical methods developed in this paper are easily
applicable to the tensor sector and give results  
consistent with\ct{WeinbTensor}.

A real-space view of cosmological perturbation dynamics
will be essential for obtaining analytic results
for neutrino perturbations, which can not be modeled by 
a fluid.
Many equations governing perturbation dynamics in the
radiation era are integrated trivially in real space.  This permits
analytic calculations that would seem hopeless in momentum 
space.  A real space analysis of cosmological perturbations
was attempted earlier in\ct{Magueijo92} and applied to CMB anisotropy 
in\cts{Baccigal1,Baccigal2}.
We follow the plane wave formalism developed in\cts{BB_GRF,BB_PRL,SB_Thesis}.

The rest of the paper is organized as follows.
In Sec.~\ref{sec_vars} we introduce a slight modification 
to the classical definition of cosmological perturbation variables.
A consequence is substantial simplification of the evolution equations, 
both for their later solution and conceptual understanding.
In Sec.~\ref{sec_radmat} we set up the notations and evolution
equations for the radiation-matter universe around the time of CMB
decoupling.  Then we study the impact of neutrinos on the evolution
of superhorizon perturbations.
In Sec.~\ref{sec_grf} we review the Green's function formalism 
and apply it to find how neutrino perturbations
influence the CMB and CDM modes that enter the horizon in the radiation era.
A reader not interested in the specifics of the analytic calculations 
can look at their results in the figures of 
Secs.~\ref{sec_radmat} and~\ref{sec_grf} and proceed toward 
the discussion of Sec.~\ref{sec_signatures}.
In Sec.~\ref{sec_signatures} we analyse the neutrino signatures in the CMB
and matter perturbation spectra and either their robustness or degeneracy
to the variation of other cosmological parameters.
In Sec.~\ref{sec_forecasts} we estimate the accuracy of
constraining the effective number of neutrino species from 
some planned or proposed CMB experiments.
We conclude in Sec.~\ref{sec_concl}.

Appendix~\ref{apx_cosmo} reviews the 
   linear cosmological perturbation theory and summarizes the
   properties of the used metric gauges.
In Appendix~\ref{apx_local} we prove that all the matter or metric
   Green's functions in the Newtonian or synchronous gauges
   vanish for growing adiabatic
   perturbations beyond the particle horizon.
Appendix~\ref{apx_ORnu} contains technical calculations
   for Sec.~\ref{sec_grf}.

All the following formulas imply the metric signature $(-1,1,1,1)$. 
Greek indices range from $0$ to~$3$; Latin from $1$ to~$3$.
Dots denote the derivatives with respect to conformal time
$d\t\equiv dt/a$, where $a$ is the cosmological scale factor.
The universe expansion rate with respect to 
conformal time is denoted by $\H\equiv {\dot a}/{a} = aH$,
where $H(z)$ is the proper Hubble expansion rate.

\section{Dynamical perturbation variables}
\lb{sec_vars}

In this paper we use predominantly the conformal Newtonian, later  
``Newtonian'', gauge~\cite{Mukh_Rept,MaBert}
and parameterize scalar metric perturbations as
\be
ds^2=a^2(\t)\lf[(-1-2\Phi)d\tau^2+(1-2\Psi)d\r^2\rt]\,.
\lb{Newt_gauge_def}
\ee
The potential $\Phi$ determines the gravitational acceleration
of free-falling objects ${\bf g}=-\bm{\Nb}\Phi$.
$\Psi$ characterizes the perturbations of spatial curvature 
in this gauge.  (This choice of potentials agrees with\ct{Mukh_Rept}. 
It is related to other frequently cited works as: 
$\Phi_{\rm there}=\Psi_{\rm here}$, $\Psi_{\rm there}=\Phi_{\rm here}$ 
for\cts{MaBert,LLbook}, and $\Psi_{\rm there}=\Phi_{\rm here}$, 
$\Phi_{\rm there}=-\Psi_{\rm here}$ for\cts{KS84,HuSugAnalyt,HuSugSmall}.)
In the presence of anisotropic stress, 
provided by neutrino perturbations, 
the potentials $\Phi$ and $\Psi$ differ from each other.

Occasionally, we invoke the synchronous, spatially flat,
comoving, and uniform density gauges.  
Their definitions and the relations between various metric 
gauges are summarized in part~3 of Appendix~\ref{apx_cosmo}.

\subsection{Coordinate particle number densities}
\lb{sbs_vars}

It appears very useful to describe perturbations of
matter\footnote{
  In Sec.~\ref{sec_vars} 
  ``matter'' refers to all the dynamical degrees of freedom,
  whether in relativistic or non-relativistic particles or fields,
  in contrast to non-dynamical scalar metric perturbations.
}
species in terms of the variables that rate of change
does not depend on the time derivatives of 
other perturbation variables.
This is not so, for example, for the usually considered proper
energy density enhancement~$\d\rho_a/\rho_a$ in the Newtonian gauge, 
nor the proper phase-space density perturbation~$\d f_a$,
nor the local CMB temperature anisotropy~$\d T(\n)/T$, 
nor the ``effective temperature'' perturbation $\d T/T+\Phi$, 
for all of which the corresponding conservation equations involve 
$\dot\Psi$ or $\dot\Phi+\dot\Psi$
terms.   Via the Poisson equation, eq.\rf{Psi_Pois} in
Appendix~\ref{apx_cosmo},
these terms, which are dominant on horizon scales,
bring the time derivatives of other matter perturbations.
This complicates the  description of perturbation evolution.

Instead, we characterize density perturbation of species~$a$
in the Newtonian gauge by a variable
\be
d_a\equiv \fr{\d n_{a,\,\rm coo}}{n_{a,\,\rm coo}} = \d_a - 3\Psi~,
\lb{d_a_def}
\ee
where 
\be
\d_a\equiv \fr{\d n_{a,\,\rm prop}}{n_{a,\,\rm prop}}
    \equiv \fr{\d\rho_a}{\rho_a+p_a}
\ee
is the proper particle number
overdensity\footnote{
  If $n$ is the density of any conserved number, its change
  in a locally inertial frame for a closed volume~$V$ equals
  $dn/n=-dV/V={d\rho}/({\rho+p})$, given 
  the energy conservation $d(\rho V)+pdV=0$.
}.  The latter is related to the energy momentum 
tensor~$T^{\mu\nu}_a$ by eq.\rf{Tmn} in Appendix~\ref{apx_cosmo}.
Unless noted otherwise, we suppose that the matter species in 
any group~$a$ do not interact non-gravitationally
with the species of the other groups; hence $T^{\mu\nu}_a$ is well defined
and covariantly conserved.  Examples of the species groups $a$ are: 
photon-baryon plasma, neutrinos, or cold dark matter (CDM).

The species mean velocity and anisotropic stress,
defined with eqs.\rf{Tmn}, for scalar perturbations can be described
by a velocity potential~$u_a$
as $v_{i\,a} = -\Nbi u_a$ and by an anisotropic 
stress potential~$\pi_a$, eqs.\rfs{Tmn}{pi_def}.
From the energy and momentum conservation equations\rf{dot_du}
and $d_a$ definition\rf{d_a_def},
\be
\dot d_a&=& \Nb^2u_a~,
\lb{dot_d}\\
\dot u_a &=& c_a^2 d_a-\chi_a u_a
               +\Nb^2\pi_a+\Phi+3c_a^2\Psi~,
\lb{dot_u}
\ee
where\footnote{
  Except for the generalized proof of superhorizon conservations in 
  Sec.~\ref{sbs_init}, 
  we will assume that for all the species~$a$ the local pressure
  $p_a$ is uniquely specified by the local energy density $\rho_a$.
  This assumption is general enough to apply to
  photon-baryon plasma, cold matter, massless or massive
  neutrinos, and constant vacuum energy.  It is not valid for
  a classical field (quintessence) or modified Hubble expansion 
  (Cardassian energy.)  Without this assumption eq.\rf{dot_d}
  should be replaced by eq.\rf{dot_d_gen}.
}
$c_a^2\equiv {d p_a}/{d\rho_a}$ 
is the ``sound speed'' and
\be
\chi_a\equiv \H\lf(1-3c_a^2\rt)
\ee
is the Hubble drag rate for the species~$a$.
Eqs.\rf{dot_d} and\rf{dot_u} can be combined into a single 
second order equation
\be
\ddot d_a + \chi_a\dot d_a - c_a^2\Nb^2d_a - \Nb^4\pi_a
          = \Nb^2(\Phi+3c_a^2\Psi)\,.
\lb{ddd}
\ee
For the special cases of CDM and tightly coupled 
photon fluid with negligible baryon density eq.\rf{ddd} reduces to
\be
 &&\ddot d_c + \H\dot d_c = \Nb^2\Phi~\qquad\qquad\qquad\rm(CDM)~;
 \lb{CDM_evol}\\
 &&\ddot d_{\g} - \fr13\,\Nb^2d_{\g} = \Nb^2(\Phi+\Psi)\quad \rm(photon~fluid)~.
 \lb{rad_fl_evol}
\ee

The variables $d_a$ can be interpreted in the Newtonian gauge
as the perturbations of the conserved particle number densities 
with respect to the coordinate differential volume, $d^3\r$, 
rather than the proper volume, $a^3(1-3\Psi)d^3\r$.  
The change of these densities with time is determined 
only by the particle flux into a unit coordinate volume,
eq.\rf{dot_d}.
It does not explicitly depend on the metric evolution, 
affecting the proper number densities~$\d_a$.
More formally, $d_a$~corresponds to the
gauge invariant quantity
\be
d_a=\d_a - 3H_L+\Nb^2\chi~,
\lb{d_a_invar}
\ee
where $H_L$ and $\chi$ parameterize the general perturbation of
spatial metric~$g_{ij}$
as given by eqs.\rfs{ds_gen}{ds_scalar}.
Therefore, the variables $d_a$ coincide with the particle number
overdensities~$\d_a$ in the gauge where the spatial metric is
unperturbed -- the ``spatially flat'' gauge of Sec.~\ref{sbs_gauges}.  
The density perturbations~$d_a$ are simply related to the
``conserved curvature'' 
perturbations of\ct{zeta_a} on the hypersurfaces of uniform 
energy density of species~$a$, $\zeta_a\equiv-H_L^{({\rm uniform},\,a)}$, 
as $d_a=3\zeta_a$.

Likewise, we eliminate $\dot\Psi$ from the 
equations of perturbation dynamics in phase space.
For this purpose we define a variable
$df_a(\t,\r,q,\n)$:
\be
df_a\equiv \d\!f_a + q\,\fr{\,\pd\!\bar f_{a}}{\pd q}\,\Psi~,
\lb{df_def}
\ee
where $\d\!f_a(\t,\r,q,\n)$ is the perturbation of the proper phase 
space distribution for species~$a$, $\bar f_{a}(\t,q)$ is 
their unperturbed density in phase space, and 
the particle comoving momentum $\bm{q}=q\n$ is defined
in Appendix~\ref{apx_cosmo} by eq.\rf{q_def}.
The linearized Boltzmann equation\rf{Boltz_lin} 
in terms of the variable\rf{df_def} reads
\be
\lf(df_a\rt)\dot{\vphantom{t}}
+\fr{q}{\eps}\,n_i\Nbi\lf(df_a\rt)
 = q\,\fr{\,\pd\!\bar f_a}{\pd q}\,n_i\Nbi\lf(\fr{q}{\eps}\,\Psi
   +\fr{\eps}{q}\,\Phi\rt)\,.
\lb{Boltz_com}
\ee
Summation over the repeated index $i=1,2,3$ is implied.
Having in mind the applications to collisionless particles,
we dropped the collision term and the terms involving the time 
derivatives of~$\bar f_a$ and~of~${\pd\!\bar f_a}/{\pd q}$.

The number of phase-space coordinates in the Boltzmann equation can be 
reduced by one,\ct{Lindquist}, when the mass of the particles 
is negligible relatively to their average kinetic energy, so that
$\eps\simeq p$ and the $f_a$ perturbations 
propagate with the same speed, the speed of light, regardless 
of the particle energy.
Defining a function
\be
D_a(\t,\r,\n)\equiv
\fr34\,\fr{\int q^2 dq\,q\,d f_a(\r,q,\n,\t)}{\int q^2 dq\,q\,\bar f_a(q)}~,
\lb{D_def}
\ee
and integrating both sides of eq.\rf{Boltz_com} over $q^3dq$, we find
\be
\dot D_a + n_i\Nbi D_a = - 3n_i\Nbi(\Psi+\Phi)~.
\lb{Bolz_int}
\ee
The variable $D_a(\t,\r,\n)$ is related to the 
energy-averaged phase space distribution perturbation
of\cts{AbbottSch86,HuSelWhtZal_ComplCMB,ZaldSelj_pol_los} 
as $F_a(\t,\r,\n)=(4/3)D_a+4\Psi$.
If the free-streaming particles had the thermal velocity distribution
at their decoupling then the temperature perturbation of the 
particles moving in a specified direction $\n$ is
$\d T_a(\t,\r,\n)/T_a=D_{\!a}/3+\Psi$.

Eq.\rf{Bolz_int} is formally solved by
\be
D_a(\t,\r,\n)&=& D_{a,\,\rm in}(\r-\n\t,\n)\,- \lb{D_losol}\\
 &&-\ 3n_i\Nbi\int_0^{\t}d\t'\lf.(\Phi+\Psi)\rt|_{\t',\,\r-\n(\t-\t')}\,.
\nn 
\ee 
In the following subsection we will show that for adiabatic initial 
conditions $D_{a,\,\rm in}(\r,\n)$ is independent of~$\n$.
It will be related to the conserved
superhorizon value of the spatial curvature perturbation~$\zeta$ 
in the uniform density gauge 
(the ``Bardeen's curvature''~\cite{zeta_orig}) as
\be
D_{a,\,\rm in}(\r,\n) = -3 \zeta\i(\r)~
\lb{D_in}
\ee
for all~$a$.

Any scalar perturbation~$D_a(\t,\r,\n)$ can be described by
scalar multipole potentials~$\{d_{l,a}(\t,\r)\}_{l=0,1,\dots}$ as
\be
D_a(\n)=\sum_{l=0}^{+\infty}(-1)^l\,(2l+1)\,P_l
                \!\lf(\fr{n_i\Nbi}{\Nb}\rt)\Nb^l d_{l,a}~,
\lb{dl_def}
\ee
where $P_l$ are Legendre polynomials.
Since $P_l(\mu)$ contains only the powers~$\mu^{l-2q}$ 
with $q=0,1,\dots\lfloor l/2\rfloor$,
the gradient operator enters the right hand side of eq.\rf{dl_def} only
through natural powers of $n_i\Nbi$ or $\Nb^2$.
The potentials $d_{l,a}$
are gauge invariant for $l\ge 2$.

As follows from eqs.~(\ref{Tmunu_expl},\,\ref{df_def},\,\ref{D_def}),
\be
\d T^{\mu}_a{}_{\nu}=4\rho_a\lf\langle 
   n^{\mu}n_{\nu}\lf(\fr13\,D_a(\n)+\Psi\rt)\rt\rangle_{\n}\,,
\lb{F2Tmunu}
\ee
where $n^0\equiv 1$, $n_0\equiv -1$, $n^i= n_i$, and 
$\langle\rangle_{\n}$ stands for $\int {d^2\Omega_{\n}}/{4\pi}$.  
Substituting 
the multipole expansion\rf{dl_def}, remembering
the definition of the variables $d_a$, $u_a$, and~$\pi_a$, 
eqs.~(\ref{d_a_def},\,\ref{Tmn}--\ref{pi_def}), 
and using that for the ultra-relativistic 
particles $p_a=\rho_a/3$, we find
\be
d_{0,a}=d_a~, \qquad  d_{1,a}=u_a~, \qquad  d_{2,a}=\fr32\,\pi_a~.
\lb{f123}
\ee

$d_{l,a}$ dynamics follows from eqs.\rfd{Bolz_int}{dl_def} 
and the relation $\mu P_l(\mu)=\fr{l+1}{2l+1}P_{l+1}+\fr{l}{2l+1}P_{l-1}$:
\be
\dot d_{l,a}=\fr{l}{2l+1}\,d_{l-1,a}
     +\fr{l+1}{2l+1}\,\Nb^2 d_{l+1,a}
     +\d_{l1}(\Phi+\Psi).\quad
\lb{dot_dl}
\ee
(The Kronecker symbol~$\d_{l1}$ in the last term
should not to be confused with a density perturbation.)
One can write a formal integral solution of these equations by
expanding the line-of-sight solution\rf{D_losol}
over the spherical harmonics.\footnote{
  \lb{mult_fs}
  Namely,
  \be
d_{l,a}(\t,\r)\!\!&=&\!\!3\lf\{-\,\fr{j_l(k\t)}{k^l}\,\zeta\i(\r)\ + \rt.
 \lb{d_losol}
\\
              &&~\quad\lf.+\int_0^{\t}d\~\t~\fr{j'_l(k\~\t)}{k^{l-1}}\,
                   \lf[\Phi(\~\t,\r)+\Psi(\~\t,\r)\rt]\rt\}_{k^2\to-\Nb^2},
\nn
   \ee
  where $j_l$ are spherical Bessel functions.
  This equation is obtained from eqs.\rfs{D_losol}{D_in} 
  and\rf{dl_def} by noting that
  for any analytic function~$f(\r)$
  \be
  f(\r-\n\t)\!&=&\!e^{-\t n_i\Nbi}f(\r)\, =
  \lb{mult_fs_eq}
  \\
            &=&\!\sum_{l=0}^{\infty}(-1)^l(2l+1)\,P_l
		 \lf(\fr{n_i\Nbi}{\Nb}\rt)i^lj_l(-i\Nb\t)\,f(\r)~.
  \nn
  \ee
  In eq.\rf{d_losol}, the operators ${j_l(k\t)}/{k^l}$ and 
  ${j'_l(k\t)}/{k^{l-1}}$, with $k^2\to-\Nb^2$, are
  well defined as their Taylor expansions involve only even powers
  of~$k$, hence, only integer powers of the Laplace operator $\Nb^2$.  
  If the perturbations in eq.\rf{d_losol}
  refer to a single spatial harmonic plane wave
  with a wave-vector~$\k$ then $-\Nb^2$ does become~$k^2$.
}

The scalar metric perturbations $\Phi$ and $\Psi$ 
are determined
from the linearized Einstein equations, eqs.\rfs{nb2_psi}{psi-phi}.
In terms of the introduced dynamical variables,
\be
\Nb^2\Psi-3\g\Psi&=&\g\lf(d+3\H u\rt)~,
\lb{Psi_eq}\\
\Psi-\Phi&=&3\g\pi~,
\lb{Phi_eq}
\ee
where $\g \equiv 4\pi Ga^2(\rho+p)$ and
\be
d=\sum_a x_a d_a~,\quad 
u=\sum_a x_a u_a~,\quad
\pi=\sum_a x_a \pi_a~,
\ee
where $x_a\equiv({\rho_a+p_a})/({\rho+p})$.
Eqs.\rfs{Psi_eq}{Phi_eq} for $\Psi$ and $\Phi$
are non-dynamical {\it constraint\/} equations.   
These constraints, however, do not limit one's
freedom of setting the initial {\it matter} distribution potentials~$d_{l,a}$.

The above equations of matter dynamics, \eg\
eqs.\rft{CDM_evol}{rad_fl_evol}{Bolz_int} for the model 
with dark matter, photon fluid, and neutrinos,
that are complemented by the elliptic gravitational 
equations\rfs{Psi_eq}{Phi_eq} form a well posed closed system.
They are the basis for our subsequent analytical analysis 
of the perturbation dynamics.

\subsection{Superhorizon conservation of the coordinate number densities}
\lb{sbs_init}

To illustrate the conservation laws prior to their general
derivation, we start from the model where 
all the scalar dynamical (matter) degrees of freedom can be
described by scalar potentials~$d_{l,a}$ that satisfy the
evolution equations of the form~(\ref{dot_d}--\ref{dot_u},\,\ref{dot_dl}).
Below, $\l$ is a characteristic comoving scale of perturbation 
variation in space.  For a harmonic mode it can be taken as $\l=1/k$.
Since in the subsequent sections we are interested in the universe evolution 
long after the inflation, we first 
assume that the comoving Hubble scale~$\H^{-1}(\t)$
grows throughout all the considered time.
Then it is natural to choose the zero of time~$\t$
at the formal limit of the equations $\H^{-1}\to 0$. 
This choice will be implied whenever we refer to 
a specific $\t$ value, in the context of post-inflationary
expansion.
All of the above assumptions will be lifted by the end of the section.

We set the initial conditions at a time $\t\i\ll\l$ as
$d_{l,a}\sim (\t\i)^l$.  Such initial conditions are natural
for {\it growing\/} modes, where perturbations
are finite for $\t\to0$. 
If the global intrinsic curvature~$K$ does not dominate the
Friedmann expansion and the universe does not inflate
($w>-1/3$) then, by eqs.\rfd{dH}{g_Fr}, 
$\H\sim 1/\t$ and $\g\sim 1/\t^2$.
Then for our initial conditions and $\t\ll\l$
we can drop all the $\Nb^2$ terms in the evolution equations
up to an error~$O(\t^2/\l^2)$.  We thus find
\be
\vsp&&\dot d_a\simeq 0~,
\lb{dot_d_super}\\
&&\dot u_a \simeq c_a^2\lf(d_a+3\H u_a+3\Psi\rt)
         -\H u_a + \Phi~,
\lb{dot_u_super}\\
\vsp&&\dot d_{l,a}\simeq {\textstyle\fr{l}{2l+1}}\,d_{l-1,a}\qquad
\mbox{(free stream. $l\ge2$)}~,
\lb{dot_dl_super}\\
\vsp&&d+3\H u + 3\Psi \simeq 0~,
\lb{Psi_eq_super}\\
\vphantom{\fr11}&&\Psi-\Phi = 3\g\pi~.
\lb{Phi_eq_super}
\ee
The corresponding solutions scale for $\t\i\ll\t\ll\l$
as $d_{l,a}\sim \t^l$ and $\Phi\sim\Psi\sim 1$.
Specifically, the coordinate particle number density
perturbations $d_a$ are {\it constant\/}, up to $O(\t^2/\l^2)$
corrections.
Adiabaticity of the initial perturbations
has {\it not\/} been assumed for this result.

If the initial conditions are {\it adiabatic\/}\footnote{\lb{def_adiab}
  We define a perturbation as ``adiabatic''
  (a curvature perturbation) if in some space-time coordinates 
  all the proper matter distributions or fields
  and their proper rate of change, 
  all smoothed over an arbitrary comoving scale~$\lambda$,  
  appear unperturbed in the limit 
  $\t/\l\to 0$, $\l=\const$. 
  (Without the assumption $\H\sim1/\t$, 
   the relevant limit is~\cite{WeinbPrivCom} $\H^{-1}(\t)/\l\to0$, 
   $\l=\const$.)
  In this, ``superhorizon'', limit 
  the space-time metric in the considered coordinates remains perturbed.
  We prove in Appendix~\ref{apx_local} that for any wave-vector~$\k$  
  there exists 
  a non-decaying perturbation mode with such properties.
  This mode satisfies the conditions of adiabaticity
  of\ct{WeinbAdiab} and\ct{LLbook}:
  As shown in this subsection, 
  under very mild natural
  assumptions about the internal matter dynamics, the corresponding comoving 
  curvature perturbation~$\R$ is constant beyond the horizon,
  up to $O(\t^2/\l^2)$ corrections.
  Since after a coordinate change\rf{gauge_var} the proper 
  energy density and pressure of species~$a$ appear perturbed
  as $\d\rho_a=\dot\rho_a\d\t$ and $\d p_a=\dot p_a\d\t$
  in the linear order,
  all the ratios $\d\rho_a/\dot\rho_a$ and $\d p_a/\dot p_a$
  in any other gauge in the above limit are equal.
}, 
from eq.\rf{gauge_tr},
the density and velocity perturbations of all the
species in the Newtonian gauge are related in the limit 
$\t/\l\to 0$, $\l=\const$ as
\be
\d_a=-3\H u_a = \d \qquad ({\rm adiabatic},~\t/\l\to 0)~,
\lb{del_a_adiab_eq}
\ee
where $\d$ is a same function of $\t$ and~$\r$
for all the species~$a$.
Also, by eq.\rf{d_a_def},
\be
\quad d_a=d  \qquad\quad ({\rm adiabatic},~\t/\l\to 0)~
\lb{d_a_adiab_eq}
\ee
with $d=\d-3\Psi$ for all~$a$.
Then the terms in eq.\rf{dot_u_super} that are proportional to $c_a^2$
cancel by eq.\rf{Psi_eq_super}.  Hence, all the velocity potentials
evolve identically on superhorizon scales 
in the leading and next to the leading orders in $\t/\l$:
\be
\quad \dot u_a \simeq  -\H u_a + \Phi  
\quad\qquad ({\rm adiabatic},~\t\ll\l)~.
\lb{u_a_adiab_eq}
\ee
The same is true for all~$d_a=d$, which remain constant by eq.\rf{dot_d_super}.
The leading order evolution of the $l\ge2$ multipoles depends on the
internal dynamics of the species.  For example, these multipoles
vanish identically for perfect fluids but they grow for neutrinos.
Nevertheless, the given definition of adiabaticity demands
that $d_{l,a} \to 0$ in the $\t\to 0$ limit.
This is consistent with our previous observation that, 
for the ``growing mode'' initial conditions,
eq.\rf{dot_dl_super} yields $d_{l,a}\sim\t^l$.

In addition to the coordinate particle number density perturbation~$d$,
two other perturbation variables are known to be constant 
for adiabatic perturbations on superhorizon scales.  These are the
spatial curvature perturbation in the comoving 
gauge
\be
\R = \Psi + \H u~,
\ee
eqs.\rfd{comov_curv_def}{comov_curv_st},\cts{Bard80,Lyth85,LLbook,WeinbAdiab}, 
and the curvature perturbation 
in the uniform density gauge $\zeta$, eq.\rf{zeta_def}~\cite{zeta_orig}.
(The sign of $\R$ and $\zeta$ in this paper coincides with
the sign of the Newtonian potentials.  Most of the references use
the opposite sign.)
Since the scalar metric perturbations are not dynamical
but are fixed by matter perturbations with constraint 
equations, one may expect a simple relation between 
the conserved matter perturbation~$d$ and the metric
perturbations~$\R$ or~$\zeta$.
Indeed, the comparison of eq.\rf{d_a_def} and eq.\rf{zeta_def}
gives immediately that
\be
d=-3\zeta~.
\lb{d2zeta}
\ee
It also follows from eqs.\rfd{curv_related}{gammaH} that
\be
\R=\zeta+O\lf(\fr{\t^2}{\l^2}\rt)~.
\lb{R_from_zeta_superhor}
\ee

Thus, up to $O({\t^2}/{\l^2})$ corrections,
for the growing adiabatic perturbations
and all the species~$a$,
\be
d_a(\t\ll\l,\r) = -3\zeta\i(\r) ~\qquad(\rm adiabatic)~,
\lb{d_a_superhor}
\ee
where $\zeta\i(\r)$ is the time-independent superhorizon value
of the curvature perturbation~$\zeta$.
Substituting this result in eq.\rf{dl_def}
and remembering that for $l\ge1$ $d_{l,a}\sim\t^l\to0$,
we also find
\be
D_a(\t\ll\l,\r,\n) = -3\zeta\i(\r) ~\qquad(\rm adiabatic)~.
\lb{D_init}
\ee
The conservation of $D_a(\n)$ or, for non-interacting
particles with non-negligible mass, of~$df_a(q,\n)$ in 
eq.\rf{df_def} is also evident from
eq.\rf{Bolz_int} or\rf{Boltz_com}, in which all the terms
with gradients can be dropped for superhorizon growing
perturbations by the same arguments as before.

Eq.\rf{ddd} assumed that local pressure of the species
is uniquely determined by their local energy density.
Without this assumption,
for all the mutually non-interacting groups~$a$ of matter
species or fields, the energy conservation
$T^{0\mu}_a{}_{;\mu}=0$ gives
\be
\dot d_a = \Nb^2u_a + 
    \fr{\dot\rho_a\d p_a-\dot p_a\d\rho_a}{(\rho_a+p_a)^2}~.
\lb{dot_d_gen}
\ee
(We emphasize that $T^{\mu\nu}_a$ and all the variables in eq.\rf{dot_d_gen}
are defined Appendix~\ref{apx_cosmo} independently of the nature
of the species self-interaction.)
As previously, $\Nb^2u_a$ can be 
dropped for growing superhorizon perturbations.  
Since energy density and pressure perturbations transform
under a gauge transformation\rf{gauge_var} as 
$\d\~\rho_a=\d\rho_a+\dot\rho_a\d\t$ and $\d\~p_a=\d p_a+\dot p_a\d\t$, 
the additional term on the right hand side
of eq.\rf{dot_d_gen} is gauge invariant.  
If for a considered group of species~$a$, which may interact among 
themselves but do not couple to the other species,
the energy density and pressure become homogeneous as
$\t/\l\to0$ in some coordinate frame, this term is initially zero 
in any frame.
We call such initial conditions for the species~$a$ ``internally adiabatic''.
For them, all the right hand side of eq.\rf{dot_d_gen} vanishes
and $d_a$ is constant beyond the horizon.  
Generalization of the arguments that follow eq.\rf{dotdz_mostgen} 
gives that $d_a$ starts changing only in the order~$O(\t^2/\l^2)$.

If all the species~$a$ are perturbed internally 
adiabatically, which is automatic for single-component perfect 
fluids, then we showed that all the~$d_a$ are constant for growing modes 
beyond the horizon.  However, the variables~$d_a$
need not be equal for different~$a$'s if the {\it overall\/}
perturbation is not adiabatic.  In this case
\be
\zeta=-\,\fr13\, d =-\,\fr13\sum_a x_a d_a
\lb{zeta_superhor}
\ee
in general changes outside of the horizon as the species enthalpy
abundances~$x_a$, eq.\rf{x_def}, vary during the
expansion.   This is essentially the curvaton 
mechanism of\ct{curvaton_orig}, see\ct{curvaton} for a 
modern version, converting isocurvature into curvature perturbations.

Most generally, any system with locally interacting 
matter and Einstein gravity possesses a covariantly
conserved energy-momentum tensor $T^{\mu\nu}$.  Therefore,
the  scalar perturbation variables~$d=\d-3\Psi$ and $u$ 
are always well defined with eqs.\rfs{Tmn}{u_def}.
From the covariant conservation $T^{0\mu}{}_{;\mu}=0$,
\be
  \dot d = \Nb^2u+\fr{\dot\rho\,\d p-\dot p\,\d\rho}{(\rho+p)^2}~.
\lb{dotdz_mostgen}
\ee
In Appendix~\ref{apx_local} we show that 
if a growing adiabatic perturbation is initially localized 
in a spatial region then all the
matter and gravitational  Newtonian gauge potentials,
including~$u$,
vanish beyond the particle horizon of this region.
Then, by the Gauss's theorem, the velocity divergence
term in eq.\rf{dotdz_mostgen} has zero integral over any 
volume enclosing the particle horizon.
For the initial conditions that are adiabatic
as defined in footnote~\ref{def_adiab}, the gauge invariant quantity 
$\dot d_{\rm non-ad}\equiv({\dot\rho\,\d p-\dot p\,\d\rho})/{(\rho+p)^2}$
tends to zero in the limit $\t/\l\to0$, $\l=\const$.
If, motivated by either the equivalence principle or analyticity
of the system dynamics,
we accept that the metric perturbations generate $\dot d_{\rm non-ad}$
only in $O(\t/\l^2)$ order (in the frame where the matter 
is initially unperturbed but so in any other frame
because of the gauge invariance of $\dot d_{\rm non-ad}$)
then in the leading and 
next to the leading orders in $\t/\l$ the variables $d$ and 
$\zeta=-d/3$ are constant.  So is constant the comoving gauge curvature 
perturbation~$\R$, eq.\rf{R_from_zeta_superhor},  
up to $O(\t^2/\l^2)$ deviations.

If a stage of inflation, defined as cosmic expansion with 
positive acceleration $d^2a/dt^2$,
hence {\it contracting\/} comoving Hubble scale $\H^{-1}(\t)$, 
is also considered, the condition $\t/\l \to 0$ should 
be replaced by $\H^{-1}(\t)/\l\to 0$,~\ct{WeinbPrivCom}.
Then this limit and the related
definition of adiabaticity in footnote~\ref{def_adiab}
in the inflating universe apply 
to the {\it future\/} rather than the past.

Finally, what
about the universe evolution that begins with an inflationary
stage and proceeds to the canonical Big Bang?
\ct{WeinbAdiab} defines adiabaticity and considers the conservation 
laws in the limit~$\l\to\infty$.
However, the initial conditions for modes with  
finite wavelengths can be different from those
for the infinitely large scales.  
To accommodate such physically viable possibilities,
we define the adiabatic initial conditions for any {\it fixed\/} 
finite spatial scale~$\l=1/k$ 
on the stage of increasing $\H^{-1}(\t)$
by formally evolving the perturbation with the post-inflationary 
equations backward in time to $\H^{-1}\to0$. 
(Similarly, by evolving forward in time to $\H^{-1}\to0$
during the inflation.)    
This approach allows one to quantify 
the primordial non-adiabaticity (admixture of isocurvature modes),
which can be probed by CMB or matter spectra,
at {\it any\/}~$k$.
The lesser the primordial non-adiabaticity and 
the ``tidal'' $O(\H^{-2}/\l^2)$ dynamical 
deviations are, the better the proved conservation laws apply.

\section{Radiation-Matter universe}
\lb{sec_radmat}

If the primordial perturbations are nearly adiabatic,\footnote{
 The following argument and its conclusion do not apply to
 isocurvature initial conditions,
 considered recently for neutrinos in\ct{Gordon:2003hw}.
}
the inhomogeneities in the neutrino background may
affect only those CMB and CDM perturbations that 
entered the horizon while the radiation fraction of the universe 
energy was non-negligible. 
As for the perturbation modes with larger wavelengths,
the number density perturbations~$d_a=-3\zeta\i$ remain frozen 
and the higher angular multipoles $d_{l\ge1,a}$ of the photon 
and matter distributions negligible until the horizon entry.
 By the time these modes 
enter the horizon and the species distributions start evolving,
the neutrino energy density perturbations are too small to have
a gravitational impact on their evolution.

Thus in much of this work we will be interested in the 
perturbation dynamics at the redshift~$z\gtrsim 10^3$.
Barring the possibilities
of noticeable early quintessence~\cite{EarlyQuint} or 
the Cardassian modification of Friedmann expansion~\cite{Cardassian,Dvali_mod},
the background expansion at that time can be described by 
a flat radiation-matter model.   The radiation energy density
is provided by CMB photons and neutrinos,
which mass becomes dynamically relevant only at 
$z\lesssim m_{\nu}/(3kT_{\nu,0})\simeq 200\,m_{\nu}/(0.1\eV)$ 
and is neglected here.
(The published WMAP~\cite{WMAPParams} $95\%$ CL limit
on the neutrino masses is $m_{\nu}<0.23\eV$.) 
The massive matter consists of cold dark matter,~$c$, 
and baryons,~$b$.  

We begin from establishing
notations convenient for the radiation-matter model.
The linearized gravitational equations, \eg\rfs{Psi_eq}{Phi_eq}
or\rfs{nb2_psi}{psi-phi}, involve the reduced enthalpy 
background density $\g\equiv 4\pi Ga^2(\rho+p)$, eqs.\rfs{ga_def}{g_Fr}.
For baryons and CDM, with $\rho_a=\rho_{a,0}/a^3$
and  negligible pressure, it equals
\be
\g_{b(c)}=\fr{4\pi G \rho_{b(c),0}}{a}
        =\fr{3H_{\rm ref}^2\om_{b(c)}}{2a}~,
\lb{g_cb}
\ee
where $H_{\rm ref}\equiv 100$\,km\,s$^{-1}$Mpc$^{-1}$ and 
$\om_{b(c)}\equiv \Om_{b(c)} h^2$.
The present WMAP constraints~\cite{WMAPParams} 
on $\om_b$ and $\om_m\equiv \om_b+\om_c$,
assuming the standard neutrino content, 
are $\om_b=0.024\pm 0.001$ and $\om_m=0.14\pm0.02$.

The reduced enthalpy of photons, $\g_{\g}=(16/3)\pi Ga^2\rho_{\g}$,
at a given redshift $1/a$ is fixed by 
today's CMB temperature~$T_{\g,0}=2.725\pm0.002\,K$,\ct{COBE_Tcmb}, 
as
\be
\g_{\g}=\fr{16\pi G \rho_{\g,0}}{3a^2}
       =\fr{2H_{\rm ref}^2\om_{\g}}{a^2}~,
\ee
where $\om_{\g}\equiv \Omega_{\g}h^2\approx 2.47{\times}10^{-5}
(T_{\g,0}/2.725\,K)^4$.
For neutrinos,
\be
\g_{\nu}=\al_{\nu}\g_{\g}~,~~~
\al_{\nu}\equiv \fr{\rho_{\nu}}{\rho_{\g}}
         \equiv \fr78\lf(\fr4{11}\rt)^{\fr43}\!\!\Neff
         \simeq 0.23\,\Neff\,,
\lb{al_nu_def}
\ee
where the standard Big Bang Nucleosynthesis predicts 
$\Neff\approx 3.04$, assuming~$3$ Standard Model neutrino 
generations and zero
neutrino chemical potential.  The effective
number of neutrino species $\Neff\not=3$, 
first, because neutrinos share some of the 
energy of~$e^+e^-$, annihilating soon after the  
neutrino decoupling peak,\ct{Dicus_NuID}.  
Second, because this energy,
most of which heats the photons after the annihilation, 
is somewhat reduced by finite temperature 
QED corrections~\cite{Heckler_NuQED,LopezTurner_NuQED}.
The physics of both effects is concisely reviewed
in\ct{Steigman_Neff}.

Of course, here we allow $\Neff$ to be a free parameter.
It characterizes the energy density of all the decoupled ultra-relativistic  
species at the considered moment, 
implied being after $e^+e^-$ annihilation but before 
CMB decoupling,
\be
\Neff\equiv {\rho_{\rm rel\ decoup}}\lf/
       \textstyle\lf[{\fr78\lf(\fr4{11}\rt)^{\fr43}\rho_{\g}}\rt]\rt..
\lb{Neff_def}
\ee

This parameter may have a non-standard value due to either
an unaccounted change of $\rho\snu$ or $\rho_{\g}$,
or due to the density of additional 
weakly interacting ultra-relativistic species~($X$).  
The latter would presumably decouple at very high temperature 
when the universe contained
more relativistic degrees of freedom than during the neutrino 
decoupling.  As the particles such as heavy leptons, 
hadrons, $W$ and $Z$ bosons, Higgs fields, 
superparticles, etc.\  become non-relativistic and annihilate, 
the entropy shared by the coupled photons, electrons, and neutrinos increases.  
Since the comoving entropy density of the decoupled species $X$
remains unchanged, their contribution to the parameter\rf{Neff_def} may become 
substantially below unity.  A light field carrying $g_X$
effective degrees of freedom, with the fermionic ones
multiplied by $7/8$, that decoupled when the remaining particles
in thermal contact had $g(T_{<X\,\rm dec})$ degrees of freedom 
contributes to the ratio\rf{Neff_def} as
\be
\D\Neff = \fr47\ g_X
       \lf[\fr{g(T_{<X\,\rm dec})}{g_{\g e\nu}}\rt]^{\fr43}~,
\ee
were $g_{\g e\nu}=43/4$.
So, for example, a hypothetical neutral Majorana fermion 
or a scalar Goldstone boson that decoupled when the 
remaining relativistic degrees of freedom were composed of 
all the fields of the minimal supersymmetric standard model 
($g(T_{<X\,\rm dec})=915/4$) would give $\D\Neff\simeq 1.7\times 10^{-2}$
and $\D\Neff\simeq 9.7\times 10^{-3}$ correspondingly.

It is sometimes convenient to use the ratio of 
comoving time~$\t$ to a characteristic time
of the radiation-matter energy equality, and the ratio
of scale factor~$a$ to its value at the equality:  
\be 
\bar\t\equiv \fr{\t}{\t_e}~,\qquad  \bar a\equiv \fr{a}{a\eq}~,
\lb{xy_def}
\ee
where
\be
a_{\rm eq}&=& \fr{\lf(1+\al_{\nu}\rt)\om_{\g}}{\om_m}\ \approx
\lb{e_eq_def}
\\
          &\approx& \fr1{3.5\cdot10^3} 
                  \lf(\fr{1+\al_{\nu}}{1.69}\rt)
                  \lf(\fr{0.3\times0.7^2}{\Omega_mh^2}\rt)\,,
\nn
\ee
and
\be
\t_e&\equiv& \fr{\t_{\rm eq}}{2(\sqrt2 - 1)}             
          = \fr1{H_{\rm ref}}\sqrt{\frac{a_{\rm eq}}{\om_m}}\ \approx
\lb{t_e_def}
\\
     &\approx& 130\,\mbox{Mpc}
             \,\sqrt{\fr{1+\al_{\nu}}{1.69}}
             \lf(\fr{0.3\times0.7^2}{\Omega_mh^2}\rt)\,.
\nn              
\ee
The Friedmann equation for the radiation-matter universe
in terms of the variables\rf{xy_def} reads
$(d\bar a/d\bar\t)^2=1+\bar a$, yielding
\be
\bar a=\bar\t+\fr14\,\bar\t{}^2~.
\lb{y_x}
\ee

We find it useful to introduce the variable
\be
r\equiv \fr{\bar\t}{\bar a} = \fr1{1+\fr14\,\bar\t}
              = \fr2{1+\sqrt{1+\bar a}}~.
\lb{r_def}
\ee
Note for reference that $\t=4\t_e(1-r)/r$, $d\t=-4\t_e dr/r^2$, 
and $a=4a_{\rm eq}(1-r)/r^2$.
In terms of~$r$,
\be
 \H=\fr{2-r}{\t}~,
\qquad \g_{\nu}=\fr{2R_{\nu}r^2}{\t^2}~,
\lb{wH_r}
\ee
where
\be
\Rnu\equiv \fr{\rho_{\nu}}{\rho_r}
            = \fr{\al_{\nu}}{1+\al_{\nu}}
\lb{Rnu_def}
\ee
is the neutrino fraction of the total radiation energy density
$\rho_r= \rho_{\g}+\rho_{\nu}$; $\Rnu\approx0.408$ for
$\Neff=3.04$.
The formulas describing the superhorizon perturbation modes,
Sec.~\ref{sec_superhor}, become very
compact if the mode evolution is parameterized by~$r$.
For example, see eq.\rf{Phi_k0} for the well known $k=0$ 
growing mode of the gravitational potential in the radiation-matter 
neutrinoless model,\ct{KS84}.
The radiation and matter domination limits of these formulas
are easily read off by setting $r$ to $1$ and~$0$ correspondingly.

\subsection{Perturbations in the radiation era}

When the universe energy density is dominated 
by photon gas and ultra-relativistic neutrinos,
for all of which $w_a=c_a^2=\fr13$,
then
\be
\H^{(r)}=\fr1{\t}~,\qquad \g^{(r)}=\fr2{\t^2}~.
\lb{radlim}
\ee
In the radiation era $x_{\g}^{(r)}=1-R_{\nu}$, $x_{\nu}^{(r)}=R_{\nu}$,
and $x_a^{(r)}$ for any non-relativistic species is negligible.
The evolution equations\rf{rad_fl_evol},\rf{Bolz_int},\rfs{Psi_eq}{Phi_eq}
in this regime become
\be
\ddot d_{\g} - \fr13\,\Nb^2d_{\g} &=& \Nb^2(\Phi+\Psi)~,\vsp
\lb{gevol_rad}\\
\dot D_{\nu} + n_i\Nbi D_{\nu} &=& - 3n_i\Nbi(\Psi+\Phi)~,\vsp
\lb{nuevol_rad}\\
\t^2\Nb^2\Psi- 6\Psi &=& 2d+\fr6{\t}\,u~,\vsp
\lb{Psi_eq_rad}\\
\Phi &=& \Psi - \fr{6R_{\nu}\pi_{\nu}}{\t^2}~.
\lb{Phi_eq_rad}
\ee

On the scales well inside the acoustic horizon, $\lambda\ll\t/\sqrt3$,
the gravitational terms on the the right hand side of 
eqs.\rfd{gevol_rad}{nuevol_rad} are negligible.
The oscillating photon acoustic modes
and the free-streaming neutrinos decouple from each other.
However, the phase and the amplitude of the acoustic oscillations 
is set by the perturbation dynamics during the horizon 
entry, when the gravity of neutrino perturbations played a significant role.
Likewise, the neutrino perturbations
affect the dark matter peculiar velocities, evolving
according to eq.\rf{CDM_evol}, as the matter received a gravitational
boost from the radiation when the fluctuations entered the horizon 
in the radiation era.

If the neutrino density is negligible ($\Rnu\to0$)
then the above equations have
compact analytic solutions in either Fourier space, 
eqs.~(\ref{Phi0k},\,\ref{d_g_rad_sol_k},\,\ref{dc_trf0}),
or real space, 
eqs.~(\ref{Php_0},\,\ref{d_g_rad_sol_x},\,\ref{d_c_rad_sol_x}).
When the neutrino gravity is appreciable, it appears difficult to track 
the evolution of the Fourier modes analytically 
through the horizon entry. 
But in Sec.~\ref{sec_grf} we succeed with  
an analytical approach in real space.

\subsection{Perturbations in the matter era}
\lb{subsec_md}

When massive matter,
with pressure and anisotropic stress being
negligible comparatively with its energy density,
becomes the dominating component then
by eqs.\rf{wH_r} and\rf{g_Fr}
\be
\H^{(m)}=\fr2{\t}~,\qquad \g^{(m)}=\fr6{\t^2}~.
\lb{matlim}
\ee
Provided the energy density perturbations are also dominated by
massive matter, giving negligible $\pi$, 
by eq.\rf{Phi_eq} the gravitational potentials $\Phi$ and $\Psi$ 
are equal.  After baryons decouple from CMB photons
at $z_d\sim1090$, we can also take $\d p\approx 0$. 
Then by eq.\rf{ddot_psi}
\be
\ddot\Phi+3\H\dot\Phi = 0~.
\ee
The corresponding non-decaying solutions 
are time independent on all scales. 
The constant $\Phi$ and~$\Psi$
modes that enter the particle horizon
after~$z_d$ are easily related to the epoch-independent
primordial curvature perturbation~$\zeta\i$.
Indeed, by applying eq.\rf{comov_curv_st} to superhorizon scales,  
where $\R=\zeta\i$ by eq.\rf{curv_related}, one finds
\be
\Phi^{(m,\l\gg\H^{-1}_{z_d})}=\Psi^{(m,\l\gg\H^{-1}_{z_d})}
                 = \fr{3}{5}\,\zeta\i~.
\lb{super_mat0}
\ee

When $\dot\Phi=\dot\Psi=0$,
we can rewrite eq.\rf{Bolz_int}
for ultra-relativistic neutrinos or 
for photons after their decoupling as
\be
\dTeff_a + n_i\Nbi\Teff_a =0~,
\lb{Teff_eq}
\ee
where
\be
\Teff_a(\t,\r,\n)\equiv\fr13\,D_{a}+\Phi+\Psi~.
\lb{Teff_def}
\ee
(If the free-streaming particles were in local thermal equilibrium 
at their decoupling then $\Teff_a(\t,\r,\n)$ is the 
relative temperature perturbation of the particles 
propagating in the direction~$\n$ plus the gravitational redshift correction: 
$\Teff_a=\d T_a(\t,\r,\n)/T_a+\Phi$.)
For superhorizon adiabatic perturbations in the matter era
$\Teff_a(\t{\,\ll\,}\l,\r,\n)=-\zeta\i+\Phi+\Psi
=\fr13\,\Phi(\r)$, eqs.\rfd{D_in}{super_mat}.
The corresponding, ``Sachs-Wolfe''~\cite{1967ApJ...147...73S}, 
solution of eq.\rf{Teff_eq} at later time~is
\be
\Teff_a(\t,\r,\n) = \fr13\,\Phi(\r-\n\t)~.
\lb{T_m_sol}
\ee
The related multipole potentials~$d_{l,a}$, eq.\rf{dl_def}, 
for a single Fourier harmonic 
$\Phi_{\k}(\r)=$\,Re\,$\lf(A\,e^{i\k{\cdot}\r}\rt)$
follow from eqs.\rfd{Teff_def}{T_m_sol} and
eq.\rf{mult_fs_eq} in footnote\rfp{mult_fs} as
\be
d_{a} = 
\lf[\fr{\sin(k\t)}{k\t}-6\rt]\Phi_{\k}~,\quad
d_{l\ge1,a} = \fr{j_l(k\t)}{k^l}~\Phi_{\k}~.~
\lb{fl_m_free}
\ee

The evolution equations for the perturbations of other species also
have simple analytic solutions when the gravitational potentials 
are time independent.
We do not write these solutions 
here because the linear perturbation dynamics  
in the matter era has been thoroughly studied in the past, and 
neutrinos, while relativistic, do not modify it.
Of course, the power spectra of perturbations in the matter era
are affected by neutrinos through 
the change of the effective initial conditions
for the modes that entered the horizon prior to matter domination.
We discuss this modification of CMB and matter power spectra 
in Sec.~\ref{sec_signatures}.

\subsection{Superhorizon Scales}
\lb{sec_superhor}

The $l\ge2$ multipoles~$d_{l,a}$ of the phase space
distributions for the non-relativistic CDM and 
baryons are negligible in the linear regime.
They are also small for CMB photons, 
isotropized by scattering prior to hydrogen recombination.
The integral solution for the multipoles
of free-streaming particles, eq.\rf{d_losol} in footnote~\ref{mult_fs},
shows that for neutrinos on superhorizon scales
$d_{l,\nu}\sim\t^l \zeta\i$. 
Particularly, the neutrino anisotropic stress potential
$\pi_{\nu}=\fr23d_{2,\nu}$ is of the order 
of~$\t^2\zeta\i$.  Then, by eqs.\rfd{Phi_eq}{wH_r},
$\Psi-\Phi=3\g_{\nu}\pi_{\nu}\sim R_{\nu}r^2\zeta\i$.
Hence, in the radiation era, when  $r\to1$,
the neutrino anisotropic stress leads to splitting 
the Newtonian gauge potentials~$\Phi$ 
and $\Psi$ even on superhorizon scales,\ct{MaBert}.
The effect disappears in the matter and later eras, after
the energy-momentum tensor of neutrinos becomes 
negligible comparatively to that of non-relativistic species.

In Sec.~\ref{sbs_init} we showed that for superhorizon 
growing adiabatic perturbations $d_a = -3\zeta\i=\const$ for all the matter
and radiation species.  The superhorizon evolution of the $l\ge1$ multipoles
$d_{l,a}$ and gravity is described by
eqs.~(\ref{u_a_adiab_eq},\,\ref{dot_dl_super}--\ref{Phi_eq_super}) 
of Sec.~\ref{sbs_init}.   In that section we also observed that
for adiabatic perturbations all the velocity potentials~$u_a$
are equal, up to $O(\t^2/\l^2)$ corrections, to the 
momentum-averaged velocity potential~$u$.  Then, by eqs.\rfd{dot_dl_super}{f123},
\be
\qquad
\dot\pi_{\nu}=\fr4{15}\,u~~\qquad(\t\ll\l)~.
\lb{dot_pi_super}
\ee
Combining this with
eqs.~(\ref{u_a_adiab_eq},\,\ref{Psi_eq_super}--\ref{Phi_eq_super},\,\ref{d2zeta})
we find a closed equation
\be
\quad
\ddot\pi_{\nu}+2\H\dot\pi_{\nu}+\fr45\g_{\nu}\pi_{\nu}
   =\fr4{15}\,\zeta\i
\qquad(\t\ll\l)~.
\lb{pi_super}
\ee
Its non-decaying numerical solution for $N\snu=3$
is plotted in \fig{fig_superhor}\,a) with the solid line.
\begin{figure*}[tb]
\includegraphics[width=16cm]{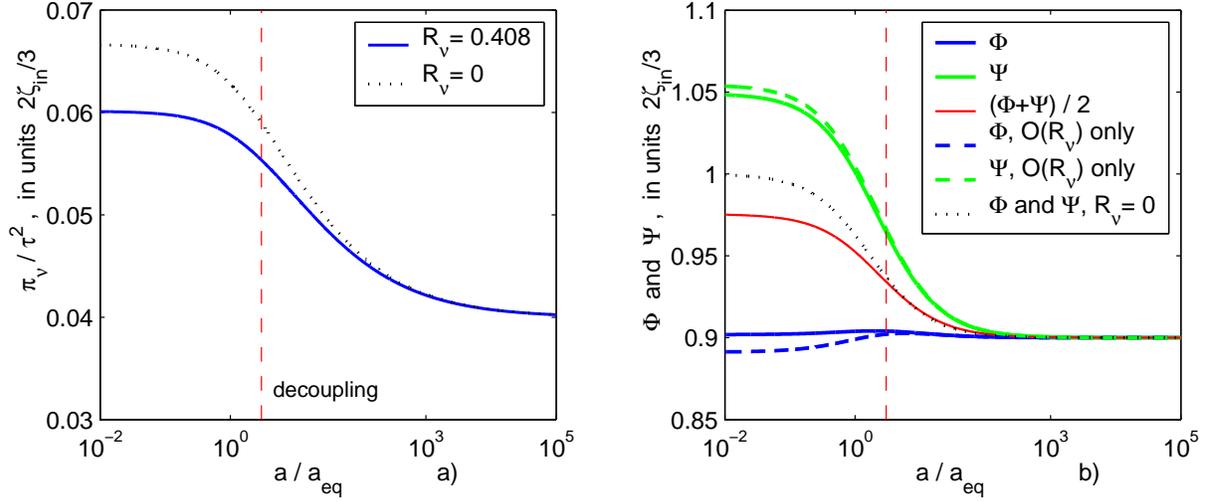}
\caption{The evolution of superhorizon adiabatic perturbations in the
 radiation-matter universe.  a) Neutrino anisotropic stress potential.
 b) The Newtonian gauge gravitational potentials.  On both plots,
 the solid curves show the full result for 3 neutrino
 species.  The dotted curves correspond to 0 neutrino species.
 The dashed curves on plot b) are the sums of the leading and
 subleading terms in the expansion of the potentials in~$\Rnu$, 
 for 3~neutrino species.
 The dashed vertical lines show $a/a\eq$ at CMB decoupling,
 given the cosmological parameters of\ct{WMAPGen}.
}  
\lb{fig_superhor}
\end{figure*}
Given the neutrino anisotropic stress potential~$\pi_{\nu}$, 
the superhorizon value of the velocity potential~$u$
can be calculated from eq.\rf{dot_pi_super}.  The 
gravitational potentials~$\Phi$ and $\Psi$ are then obtained 
from eqs.\rfd{u_a_adiab_eq}{Phi_eq}.  The potentials 
corresponding to the $N\snu=3$
numerical solution of eq.\rf{pi_super}
are plotted in \fig{fig_superhor}\,b) with the solid lines.

We observe that, first, coincidentally, 
the Newtonian potential~$\Phi$ is almost unchanged 
during the radiation-matter transition if $N\snu\approx 3$.
Second, when radiation is dynamically significant,
the sum $\Phi+\Psi$ is smaller
in a universe with a larger effective number of neutrinos.
This sum governs the propagation of CMB photons,
as seen from eq.\rf{ddd} with $c_a^2=\fr13$.
The following analytical analysis quantifies these observations.

In the {\it radiation\/} era $\H^{(r)}=1/\t$, $\g_{\nu}^{(r)}=2R_{\nu}/\t^2$,
and the growing mode solution of eq.\rf{pi_super} is
\be
\pi_{\nu}^{(r,\,\t\ll\l)}=\fr{\t^2}{15+4R_{\nu}}\,\fr{2\zeta\i}{3}~.
\lb{pi_nu_rad}
\ee
Hence, from eqs.\rft{dot_pi_super}{u_a_adiab_eq}{Phi_eq},
\be
&&\Phi^{(r,\,\t\ll\l)} = \fr1{1+\fr4{15}\,R_{\nu}}\lf(\fr{2\zeta\i}{3}\rt)~,\qquad
\lb{super_rad_Phi}
\\
&&\Psi^{(r,\,\t\ll\l)} = \lf(1+\fr25\,R_{\nu}\rt)\Phi^{(r,\,\t\ll\l)}~.
\lb{super_rad}
\ee
Relation\rf{super_rad} between the potentials in the 
radiation era was previously derived in\ct{MaBert}.

In the {\it matter\/} era $\H^{(m)}=2/\t$ and the $\g_{\nu}\pi_{\nu}$
term in eq.\rf{pi_super} is negligible.  Then
\be
\pi_{\nu}^{(m,\,\t\ll\l)}=\fr{\t^2}{25}\,\fr{2\zeta\i}{3}~
\lb{pi_nu_mat}
\ee
and we obtain the conventional result:
\be
\Phi^{(m,\,\t\ll\l)} = \Psi^{(m,\,\t\ll\l)} = \fr{3}{5}\,\zeta\i~,
\lb{super_mat}
\ee
\cf eq.\rf{super_mat0}.

In the intermediate regime eq.\rf{pi_super} 
has no simple exact solution.  
But the physics of the superhorizon perturbation 
dynamics in the presence of neutrino 
anisotropic stress can we 
studied analytically by expanding the solution in the powers of~$R\snu$.
The calculations in the zeroth and the first orders
are straightforward and are given below.

In the zeroth order, 
\ie\ when the neutrino fraction $\Rnu$ is negligible, the gravitational potentials 
$\Phi$ and $\Psi$ are equal.  Then using eqs.\rfd{u_a_adiab_eq}{Psi_eq_super}
and remembering that on the superhorizon scales
$d=-3\zeta\i$ we have
\be
\fr1{a^2}\lf(a^2u^{(R_{\nu}\to0)}\rt)\odot=\zeta\i~.
\ee 
This relation is easily integrated when $r$ of 
eq.\rf{r_def} is taken for the evolution variable:
\be
u^{(R_{\nu}\to0)} = \fr{\t\lf(6+3r+r^2\rt)}{30}\,\zeta\i~.
\lb{u_nu0k0}
\ee
The gravitational potentials then follow from eq.\rf{Psi_eq_super}
as
\be
\Phi^{(R_{\nu}\to0)}=\Psi^{(R_{\nu}\to0)}
   = \fr1{10}\lf(9+\fr{r^2}2+\fr{r^3}{2}\rt)\fr{2\zeta\i}{3}~.
\lb{Phi_k0}
\ee
This fluid limit solution, known
in more lengthy forms before from\ct{KS84},
is plotted in Fig.~\ref{fig_superhor}\,b) with the dotted line.
The anisotropic stress potential~$\pi_{\nu}$ of a trace
amount of neutrinos can be found by the integration of
eqs.\rfd{dot_pi_super}{u_nu0k0} as
\be
\pi_{\nu}^{(R_{\nu}\to0)}
  = \fr{\t^2 f_{\pi}(r)}{15}\,\fr{2\zeta\i}{3}~,
\lb{pi_nu0}
\ee
where
\be
f_{\pi}(r)\equiv \fr15\,\lf[\fr{3-r^2}{1-r}+\fr{2r^2\ln r}{(1-r)^2}\rt]~.
\nn
\ee
The function~$f_{\pi}(r)$ is plotted in  \fig{fig_f_r}\,a).
Its radiation ($r\to 1$) and matter ($r\to 0$) 
era limits are $1$ and $\fr35$ correspondingly. 
In Fig.~\ref{fig_superhor}\,a) we compare 
the leading order solution $\pi_{\nu}^{(R_{\nu}\to0)}/\t^2$,
dotted curve, with the previously found
numerical solution for $N_{\nu}=3$.
\begin{figure*}[tb]
\includegraphics[width=16cm]{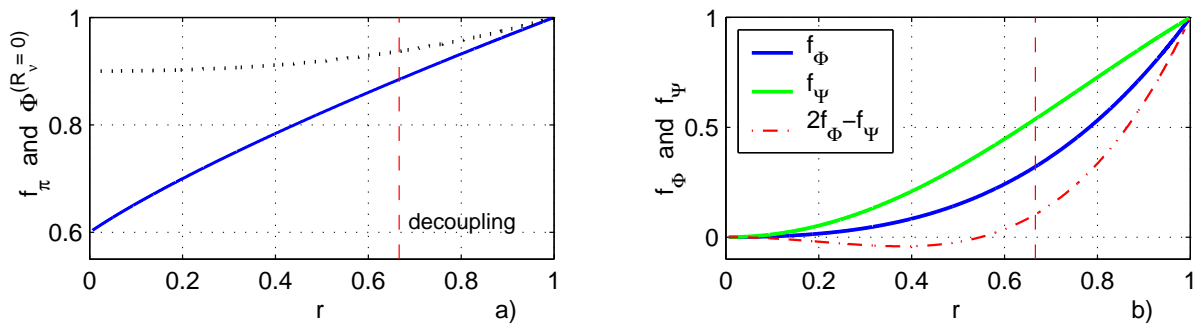}
\caption{ The functions $f_{\pi}$, $f_{\Phi}$, $f_{\Psi}$, and 
  the combination $2f_{\Phi}-f_{\Psi}$ that appear in the $O(\Rnu)$ order of
  superhorizon perturbation
  evolution, as considered in the main text.  
  The evolution variable $r$ is defined by eq.\rf{r_def}.
  The radiation density domination corresponds to $r\to1$, the matter 
  density domination to $r\to0$.
}  
\lb{fig_f_r}
\end{figure*}

To determine the $O(\Rnu)$ terms in
the gravitational potentials, we rewrite
eqs.\rfd{dot_pi_super}{pi_super} as
\be
\fr1{a^2}\lf(a^2u\rt)\odot = \zeta\i - \fr{6\Rnu r^2}{\t^2}\,\pi_{\nu}~,
\lb{u_eq_super}
\ee
where we substituted the result\rf{wH_r} for $\g_{\nu}$.
The $O(\Rnu)$ solution for the velocity potential is obtained by 
using the $O(1)$ solution\rf{pi_nu0} for $\pi_{\nu}$
on the right hand side of eq.\rf{u_eq_super}.
After its integration,
\be
u=u^{(R_{\nu}\to0)}-\fr{4\t\zeta\i}{45}\,f_u\Rnu
         +O(\Rnu^2)~,
\ee
with
\be
f_u(r)\equiv \fr{3r^2}{10}\lf[
         5\,\fr{(2-r)f_{\pi}-1}{(1-r)}+1-r\rt]. 
\nn 
\ee
Hence, by eqs.\rfs{Psi_eq_super}{Phi_eq_super},
\be
 \ba{lcr}
\Phi &=& \dst\Phi^{(R_{\nu}\to0)} - \fr{8\zeta\i}{45}\,f_{\Phi}\Rnu+O(\Rnu^2)~,\Vsp \\
\Psi &=& \dst\Phi^{(R_{\nu}\to0)} + \fr{4\zeta\i}{45}\,f_{\Psi}\Rnu+O(\Rnu^2)~,\Vsp 
 \ea
\lb{OR_Grav}
\ee
where
\be
f_{\Psi}(r)\equiv (2-r)\,f_u~, \qquad
f_{\Phi}(r)\equiv \fr{3r^2f_{\pi}-f_{\Psi}}{2}~. 
\nn
\ee
All the functions $f_u$, $f_{\Phi}$, and $f_{\Psi}$ 
tend to~$1$ in the radiation era limit $r\to 1$,
and to~$0$ in the matter era limit $r\to 0$.
$f_{\Phi}(r)$ and $f_{\Psi}(r)$ 
are plotted with solid curves in \fig{fig_f_r}\,b).

The dashed lines in \fig{fig_superhor}\,b) show
the sums of the leading
and subleading terms in the analytic solutions\rf{OR_Grav} 
with $\Rnu$ set to its standard value $0.408$, 
corresponding to $\Neff=3.04$.
As seen from the plots, the $O(\Rnu)$ approximations
describe the main features of the numerical solutions rather well.
The about $11\%$ smaller than predicted splitting between the 
potentials $\Psi$ and $\Phi$ in the radiation era 
corresponds to $4R_{\nu}/15\approx 11\%$ smaller actual value of 
the anisotropic stress $\pi_{\nu}$ than it is given by 
the leading order formula\rf{pi_nu0}, \cf \fig{fig_superhor}\,a).

By eqs.\rfd{rad_fl_evol}{Bolz_int}, both after the photon decoupling 
and when the baryon loading is negligible prior to the decoupling, 
the photon dynamics is affected only by the sum~$\Phi+\Psi$.
This sum depends on the neutrino abundance as 
\be
\fr{\pd(\Phi+\Psi)}{\pd\Rnu}=
   -\,\fr{4\zeta\i}{45}\lf[2f_{\Phi}(r)-f_{\Psi}(r)\rt]+O(\Rnu)~.
\ee
The combination $2f_{\Phi}-f_{\Psi}$ is plotted in
\fig{fig_f_r}\,b) with the dash-dotted line.
It vanishes at $r\simeq 0.55$, corresponding to 
$a/a\eq\simeq 6.0$, and has a small negative value 
at smaller redshifts.

The photon density perturbation $d_{\g}$ on superhorizon scales 
remains constant and independent of $\Phi$ and $\Psi$ evolution.  
Thus on the scales entering the particle horizon 
at the redshifts $z\lesssim z\eq/6$
(for $\omega_m=0.14$ and 3~neutrinos, these scales exceed the
acoustic horizon at recombination threefold or more) 
the potential variation and the induced by it ISW effect 
in the CMB temperature anisotropy are little affected
by neutrino perturbations.  Even more so, 
the background and the adiabatic perturbations of
relativistic species play no role in the {\it late\/} ISW effect,
caused by the global potential decay during the universe transition 
from matter to dark energy domination.
By this time, their energy density is dynamically irrelevant.

\section{Studying radiation era with Green's functions}
\lb{sec_grf}

The evolution of cosmological perturbations 
in the linear regime may be studied by superposing 
perturbative solutions (Green's functions)
that are localized in real space.
It is convenient to consider the Green's functions
that vary with only one spatial coordinate~\cite{BB_GRF}, 
say~$x$.
They are related to the Fourier space perturbation modes
by one-dimensional Fourier transformation.
For example, for the curvature perturbation~$\zeta$,
\be  
\zeta(\t,x)= \int^{+\infty}_{-\infty}\fr{dk}{2\pi}\,e^{ikx}\,
                  \zeta(\t,k)~.
\lb{phi1k}
\ee 
Normalizing the Fourier modes to 
\be
\zeta(\t\to 0,k)= \zeta\i~,
\lb{ic_k}
\ee
where $\zeta\i$ is a $k$-independent constant,
we see from eq.\rf{phi1k} that
\be
\zeta(\t\to 0,x)= \zeta\i\,\dd(x)~
\lb{ic1}
\ee
($\dd(x)$ denotes the Dirac delta function.)
Thus the considered Green's function describes
the linear evolution of a sheet-like curvature perturbation 
that was created on the whole plane $x=0$ and 
is independent of the $y$ and $z$ coordinates.

The initial ratios of the perturbations for different
species should be specified as well.
In our analysis the initial conditions are assumed 
adiabatic, but the Green's function method can be
generalized to incorporate admixture of 
isocurvature perturbations.
The right hand side of eq.\rf{ic_k}
could also be chosen as~$A(ik)^n$ where $A$ is constant and 
$n$ is assumed natural.  The resulting Green's functions would 
also be initially localized, with 
$\lim_{\t\to 0}\zeta(\t,x)=A\,\dd^{(n)}(x)$.
These initial conditions are even or odd with respect to the parity
transformation $x\to-x$ if $n$ is even or odd correspondingly.
If the initial conditions for the relative species perturbations
have the same parity, this parity will be preserved for all~$\t$.
It is convenient to impose {\it even} initial conditions,
as implied by eq.\rf{ic1}, for {\it adiabatic\/} perturbations 
and {\it odd} for {\it isocurvature\/} 
ones.

We discuss only the ``growing'' mode
Green's functions, corresponding to growing Fourier modes in
eq.\rf{phi1k}.  The decaying solutions of the evolution equations
are irrelevant if the primordial perturbations were generated
many e-folding before the scales of our interest entered the 
horizon.

The following two observations prove extremely handy
in calculating the Green's function.  First, applying the
inverse Fourier transformation to eq.\rf{phi1k} 
and setting $k$ to $0$, one finds the following simple connection
between the integral of a Green's function over all the space
and the superhorizon Fourier modes of the same variable:
\be
\int^{+\infty}_{-\infty}\!{dx}~
                  \zeta(\t,x) = \zeta(\t,k\to0)~,
\lb{sum_rules}
\ee
or an analogous relation for any other perturbation variable.
As it was shown in Sec.~\ref{sbs_init}, for adiabatic
perturbations the right hand side of a sum rule such as eq.\rf{sum_rules}
is time independent for the curvature or density perturbations 
$\zeta$, $\R$, $d_a$, $D_a$, or $df_a$.  
It vanishes for the growing adiabatic perturbations of all
the $l>0$ multipole potentials.
In the radiation era, the $k\to0$ limit for any other 
perturbation, \ie\ involving gravitational potentials,
is trivially calculable for any neutrino density from the results
of Sec.~\ref{sec_superhor}.

Second, the adiabatic Green's functions with even initial 
conditions\rf{ic1} identically vanish beyond the particle horizon
of the original perturbation, $|x|>\t$, 
for all the considered perturbation variables in the Newtonian gauge,
including the potentials $\Phi$ and $\Psi$.
This result is proven in Appendix~\ref{apx_local}.
This is a non-trivial statement, taking into account that 
the $l\ge1$ multipole and gravitational potentials 
are not locally measurable physical quantities
and their dynamics is not necessarily causal.\footnote{
  For gravitationally interacting {\it perfect fluids\/}  
  the linear evolution of the Newtonian gravitational potentials
  turns out to be causal, as it can be shown by 
  generalizing the formalism of\ct{BB_GRF} to arbitrary fluids.
  This is not true in general.
  As a simple counterexample, consider an absolutely {\it inelastic} 
  collision at $x=0$ of two identical sheets  
  of ultra-relativistic particles that are orthogonal 
  to the $x$~axis and move 
  toward each other with opposite velocities.  
  At the moment of the collision the system anisotropic stress at $x=0$ 
  disappears. This changes $\Psi-\Phi$ instantly throughout all the space.
}

\subsection{Green's functions for phase space distributions}

Neutrinos and decoupled photons are described by their
distributions $f_a(\t,\r,q,\n)$ in the 
phase space $(\r,\bm{q}{\,\equiv\,}q\n)$. 
A {\it scalar\/} perturbation of~$f_a$ 
that does not depend on $y$ and $z$ coordinates must also be independent of 
$n_y$ and $n_z$: $\d f_a=\d f_a(\t,x,q,\mu)$, where $\mu\equiv n_x$.
For ultra-relativistic free streaming particles, the energy-averaged distribution 
$D_a(\t,x,\mu)$, eqs.\rfd{D_def}{df_def}, 
satisfies the transport equation
\be
\dot D_a + \mu\Nb D_a =  -3\mu\Nb(\Psi+\Phi)~,
\lb{Bolz_int_real}
\ee
where $\Nb\equiv \pd/\pd x$, \cf eq.\rf{Bolz_int}.
The corresponding multipole potentials~$d_{l,a}(\t,x)$,  
eq.\rf{dl_def}, equal
\be
(-1)^l\Nb^l d_{l,a} = \int_{-1}^1\fr{d\mu}{2}\,P_l(\mu)\,D_a(\mu)~.
\lb{fl_inv}
\ee

To illustrate the application of Green's functions to neutrino
dynamics, we first 
consider the free streaming of massless particles in a time-independent 
gravitational potential, as it is the case in the matter era,
Sec.~\ref{subsec_md}.
The transport equation satisfied by 
the effective temperature Green's function
$\Teff_a(\t,x,\mu)\equiv \fr13\,D_{a}+\Phi+\Psi$, 
eq.\rf{Teff_eq}, becomes
\be
\dTeff_a + \mu\Nb\Teff_a = 0~. 
\lb{free_eq_real}
\ee
Given the initial conditions 
$\Teff_a(0,x,\mu) = \Theta\i\dd(x)$,
it is solved by
\be
\Teff_a(\t,x,\mu) = \Theta\i\dd(x-\mu\t)~.
\ee 
The multipole potentials 
$d_{l,a}$ corresponding to $D_{a}=3\lf(\Teff_a-\Phi-\Psi\rt)$
follow immediately from eq.\rf{fl_inv}.
Remembering the definition of the Legendre polynomials
\be
P_l(\mu)\equiv  \fr1{2^l l!}\,\fr{d^l{(\mu^2-1)^l}}{d\mu^l}~,\qquad 
\nn   
\ee
we find
\be
d_{l,a}\! = \fr{3\Theta\i\t^{l-1}}{2^{l+1}l!}
      \lf[1-\lf(\fr{x}{\t}\rt)^{\!2}\rt]^l\!
              \theta(\t-|x|) - 3\d_{l0}(\Phi+\Psi)\,,~~
\lb{fl_free_grf}
\ee
where $\theta$ is the Heaviside step function.
In the matter era $\Phi=\Psi=\Phi\i\dd(x)$,
$\Theta\i=\Phi\i/3$, and the Fourier transform of the above 
formula reproduces the modes of eq.\rf{fl_m_free}.

\subsection{Neutrino distributions}
\lb{sbs_grf_nu}

In the {\it radiation\/} era 
the evolution equations\rfs{gevol_rad}{Phi_eq_rad} can be converted
from partial into ordinary differential equations with respect to
a dimensionless variable 
\be
\chf\equiv \fr{x}{\t}~.
\ee
Indeed, the growing modes of such perturbations
as $d_{\g}$, $D_{\nu}$, $\Phi$, or $\Psi$ during 
radiation domination have the form $f(\t,k)=f(k\t)$.
The corresponding Green's functions, \cf eq.\rf{phi1k}, 
scale as
\be
 \ba{ll}\dst
\Phi(\t,x)=\fr1{\t}\,\bar\Phi(\chf)~,&\dst \Vsp
\Psi(\t,x)=\fr1{\t}\,\bar\Psi(\chf)~,\\ \dst
d_{\g}(\t,x)=\fr1{\t}\,\bar d_{\g}(\chf)~,&\dst \Vsp
D_{\nu}(\t,x,\mu)=\fr1{\t}\,\bar D_{\nu}(\chf,\mu)~.
 \ea
\lb{selfsim} 
\ee
All the first or second partial derivatives of
a function $f(\t,x)=\bar f(\chf)/\t$ are complete $\chf$~derivatives, 
which we denote by primes, times some power of~$\t$:
\be
 \ba{ll}
 \dot f = - (\chf\bar f)'/\t^2~,&\quad      \Nb f = \bar f'/\t^2~,\Vsp\\
\ddot f = (\chf^2\bar f)''/\t^3~,&\quad \Nb\dot f = - (\chf\bar f)''/\t^3~,\\
                                 &\quad \Nb^2  f = \bar f''/\t^3~.\Vsp
 \ea	
\lb{selfsim_der} 
\ee
The powers of~$\t$ can be canceled out of all the terms
in the evolution equations\rfs{gevol_rad}{Phi_eq_rad}. 
For future references, let us note that
\be
f(\t,k)=\int_{-\infty}^{+\infty}\!\!
                d\chf\,e^{-ik\t\chf}\,\bar f(\chf)~.
\lb{FTbarf}
\ee

We define 
\be
\Phi_+\equiv \fr{\Phi+\Psi}{2}~,\qquad
\Phi_-\equiv \fr{\Psi-\Phi}{2}~.
\lb{Phpm_def}
\ee
The gravitational potential $\Phi_-$ is sourced directly
by neutrino anisotropic stress as described by eq.\rf{Phi_eq_rad}.  
On the other hand,
the motion of photons in the radiation era 
and of neutrinos, eqs.\rfd{gevol_rad}{nuevol_rad}, 
is affected by~$\Phi_+$ only.

Applying differentiation rules\rf{selfsim_der} to
the neutrino transport formula\rf{Bolz_int_real}, 
we obtain an easily integrable equation
\be
\lf[(\chf-\mu)\,\bar D_{\nu}\rt]' = 6\mu\bar\Phi_+'~.
\ee
Since the Green's functions vanish for $|\chf|>1$,
\be
(\chf-\mu)\,\bar D_{\nu} = 6\mu\bar\Phi_+~.
\lb{oiewcnbdf}
\ee
Eq.\rf{oiewcnbdf} does not constrain $\bar D_{\nu}$
at $\chf=\mu$.  It is satisfied~by
\be
\bar D_{\nu}(\chf,\mu)= p_{\nu}(\mu)\,\dd(\chf-\mu)
                        +\fr{6\mu}{\chf-\mu}\,\bar\Phi_+(\chf)~
\lb{Theta_grf_rad}
\ee
with any function $p_{\nu}(\mu)$.  (Even when $\Phi_+$
and so the right hand side of eq.\rf{Bolz_int_real}
are identically zero, there are
non-zero $D_{\nu}$~solutions
that describe free streaming neutrinos in Minkowski
space.)

The function $p_{\nu}(\mu)$ in eq.\rf{Theta_grf_rad} 
must be fixed by the initial conditions.
Considering $\t k\to0$ limit of relation\rf{FTbarf}
and remembering eq.\rf{D_init}, 
for any $|\mu|\le1$ we find  
\be
\int_{-1}^1 d\chf\,\bar D_{\nu}(\chf,\mu)
   = D_{\nu}(\t k\to 0,\mu) = -3\zeta\i~.
\lb{p_fixing}
\ee
Substituting solution\rf{Theta_grf_rad} 
into the left hand side of the above identity,
\be
p_{\nu}(\mu)= -3\zeta\i
    -\int_{-1}^1 d\chf\,\fr{6\mu}{\chf-\mu}\,\bar\Phi_+(\chf)~.
\lb{f_nu_eq}
\ee

For the multipoles
\be
\bar D_{l,\nu}(\chf)\equiv \int_{-1}^1\fr{d\mu}{2}\,P_l(\mu)\, \bar D_{\nu}(\chf,\mu)
\nn
\ee
eqs.\rfd{Theta_grf_rad}{f_nu_eq} give:
\be
&&\!\!\!\!\!\!\!
\bar D_{l,\nu}(\chf)=-3\lf[\fr12\,\zeta\i P_l(\chf)\,+\rt.
\lb{D_l_sol_real}
\\
&&\!\!\!\!\!
\lf.+\int_{-1}^1 d\chf'~\fr{\bar\Phi_+(\chf')\,\chf\,P_l(\chf)
        +\bar\Phi_+(\chf)\,\chf' P_l(\chf')}
        {\chf'-\chf}
       \rt]\theta(1-|\chf|)\,.
\nn
\ee
In \fig{fig_grf}\,a) the solid line shows
the neutrino density perturbation $\bar D_{0,\nu}(\chf)=\bar d_{\nu}(\chf)$
that is obtained from this equation and
the potential~$\bar\Phi_+$ in the limit $\Rnu\to0$,
eq.\rf{Php_0}, when the integrals in eq.\rf{D_l_sol_real}
are easily taken.

\subsection{A note on generalized functions}

The expressions under the integrals in eq.\rf{f_nu_eq}
or\rf{D_l_sol_real} are singular at $\chf'=\chf$.  
The value of the integrals depends on how the singularity 
is treated during the 
integration.   Physically, this ambiguity corresponds to
resolving the last, divergent, term
in eq.\rf{Theta_grf_rad} outside of an interval 
$\chi\in [\mu-\eps_1,\mu+\eps_2]$
and approximating the $\bar D_{l,\nu}$ structure 
inside the interval by the first, $\delta$-function term
in eq.\rf{Theta_grf_rad}.
The most direct approach is to take the integral
Cauchy's principal values, implying $\eps_1=\eps_2\to0$.

Soon we will encounter the integrals of even more 
divergent expressions, such as $x^{-2}$ or $x^{-2}\ln x$,  
for that even the Cauchy's principal value  
does not exist.  Nevertheless, we can proceed with their
meaningful calculation if all the singular expressions
are understood as {\it generalized functions\/}.
A detailed mathematical treatment of the
latter can be found in\ct{Jones_GenFunc}.
The physical meaning of these calculations is clarified
by the following two theorems.  The first one
states that any generalized function is a finite order 
generalized derivative of a continuous function.\footnote{
  To be precise, any finite order derivative of a continuous 
  function~$f(x)$ with $|f|$ bounded
  as $|x|\to\infty$ by a finite power of~$|x|$
  defines a generalized function and, conversely,
  any generalized function can be presented as 
  such a derivative.
}
The second that differentiation of a generalized 
function~$f(x)$ multiplies its Fourier 
components~$f(k)$ by~$ik$.
Green's functions can be formally defined as Fourier 
integrals of perturbation modes, eq.\rf{phi1k}.
Generalized functions provide a consistent, elegant
formalism for their manipulation even when the integrals
diverge in the Riemann's sense.   One could avoid the divergences 
by working only with sufficiently smooth potentials of 
singular real space perturbations, \eg, considering $\Phi$
instead of $\d\rho\propto\Nb^2\Phi$, on small scales.
But the use of generalized Green's functions simplifies
and streamlines the calculations.

\begin{table}[t]
\begin{center}
\begin{tabular}{c|c}
$  f(x)    $&$  \fp\int_{x_1}^{x_2}dx\,f(x)  \Vsp$\\
\hline 
$\quad \dst  \fr1{(x-a)^{n+1}}\,,~~n=1,2,\dots    \quad $&$\dst \qquad 
                     -\lf.\fr{1}{n\,(x-a)^n}\,\rt|_{x_1}^{x_2}   \qquad\Vsp$\\
$\dst   \fr1{x-a} \qquad   $&$\dst  
                    \lf.\ln|x-a|\,\vsp\rt|_{x_1}^{x_2}   \Vsp$\\
$\dst   \fr{\ln|x-a|}{(x-a)^{n+1}}\,,~~n=1,2,\dots     $&$\dst
   -\lf.\fr{\ln|x-a|+\fr1n}{n\,(x-a)^n}\,\rt|_{x_1}^{x_2}   \Vsp$\\
$\dst   \fr{\ln|x-a|}{x-a} \qquad     $&$\dst    
                     \fr12\lf.\ln^2|x-a|\,\vsp\rt|_{x_1}^{x_2}  \Vsp$
\end{tabular}
\caption{The integrals of singular generalized functions.
   Additional valid formulas are obtained by simultaneously 
   multiplying the expressions on the left and on the right 
   by $\sign(x-a)$.}
\lb{FPtable}
\end{center}
\end{table}
Table~\ref{FPtable} gives the definite integrals of several
singular functions, interpreted as generalized functions.
This table is easily understood by noting that the
generalized functions corresponding to the expressions
on the left are defined as the derivatives of 
the less singular expressions on the right.
Of course, the generalized integration agrees with
the conventional one on any interval on that
the Riemann integral exists.

\begin{table*}[t]
\begin{center}
\begin{tabular}{c|c}
$  \bar f(\chi)=\int_{-\infty}^{+\infty}\!
                \fr{d\f}{2\pi}\,e^{i\f\chi}\,f(\f)    $&$ 
    f(\f)=\int_{-\infty}^{+\infty}\!
                d\chi\,e^{-i\f\chi}\,\bar f(\chi)   \Vsp$\\
\hline
$  \d^{(n)}(\chi-a)\,,\ n=0,1,2,\dots     $&$    (i\f)^n e^{-ia\f}   \Vsp$\\
$   \sign\chi        $&$   \dst \fr{2}{i\f}  \Vsp$\\
$    (\chi-a)^{-n},\ n=1,2,\dots     $&$  
                   \fr{(-i)^n\pi}{(n-1)!}\,\f^{n-1}\sign\f~e^{-ia\f} \Vsp$\\
$    \chi^{-n}\sign\chi\,,\ n=1,2,\dots   
      $&$   
 -\,\fr{2(-i)^{n-1}}{(n-1)!}\,\f^{n-1}\lf[\ln|\f|-\psi(n)\rt] \Vsp$\\
$  \ln|\chi|     $&$    -\pi\lf[\fr1{|\f|}+2\g\d(\f)\rt]  \Vsp$\\
$  \chi^{-n}\ln|\chi|\,,\ n=1,2,\dots       $&$  
   -\,\fr{(-i)^n\pi}{(n-1)!}\,\f^{n-1}\sign\f\lf[\ln|\f| - \psi(n)\rt]  \Vsp$\\
$  \chi^{-n}\sign\chi\ln|\chi|\,,\ n=1,2,\dots       $&$  
   \fr{(-i)^{n-1}}{(n-1)!}\,\f^{n-1}\lf\{\lf[\ln|\f|-\psi(n)\rt]^2-\fr16\pi^2-\psi'(n)\rt\} \Vsp$
\end{tabular}
\caption{The Fourier transforms of singular generalized functions.
  The derivation of these results can be found in\ct{Jones_GenFunc}.
  The values of~$\psi(n)$, the logarithmic derivative of the gamma function,
  for a natural argument
  follow recursively from $\psi(n+1)=\fr1n+\psi(n)$ and 
  $\psi(1)=-\g$, where $\g\simeq 0.5772$ is the Euler's constant.
  In every case, shifting the transformed function argument by a constant~$a$,
  as displayed in the first and third lines, multiplies the Fourier image
  by~$e^{-ia\f}$.}
\lb{FTtable}
\end{center}
\end{table*}
Table~\ref{FTtable} lists the Fourier transforms 
of singular generalized functions.
It is useful to remember that the Fourier image of an even
real function is even and real and of an odd real function
is odd and imaginary.

\subsection{Gravitational potentials}
\lb{sbs_grf_pot}

Now we turn to the linearized Einstein equations and solve them
consistently with the dynamical equations for the relativistic
matter.
First, we differentiate eq.\rf{Phi_eq_rad}  
twice to obtain
\be
\bar\Phi_-''= 2\Rnu\bar D_{2,\nu}~,
\lb{eqcls}
\ee
where we applied the last of eqs.\rf{f123}.
The perturbation~$\bar D_{2,\nu}(\chf)$ on the right hand side
is given in terms of the potential $\bar\Phi_+$
by eq.\rf{D_l_sol_real} with $l=2$.

Second, we note that for 
adiabatic perturbations in the radiation era $\d p /\d\rho =1/3$.
Then we can easily eliminate all the matter perturbations from
Einstein eqs.~(\ref{nb2_psi},\,\ref{ddot_psi}--\ref{psi-phi})
to find
\be
\ddot\Psi-\fr23\,\Nb^2\Psi+\fr13\,\Nb^2\Phi
+\fr1{\t}\lf(3\dot\Psi+\dot\Phi\rt)=0~.
\ee
Using eqs.\rf{selfsim_der}, we obtain a relation
that can be integrated once trivially, giving
\be
\lf(\chf^2-\fr13\rt)\bar\Phi'_+
+\lf(\chf^2-1\rt)\bar\Phi'_- 
-2\chf\bar\Phi_+=0~.
\lb{eqgrav}
\ee

The general solution of the above equation is
\be
&&\qquad\fr{\bar\Phi_+(\chf)}{\chf^2-\fr13}=\const - F_-(\chf)~,
\lb{Php_sol_indef}
\\
&&F_-(\chf)\equiv \fp\int_{-1}^{\chf}d\chf'\,
        \frac{\chf'^2-1}{\lf(\chf'^2-\frac13\rt)^2}\,\bar\Phi_-'(\chf')~.
\lb{Fm_def}
\ee
The integration constant on the right hand side of eq.\rf{Php_sol_indef}
is unambiguously defined for all~$\chf$ if only we specify
how the integral in the second equation is understood
for $\chf>-\fr1{\sqrt3}$, when the integration path encounters 
singularities at $\chi'=\pm\fr1{\sqrt3}$. 
As discussed in the preceding subsection, we treat 
the singular expression 
under the integral as a generalized function.
Then, given a certain $\bar\Phi_-'$, one 
can integrate the singular terms using Table~\ref{FPtable}.

The constant in eq.\rf{Php_sol_indef} may differ among the 
$\chf$ intervals $(-\infty,-\fr1{\sqrt3})$, $(-\fr1{\sqrt3},\fr1{\sqrt3})$,
and $(\fr1{\sqrt3},\infty)$.  For example, if $R_{\nu}=0$ 
then $\bar\Phi_-=0$, $F_-(\chf)=0$
and $\bar\Phi_+(\chf)$ vanishes for $|\chf|>\fr1{\sqrt3}$ but
not for $|\chf|<\fr1{\sqrt3}$, \cf eq.\rf{Php_0}.
 As shown in Appendix~\ref{apx_local},
with the initial conditions that are adiabatic and satisfy eq.\rf{ic1},
the metric must remain unperturbed beyond the particle horizon
$|\chf|=1$.  From this we immediately conclude that 
that $\const=0$ in the interval $(-\infty,-\fr1{\sqrt3})$.
Taking into account that $F_-(\chf{\,\ge\,}1)=0$
because $\bar\Phi_-'(\chf)$ is odd, we see that the constant also 
vanishes for $\chf\ge \fr1{\sqrt3}$.
Denoting the value of the constant in the interval 
$(-\fr1{\sqrt3},\fr1{\sqrt3})$ by $ p_{\Phi}$, we thus  have
\be
\bar\Phi_+(\chf)=\lf({\chf^2-\frac13}\rt)\lf[
       p_{\Phi}\,\theta\lf(\fr1{\sqrt3}-|\chf|\rt) - F_-(\chf)\rt]\!.~~
\lb{Php_sol}
\ee
$p_{\Phi}$ may be calculated from the 
results\rfs{super_rad_Phi}{super_rad} 
for superhorizon Fourier modes in the radiation era 
that give
\be
\int_{-1}^1 d\chf\,\bar\Phi_+(\chf)
   =\fr{1+\fr15\Rnu}{1+\fr{4}{15}\Rnu}\,\fr{2\zeta\i}{3}~.
\lb{Phi_p_norm}
\ee
In the following subsection we obtain another, 
equivalent but easier to apply, condition fixing~$p_{\Phi}$.

The system of the integro-differential 
equations\rf{eqcls} and\rfs{Fm_def}{Php_sol}
for the pair $(\bar\Phi_+,\bar\Phi'_-)$ can be solved by iterations
starting from the solution in the limit $\Rnu\to0$.
The latter is
\be
&&\bar\Phi_-^{(\Rnu\to0)}=0~,
\lb{Phm_0}\\ 
&&\bar\Phi_+^{(\Rnu\to0)}=
     \fr{3\sqrt3\,\zeta\i}{2}
    \lf({\fr13-\chf^2}\rt)\theta\!\lf(\fr1{\sqrt3}-|\chf|\rt)\!,\vsp \qquad\quad
\lb{Php_0}
\ee
as immediately follows from 
eq.\rf{eqcls} and eqs.~(\ref{Php_sol}--\ref{Phi_p_norm}).
Note that the Fourier transform\rf{FTbarf}
of the last Green's function matches the well known potential 
modes in tightly coupled radiation fluid:
\be
\quad
\Phi^{(\Rnu\to0)}_+(\t,k)
              =2\zeta\i\lf(\fr{\sin\fs}{\fs^3}-\fr{\cos\fs}{\fs^2}\rt)\,,
\lb{Phi0k}
\ee
with $\fs= {k\t}/{\sqrt3}$.

The next iteration
step, giving the potentials in $O(\Rnu)$ order, is performed
in Appendix~\ref{apx_ORnu}.  \fig{fig_grf}\,b) shows the 
zeroth order $\bar\Phi_+$ (dotted) 
and the $O(\Rnu)$ corrected 
$\bar\Phi_+$ (solid) and $\bar\Phi_-$ (dashed)
potentials for $\Neff=3.04$.
\begin{figure*}[tb]
\mbox{
\includegraphics[width=7.5cm]{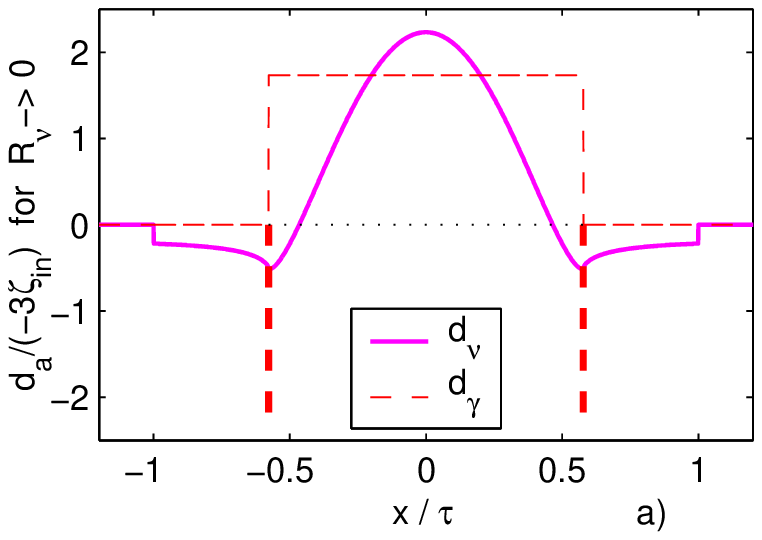}\qquad\quad
\includegraphics[width=7.5cm]{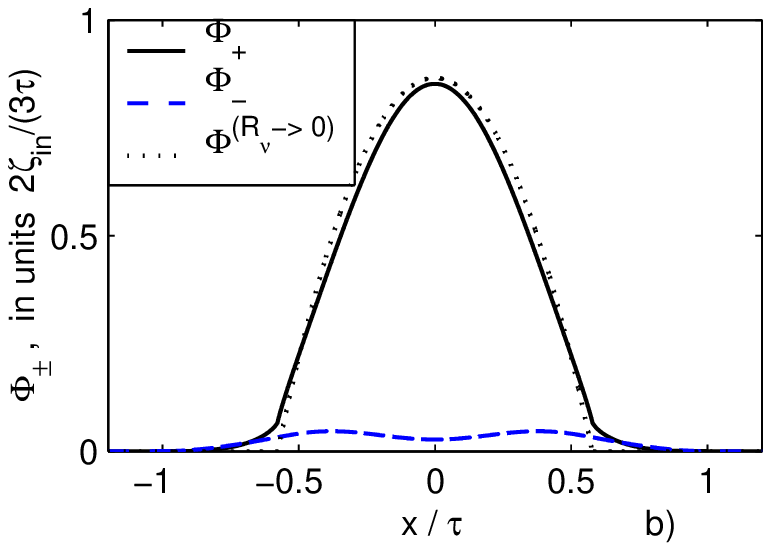}
}
\caption{a) Adiabatic 
      Green's functions for neutrino (solid) 
      and photon (dashed) number density perturbations
      in the radiation era. 
      The neutrino fraction, $\Rnu$, of the radiation density is
      assumed infinitesimal.
      b) Adiabatic  Green's functions for 
      the gravitational potentials
      $\Phi_{\pm}\equiv(\Psi\pm\Phi)/2$ in the radiation era.
      The solid and dashed curves are the sums of the $O(\Rnu^0)$
      and $O(\Rnu)$ terms for three neutrino species.
      The dotted line is $\Phi_+=\Phi$ for $\Rnu\to0$.}  
\lb{fig_grf}
\end{figure*}

\subsection{Neutrino effect on CMB perturbations}
\lb{sec_nuphot}

The Green's function for the photon density perturbation 
can be easily found in terms of the gravitational potential~$\Phi_+$.
From eqs.\rfd{gevol_rad}{selfsim_der} we obtain
\be
\lf(\chf^2-\frac13\rt)\bar d_{\g}
= 2\bar\Phi_+~.
\ee
The general solution of this equation is
\be
\bar d_{\g}
 = p_{\g}\,\dd\!\lf(|\chf|-\frac1{\sqrt3}\rt)
   + \frac{2\bar\Phi_+}{\chf^2-\fr13}~.
\lb{dgam_gen}
\ee
The delta function pre-factor $p_{\g}$ is fixed by the condition
$\int_{-1}^1 d\chf\,\bar d_{\g}(\chf)=d_{\g}(k\t\to0)=-3\zeta\i$,
giving
\be
p_{\g}=-\fr32\,\zeta\i
       -\int_{-1}^1 d\chf\,\frac{~\bar\Phi_+(\chf)}{\chf^2-\fr13}~. 
\lb{p_g_norm}
\ee

The pre-factor~$p_{\g}$ can be related to the
constant~$p_{\Phi}$, appearing in eq.\rf{Php_sol}, by applying  
the ``Poisson law''\rf{Psi_eq_rad}.
When eq.\rf{Psi_eq_rad} is considered as 
a relation among the Green's functions,
the only delta-function singularity appearing on its right hand side
is the one provided by the photon
density perturbation\rf{dgam_gen}.  As for the left hand side,
where $\Psi=\Phi_+ + \Phi_-$, the only delta-function 
comes from the double derivative of the term
$\lf({\chf^2-\frac13}\rt)p_{\Phi}\,\theta\lf(\fr1{\sqrt3}-|\chf|\rt)$
in eq.\rf{Php_sol}.  The equality
of these contributions requires
\be
p_{\Phi}=-\sqrt3(1-\Rnu)p_{\g}~.
\lb{p_g2Phi}
\ee
Substituting eq.\rf{Php_sol} in\rf{p_g_norm}
and eliminating $p_{\Phi}$ with the relation above,
we obtain
\be
p_{\g}=\fr1{1-2\Rnu}\lf[\fr32\,\zeta\i
       -\int_{-1}^1 d\chf\,F_-(\chf)\rt]~.
\lb{p_g_sol}
\ee
Calculating~$p_{\Phi}$ from the last two equations is somewhat 
easier than from eq.\rf{Phi_p_norm}.

Now we have all the analytic tools to analyze how neutrinos affect
CMB perturbations.  The evolution of metric perturbations 
{\it without neutrinos\/} is given
by eqs.\rfs{Phm_0}{Php_0}.  Then the photon density 
Green's function follows from eqs.\rfd{dgam_gen}{p_g_sol} as
\be
\ba{r}
\bar d_{\g}^{(\Rnu\to0)}
 = -3\zeta\i\lf[\sqrt3~\theta\!\lf(\fr1{\sqrt3}-|\chf|\rt)\,-\rt.~~\Vsp\\\
   \lf.-\,\fr12~\dd\!\lf(|\chf|-\frac1{\sqrt3}\rt) 
      \rt] \Vsp~.
\ea
\lb{d_g_rad_sol_x}
\ee
Its Fourier transform\rf{FTbarf} leads
to the photon density Fourier modes in the radiation era:
\be
d_{\g}^{(\Rnu\to0)}(\t,k)
     =-3\zeta\i\lf(
	\fr{2\sin\fs}{\fs}-\cos\fs
       \rt)\,,
\lb{d_g_rad_sol_k}
\ee
with $\fs= {k\t}/{\sqrt3}$.
In particular, without neutrinos the photon density modes
oscillate under the acoustic horizon ($\fs\gg1$)
as a pure $\fs$~cosine.

The predictions for 
both the phase and the amplitude of 
the photon mode oscillations differ when the gravity 
of neutrino perturbations is taken into account.  
The oscillations of the Fourier modes on subhorizon scales  
are described by the singular terms in the real 
space Green's functions.  For the photon density\rf{dgam_gen}
these are the $\d$-function and 
$(\chf\pm\fr1{\sqrt3})^{-1}$ singularities at $\chf=\pm\fr1{\sqrt3}$:
\be
\bar d_{\g}(\chf)= p_{\g}\,\dd\!\lf(|\chf|-\fr1{\sqrt3}\rt)
           + \frac{2r_{\g}}{\chf^2-\fr13}+\dots~,
\lb{p_g_sing_gen}
\ee 
where
\be
r_{\g}= \bar\Phi_+(1/{\sqrt3})~
\lb{r_g}
\ee
and the dots stand for more regular terms.
The Fourier transform of eq.\rf{p_g_sing_gen} follows from 
the first and third lines of Table~\ref{FTtable}, where $n$ is 
set to $0$ and~$1$, as
\be
d_{\g}(\t,k)=2\lf(p_{\g}\cos\fs-r_{\g}\pi\sqrt3\sin\fs\rt)\!+O(\fs^{-1})\,.
  \,\quad
\lb{p_g_lg_gen}
\ee
A non-zero phase shift with respect to the $\cos\fs$
oscillations is generated whenever~$r_{\g}\not=0$.
By eq.\rf{r_g} this can happen for adiabatic perturbations 
if only some perturbations propagate {\it faster\/} than the 
sound speed in the photon fluid, and thus are able to
generate metric perturbations beyond the acoustic horizon.  
This is the case for the neutrino perturbations, 
propagating with the speed of light,
\fig{fig_grf}\,a).

The values of $p_{\g}$ and $r_{\g}$ in eq.\rf{p_g_sing_gen} are calculated 
in $O(\Rnu)$ order in Appendix~\ref{apx_ORnu}. With its
results\rf{pg1} and\rf{rg1}, the mode\rf{p_g_lg_gen}
can be presented as
\be
d_{\g}(\t,k)= 3\zeta\i(1+\D_{\g})\cos\lf(\fs+\d\f\rt)+O(\fs^{-1})\,,\quad
\lb{dg_trf_cor}
\ee
where
\be
\ba{rcl}
\D_{\g}&\simeq& -\,0.2683\Rnu+O(\Rnu^2)~,
\qquad \vsp \\
\d\f&\simeq& 0.1912\,\pi\Rnu+O(\Rnu^2)~.\vsp
\ea
\lb{Aph_res}
\ee
As demonstrated in \fig{fig_comp}\,a),
our theoretical predictions
are in excellent agreement with numerical calculations 
for the radiation era,
at the redshift~$z=10^7$, obtained with CMBFAST~\cite{CMBFAST}.
\begin{figure*}[tb]
\includegraphics[width=16.5cm]{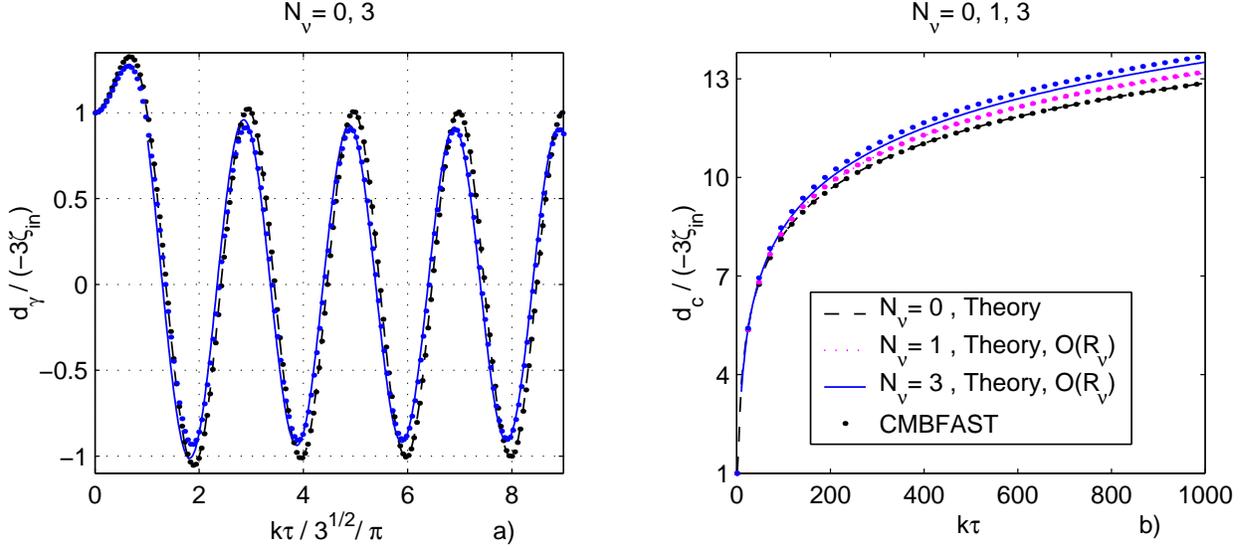}
\caption{a) Numerically calculated photon number density 
perturbation $d\sg$ in the radiation era for 0 and 3 
neutrino species $N\snu$ (dots) versus the theoretical
prediction\rf{d_g_rad_sol_k} for $N\snu=0$ (dashed) and 
its rescaled and phase shifted asymptotic form\rfd{dg_trf_cor}{Aph_res} 
for $N\snu=3$ (solid).
b) Similar comparison for the dark matter density perturbation~$d_c$
and $N\snu=0,1,3$.  The theoretical predictions are given 
by eqs.\rfs{dc_trf0}{Ac_res}.  In all the cases 
the~$O(\Rnu^2)$ terms in the analytical formulas are neglected.
}  
\lb{fig_comp}
\end{figure*}
The numerical calculations show that 
the $O(\Rnu^2)$ corrections contribute 
to $\D_{\g}$ and $\d\f$ less than~$10\%$ 
when $N_{\nu}\sim3$.

Our analytical results\rfs{dg_trf_cor}{Aph_res} can be
compared with numerical fit~(B7) of\ct{HuSugSmall}, giving
for the photon overdensity in the radiation era
\be 
d_{\g}^{\rm (HS\ fit)}(\t,k)\approx 
     \fr{9\Psi(\t k\to 0)\cos\fs}{2\lf(1+\fr25\,\Rnu\rt)}
    = \fr{3\zeta\i\cos\fs}{\lf(1+\fr4{15}\,\Rnu\rt)}
\,.~~~
\lb{dgHuSug}
\ee
This formula misses the phase shift due to neutrino 
perturbations but describes the decrease of the 
oscillation amplitude with the increase of neutrino fraction 
remarkably well.
In the $O(\Rnu)$ order it coincides with the result\rf{Aph_res} by 
better that~$1\%$.

Although the primordial magnitude of the cosmological perturbations
is unmeasurable directly, one can detect the amplitude change of 
the CMB acoustic oscillations, predicted by the first of eqs.\rf{Aph_res},
by comparing CMB and dark matter density fluctuations.  
The latter, however, are themselves affected by neutrinos.
In the next section we find the leading, $O(\Rnu)$, corrections
to the CDM density perturbation modes entering the horizon in
the radiation era.

\subsection{Neutrino effect on CDM perturbations}
\lb{subsec_CDMrad}

The evolution of the CDM coordinate density perturbation~$d_c$
is given by eq.\rf{CDM_evol}. In the radiation era it becomes
\be
\ddot d_c + \fr1{\t}\,\dot d_c = \Nb^2\Phi~.
\lb{dc_evol_rad}
\ee
Using the CDM Green's function ansatz
$d_c(\t,x)=\bar d_c(\chi)/\t$ and the differentiation 
rules\rf{selfsim_der}, canceling the common factors~$1/\t^3$,
and integrating the resulting equation once
trivially, we find
\be
(\chi^2\bar d_c)' - \chi\bar d_c = \bar\Phi'~.
\lb{dc_real_eq}
\ee
In Appendix~\ref{apx_ORnu} we show that in the $\Rnu\to 0$ limit
and with the adiabatic initial conditions this equation gives for
$\chi\not= 0$
\be
\bar d_c^{(\Rnu\to0)}=\!-3\zeta\i\lf(\sqrt3-\fr{1}{|\chi|}\rt) 
                  \theta\!\lf(\fr1{\sqrt3}-|\chi|\rt)\,.
\lb{d_c_rad_sol_x}
\ee
In Fourier space
\be
d_c^{(\Rnu\to0)} \!(\t,k)\!=
   \! -6\zeta\i\!\lf(\ln\fs\!+\g-\fr12-\ci\fs\!+\!\fr{\sin\fs}{\fs}\rt)\!,~~~
\lb{dc_trf0}
\ee
with $\fs={k\t}/{\sqrt3}$.

A finite neutrino fraction of the radiation energy density 
$\Rnu$ affects the gravitational potential on the right hand side
of eq.\rf{dc_evol_rad} and so the matter density 
perturbation~$d_c$.  On the scales well inside the acoustic 
horizon  ($\fs\gg1$), when the potential term in eq.\rf{dc_evol_rad} 
can be dropped, the general solution for CDM Fourier modes
should be of the form 
\be
d_c(\t,k)= -6\zeta\i(1+\D_c)\lf(\ln\fs+\g-\fr12+\d c\rt)\!+~~~ 
\lb{dc_trf_cor}
\\
+\ O(\fs^{-1})\ .  
\nn
\ee
The values of the integration constants $\D_c$ and $\d c$ are determined
by the mode dynamics during the horizon entry and are sensitive
to the gravity of neutrino perturbations.
The real space calculations in Appendix~\ref{apx_ORnu} give that
\be
\ba{rcl}
\D_c&\simeq& 0.2297\Rnu + O(\Rnu^2)~,\qquad \vsp\\
\d c &=& -0.6323\Rnu + O(\Rnu^2)~.\vsp
\ea
\lb{Ac_res}
\ee
The density perturbation\rfs{dc_trf_cor}{Ac_res} in $O(\Rnu)$
order is compared with the radiation era ($z=10^7$)
CMBFAST calculations in \fig{fig_comp}\,b).
When $k\t\gg1$, the $O(\Rnu)$ analytical results underpredict $d_c$
variation for $N_{\nu}$ change from~$0$
to~$1$ by $11.6\%$ and from~$0$ to~$3$ by $23\%$.
Since in the second case $\Rnu$ changes by $2.1$ times 
more than in the first one,
the twofold increase in the relative error is consistent 
with the origin of this deviation from the $O(\Rnu^2)$ corrections.

\section{Neutrino signatures in CMB and matter spectra}
\lb{sec_signatures}

Decoupled neutrinos affect observable cosmological probes 
both by the gravity of their {\it perturbations\/} and by
the change of the cosmological expansion rate due to
the contribution of the neutrino {\it background\/} to the universe 
energy density.  The first effect is prominent when cosmological perturbation modes
enter the horizon in the radiation era.   The corresponding
modifications of the photon and CDM perturbations 
were found in the previous section.  The perturbations 
remain to be propagated to the later epochs and
related to observable statistical power spectra.
These tasks are addressed in the current section.

\subsection{CMB power spectra} 
\lb{sbs_Cl}

\subsubsection*{\it Theory overview}

While photons are tightly coupled, their
number density perturbation~$d_{\g}$ 
satisfies the equation
\be
&&\ddot d_{\g} + \fr{\H R_b}{1+R_b}\,\dot d_{\g}
   - c_s^2\Nb^2d_{\g}=
\lb{ddd_g}
\\
&&\qquad \qquad \qquad = 2\t_d\Nb^2\dot d_{\g} + 
\Nb^2\lf(\Phi+\fr{\Psi}{1+R_b}\rt).
\nn
\ee
It follows from eq.\rf{ddd} applied to the 
photon-baryon fluid with
\be
 \ba{rcl}
  c_s^2(\t)&=&\dst \fr{\,d p_{\g b}}{\,d\rho_{\g b}} 
  =\fr1{3(1+R_b)}~,\\ \\
  R_b(\t)&\equiv& \dst \fr{3\rho_b}{4\rho_{\g}}  
  \simeq 0.6~\fr{a(\t)}{10^{-3}}\,\lf(\fr{\om_b}{0.02}\rt)\,,
 \ea
\lb{R_def}
\ee
Ref.~\cite{HuSugAnalyt}.
The Silk damping term $2\t_d\Nb^2\dot d_{\g}$
in eq.\rf{ddd_g} owes its origin to both partial photon
diffusion and to the lagging of baryons behind 
photons in imperfectly coupled photon-baryon 
plasma. While these effects are minor,
the damping coefficient equals~\cite{Kaiser83}
\be
\t_d(\t)=\fr{\t_c(\t)}6\lf[1-\fr{14}{15(1+R_b)}
                             +\fr1{(1+R_b)^2}\rt],~~
\lb{tc_def}
\ee
where
\be
\qquad \t_c(\t)\equiv\fr1{an_e\sigma}_{\!\rm Thompson}~.
\ee
The above result for $\t_d$ and eq.\rf{ddd_g} are 
valid while $\t_c\ll\min(1/k,\t)$.  
The damping increases substantially,\ct{Mukh_Rept}, 
within the CMB last scattering surface,
where the imperfect fluid approximations fail.

The general solution of eq.\rf{ddd_g} can be obtained
for subhorizon Fourier modes, $k\t\gg1$,
using WKB approximation,\cts{HuSugAnalyt,WeinbSmall}.
For the monopole of the photon effective temperature 
perturbation
\be
\Teff_{0,\g}\equiv \fr{\langle\d T_{\g}(\t,\r,\n)\rangle_{\n}}{T_{\g}}+\Psi
        =\fr13\,d_{\g}+\Phi+\Psi~,
\lb{Teff_CMB}
\ee
it gives
\be
\Teff_{0,\g}\simeq A\,\fr{\dst e^{-k^2x^2_S}}{(1+R_b)^{1/4}}
\cos\lf(kS+\d\f\rt)-R_b\Phi~,
\lb{Tsol_WKB}
\ee
where the size of the acoustic horizon~$S$ 
and the Silk damping length~$x_S$ equal 
\be
S(\t)\equiv \int_0^{\t}c_s\,d\t'~,\qquad
x_S^2(\t)\equiv \!\int_0^{\t}\!\t_d\,d\t'~.
\lb{S_def}
\ee
The solution\rfs{Tsol_WKB}{S_def} 
takes into account that on the subhorizon scales 
photon-baryon fluid and neutrinos contribute negligibly
to the gravitational potentials~$\Phi$ and $\Psi$,
primarily generated by CDM.
Hence, these potentials 
do not vary substantially over a single period 
of acoustic oscillations.  It also assumes that 
$\t_d\ll 1/k$, which is a necessary condition 
for the validity of eqs.\rf{ddd_g} and\rf{tc_def}.

The photon-baryon plasma velocity potential, affecting the CMB
anisotropy through the Doppler effect,  is easily found from
eq.\rf{dot_d}.  For Fourier modes,
\begin {widetext}
\be
u_{\g}=-\,\fr{\dot d_{\g}}{k^2} 
        \simeq -\,\fr{3\dot\Theta^{\rm eff}_{0,\g}}{k^2}
        \simeq \fr{A\sqrt3}{k}\fr{\dst e^{-k^2x^2_S}}{(1+R_b)^{3/4}}
        \sin\lf(kS+\d\f\rt)~,
\lb{Doppler_WKB}
\ee
where the last two equalities are valid within the
approximations that were applied to WKB result\rf{Tsol_WKB}.

The present CMB temperature anisotropy observed in the direction $\n$,
\be
\fr{\d T(\n)}{T} 
          \equiv \Theta(\n)\equiv 
          \sum_{l=0}^{\infty}\sum_{m=-l}^{l}\Theta_{lm}Y_{lm}(\n)~,
\ee
can be written as the line of sight
integral~\cite{CMBFAST}:
\be
\Theta(\n)=\int_0^{\t_0}\!d\t\lf[\,
             \dot g \lf(\Teff_{0,\g}-v_{\g}^i n_i+Q^{ij}n_in_j\rt)\rt.+
\lf. g\lf(\dot\Phi+\dot\Psi\rt)\rt]_{\t,\,\r=(\t_0-\t)\n}~.
\lb{los_zeta}
\ee
Here, for scalar perturbations, 
$\Teff_{0,\g}$ is given by eq.\rf{Teff_CMB},
$v_{\g}^i=-\Nb^i u_{\g}$, and
$Q^{ij}=\lf(\Nb^i\Nb^j-\fr13\d^{ij}\Nb^2\rt)q$,
where $q$,  negligible during the tight coupling, 
is a linear combination of $\pi_{\g}$ and 
multipole potentials that describe photon polarization.
The perturbations $\Teff_{0,\g}$, $u_{\g}$, and $q$
are evaluated along the line of sight
$\r(\t)=(\t_0-\t)\n$,
assuming the observer is positioned at the origin.
The ``integral visibility function''~$g(\t)$ is the probability 
for a CMB photon to propagate unscattered
from the time~$\t$ to the present time~$\t_0$: 
\be
g(\t)= \exp\lf[-\int^{\t_0}_{\t}\!\fr{d\t'}{\t_c(\t')}\rt]~.
\lb{visfun_def}
\ee

Expanding the expression under the line of sight integral\rf{los_zeta} 
over Fourier harmonics in a flat universe, we have
\be
\Theta(\n)=\!\!\int_0^{\t_0}\!d\t\int\fr{d^3\k}{(2\pi)^3}\,
  \zeta\i(\k)\, T_{\Theta}(\t,k)\, e^{i(\t_0-\t)\,\n\cdot\k}~.\quad
\ee
The transfer function $T_{\Theta}$ in the above equation
is constructed from the perturbation Fourier modes,
normalized by $\zeta(\t{\to0},\k)=1$, as
\be
T_{\Theta}(\t,k)\equiv
        \dot g \lf[\Teff_{0,\g}+u_{\g}\,\fr{\pd}{\pd\t_0}
        +q\lf(\fr{\pd^2}{\pd\t_0^2}+\fr13\,k^2\rt)\rt]
              +g\lf(\dot\Phi+\dot\Psi\rt)~.
\lb{T_Theta}
\ee

Given that for primordial fluctuations 
$\langle\zeta\i(\k)\,\zeta\i^*(\k')\rangle
=(2\pi)^3\dd^{(3)}(\k-\k')\,P_{\zeta}(k)$
and that a plane harmonic projects onto a spherical 
one as $\int d^2\Omega_{\n}\,Y^*_{lm}(\n)\,e^{i\n\cdot{\bf x}}
= 4\pi Y_{lm}^*(\^{\bf x})\,i^lj_l(x)$,
the CMB temperature auto-correlation 
function $C^{TT}_l\equiv\langle|\Theta_{lm}|^2\rangle$
becomes
\be
C^{TT}_l = \fr{2}{\pi}\int k^2dk~P_{\zeta}(k)\,
     \lf|\int_0^{\t_0}\!d\t\,
     T_{\Theta}(\t,k)\,j_l(k(\t_0-\t))\rt|^2~.
\lb{Cl_def}
\ee
\end {widetext}
When CMB polarization or other cosmological anisotropies are
accessible, additional two-point correlations
can be considered, such as $C_l^{EE}$ or $C_l^{TE}$ 
for the linear polarization component~$E$
generated by scalar perturbations,\ct{SeljZald_pol_sign}.   
The observed CMB polarization can be expressed
similarly to eq.\rf{los_zeta} as a line of sight 
integral~\cite{ZaldSelj_pol_los}
over the perturbations that source the photon 
polarization.
Likewise, the corresponding $C_l$'s are given by a $dk$ integral of
the product of the power spectrum $P_{\zeta}(k)$ 
and two time convolutions of perturbation variables
with $j_l(k(\t_0-\t))$ or its derivatives.

Any contribution to the correlation functions~$C_l$ 
from the time of last scattering is characterized by~$\t\ll\t_0$.
The corresponding Bessel functions vanish exponentially
when their argument $k(\t_0-\t)\simeq k\t_0$ is less than~$l$,
given $l\gg1$.
Thus the related to the acoustic oscillations constituents
of $C_l$ are essentially affected by only the 
modes that enter the horizon well before the radiation-matter equality
if $l\gg \t_0/S(\t_e)\approx 230$, 
with $\t_e$ given by eq.\rf{t_e_def} and WMAP best fit
parameters~\cite{WMAPGen}.

\subsubsection*{Neutrino signatures and degeneracies}

For the modes that enter the horizon in the radiation era, 
both the WKB solution\rfs{Tsol_WKB}{S_def} 
and  the radiation era 
solution\rf{dg_trf_cor} should be valid over a positive 
time interval $1/k\ll\t\ll\t_e$.  
Then the comparison of eq.\rf{Tsol_WKB}, where 
$R_b$ and $x_S^2$ are negligible in the radiation era, 
with formulas\rfd{Teff_CMB}{dg_trf_cor} 
shows that in eq.\rf{Tsol_WKB}:

a) The phase shift $\d\f$ is given by eq.\rf{Aph_res};

b) The integration constant~$A$ equals
\be
A=\zeta\i(1+\D_{\g})~.
\lb{A_rescale}
\ee

\noindent
The Fourier modes of plasma velocity potential\rf{Doppler_WKB} 
acquire the same phase shift and same
multiplicative change of the amplitude relative to the 
neutrinoless case.

The phase shift of the photon
acoustic oscillations on subhorizon scales 
can {\it not\/} be produced 
from adiabatic primordial fluctuations
by any dynamics involving only
the photon-baryon plasma and non-relativistic species.
This is seen easily in real space by noting that 
the acoustic oscillations correspond to small-scale, 
appearing singular on the Hubble scale, features in the
photon Green's function.
We found in Sec.~\ref{sec_nuphot} that in the 
radiation-dominated universe where no 
perturbations can propagate faster than the 
acoustic speed the only such features 
would be the delta function spikes in eq.\rf{d_g_rad_sol_x}
at $x=\pm\t/\sqrt3$.

The spikes would continue propagating away from 
the perturbation origin with the speed~$c_s(\t)$
of eq.\rf{R_def} 
past the radiation era until the recombination,
as, by the general equivalence principle, their local acoustic dynamics
could not be altered by the gravity of perturbations of 
other species. 
By the time of recombination, the singular part of the photon
density Green's function would have the form
\be
d_{\g,\,\rm sing}(\t,x)=
    D(\t)\,\d_D\!\lf(|x|-S(\t)\rt)\,.
\ee
The amplitude $D(\t)$ depends on the expansion rate of the
photon background.  The calculations of\ct{BB_GRF} in real space 
show that $D(\t)\propto e^{-k^2x_S^2}/(1+R_b)^{1/4}$.
The singularity $\d_D(|x|-S)$ becomes $2\cos(kS)$ in Fourier space, \ie, 
without neutrinos we would recover
the oscillating part of the WKB solution\rf{Tsol_WKB} with $\d\f=0$.

Although the observed period of $C_l$ oscillations 
depends on a number of cosmological parameters, 
such as $\om_b=\Om_bh^2$, $\om_m=\Om_mh^2$, or the total~$\Om\stot$, 
the oscillation phase, ideally, can be extracted independently of 
the period.
In practice, caution is required. 
Indeed, even in the neutrinoless model the location of 
the acoustic peaks in $C_l$ 
is not exactly proportional to the wave number~$k$ of the 
extrema of the effective temperature
transfer function\rf{Tsol_WKB}.
This is caused by a variety of effects, listed in 
Sec.~8.3.2 of\ct{tc_notes}.
Most of them vanish in the~$l\to\infty$ limit but
only as negative powers of~$l$ and are still
sufficiently important to modify
quantitative predictions for~$C_l$ phase even at $l\sim3000$.
Two of the effects remain finite for arbitrarily high~$l$. 
First, the Bessel function in eq.\rf{Cl_def}, 
or a similar equation for polarization, 
is exponentially cut off for $l>kr$, 
where $r\equiv \t_0-\t\dec$ 
in the flat model.  But $\langle|j_l|^2\rangle$
decreases slowly, by a power law, for $l<kr$.
As a consequence, an extremum of the temperature perturbation transfer 
function\rf{Tsol_WKB} at a certain~$k$ contributes most to the $C_l$ 
with $l$ somewhat {\it less\/} than~$kr$.
This shifts $C_l$
peaks toward lower~$l$'s: $l_{n\rm 'th\,peak}=(\pi r/S(\t\dec))(n-\phi_n)$,
with $\phi_n>0$.  Fortunately, the corresponding shift~$\phi_n$ 
for $C^{TT}_l$ approaches a model independent constant 
value $1/8$,\ct{tc_notes}.
Second, the phase of $C_l$ oscillations is obscured by 
the rapid decrease of $C_l$ magnitude for $l\gtrsim10^3$
due to the Silk damping and smoothing of the anisotropies 
by the non-zero width of the last scattering surface.
Nevertheless, the numerical analysis of Sec.~\ref{sec_forecasts}
shows the robustness of the phase under the change of any 
standard parameter other than the neutrino density.
The phase shift signature enables one to constraint $\Neff$ 
tightly when a sufficiently large~$l$ interval is accessible.

The rescaling of the photon-baryon oscillation amplitude
by $1+\D_{\g}<1$ for $k\gg1/S(\t_e)$ causes the same 
rescaling of all the photon phase space density and polarization 
multipoles, which develop during the decoupling.  
The non-oscillating
part of the gravitational potentials, also affecting the CMB anisotropy
by eq.\rf{Tsol_WKB} and\rf{los_zeta}, is however generated by CDM density 
perturbations and rescales differently, by~$1+\D_c>1$, 
Sec.~\ref{subsec_CDMrad}.   
In the {\it square\/} of the transfer function
in eq.\rf{Cl_def} some terms oscillate in~$k$ with the period 
$\D k=\pi/S$.  They come from and only from the product of two oscillating 
perturbations and must involve the factor $(1+\D_{\g})^2<1$.
There are terms that oscillate with the period $\D k=2\pi/S$.
They are produced by the cross-product of the oscillating photon
and non-oscillating CDM generated contributions and thus get
multiplied by $(1+\D_{\g})(1+\D_c)$, which by eqs.\rfd{Aph_res}{Ac_res}
is very close to~$1$.   It is hard to find a similar factor for the
sum of non-oscillating terms without a detailed calculation.
These terms consist of both the CDM--CDM contributions, 
scaling as $(1+\D_c)^2$, and of the non-oscillating parts of 
$\g$--$\g$ terms from the squares like $(\cos\f)^2=\fr12+\fr12\cos2\f$, 
scaling as $(1+\D_{\g})^2$.

Thus we expect that for $l\gg r/S(\t_e)\approx 230$ 
the acoustic oscillations of $C_l$'s
are {\it decreased\/} by the gravity of neutrino perturbations
by the factor $(1+\D_{\g})^2$.  But the magnitude of the
$C_l$ oscillations depends on other physical quantities
as well.  

First of all, on the primordial power spectrum~$P_{\zeta}(k)$.
In principle, this unknown could be excluded by comparing the 
CMB anisotropies 
with the matter density perturbations.  The latter 
scale differently under $\Neff$ variation,
as seen in Sec.~\ref{subsec_CDMrad} 
and in the next subsection.  
However, given the large experimental error on the 
matter perturbations, a better 
method to exclude~$P_{\zeta}$ is to compare the height of the initial 
acoustic peaks, entering the horizon closer to the matter era
and hence less affected by neutrino perturbations, to the height
of the subsequent peaks.   This signature is degenerate with the power
spectral index $n_s\equiv d\ln P_{\zeta}/(d\ln k)+4$ 
and its running in~$k$, especially for only a limited
number of observed peaks~\cite{Bowen}.

Second, the scaling $(1+\D_{\g})^2$ is directly applicable only if
one compares the models in which the photon subhorizon dynamics 
is identical in angular and redshift coordinates\footnote{
  A universe with larger neutrino density expands faster, 
  at least, in the radiation era.  The corresponding photon 
  temperature must, therefore, decrease faster in the time coordinate.
}
since the radiation era until the present.
Here is the list of the dimensionless quantities that 
characterize the cosmological background expansion and the 
local photon-baryon dynamics:\footnote{
\lb{CMB-CDM_par}
  We yet ignore the somewhat relevant to CMB ratio $\rho_b/\rho_m$.
  As seen in the next subsection, this ratio is very important
  for matter evolution.  But it affects CMB anisotropies rather mildly,
  through the non-oscillating CDM potential term in eq.\rf{Tsol_WKB}
  or through the CMB lensing by matter structure. 
}  
1. The ratio of the universe expansion rates at any redshift~$z$
and at the present:~$H(z)/H_0$.  It shows how distances
are mapped to angles: $(z,k/h)\to l$.
It also affects the photon and baryon
perturbation dynamics, considered in
the dimensionless variables $dz=-H(z)\,d\t$ and $H_0dx$.
This ratio,
\be
{H(z)}/{H_0}=\lf[\Om\,{\rho(z)}/{\rho(0)}+(1-\Om)/{a^2}\rt]^{1/2},~~
\ee
is specified by today's total density 
parameter $\Om\stot\simeq \Om_m+\Om\sd$, 
$\Om_m$, the redshift of the radiation-matter equality 
$1+z\eq=\rho_{m,0}/\rho_{r,0}$, and
the dark energy ``equation of state'' 
$w\sd(z)\equiv p\sd/\rho\sd$.
2. The ratio of baryon and photon densities, $\rho_{b}/\rho_{\g}$,
controlling the photon-baryon plasma dynamics.
3. The ratio of the photon free-flight time~$\t_c(\t)$,
eq.\rf{tc_def}, to the particle horizon size~$\t$.
This ratio determines the integral visibility function~$g(z)$,
eq.\rf{visfun_def}, and the important Silk damping
scale, eq.\rf{S_def}.

The present photon density $\rho_{\g,0}$ is well constrained by
the COBE measurement of CMB temperature $2.725\pm0.002\,K$,\ct{COBE_Tcmb}.
It is expected to redshift predictably as $\rho_{\g}=(1+z)^4\rho_{\g,0}$ 
deep into the radiation era.  The total radiation
density depends on $\Neff$ as $\rho_r=\rho_{\g}(1+\alnu)$,
where $\alnu\simeq 0.23\,\Neff$, eq.\rf{al_nu_def}.  Given this, 
two models with different~$\Neff$ will
have same ratios $\rho_{b}/\rho_{\g}$
and $H(z)/H_0$, and the related $\rho_{m,0}/\rho_{r,0}=z\eq+1$,
if these models have the same $\om_b=\Om_bh^2$, 
$\Om\stot$, $\Om_m$, and $w\sd(z)$,
but their Hubble constants scale as $h\propto\sqrt{1+\alnu}$.

\begin{figure*}[tb]
\includegraphics[width=15cm]{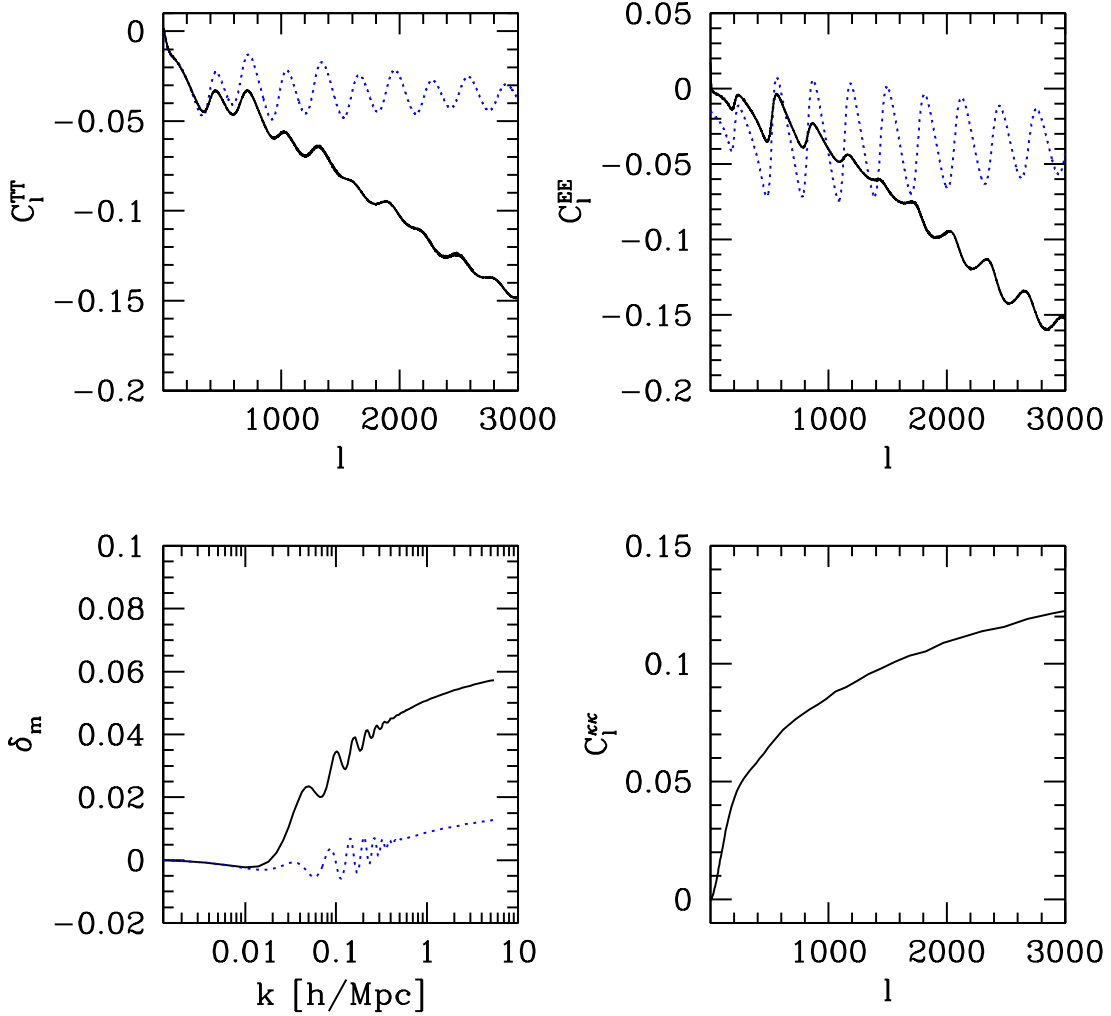}
\vspace{-7mm}
\caption{The relative change in $C_l^{TT}$ (top left), $C_l^{EE}$ (top right), 
matter density perturbation 
$\d_m(k/h)$ (bottom left), and $C_l^{\kappa \kappa}$ (bottom 
right) when $N_{\nu}$ varies from 2.5 to 3.5. The solid curves show 
the changes when all the other parameters, listed in 
Sec.~\ref{sec_forecasts}, are fixed. 
The top two panels also show the change for fixed recombination history and 
equivalent Silk damping (dotted, blue). 
The dotted, blue curve on
the bottom left panel gives the change in~$\d_m$
when $\omega_b/\omega_m$ is held fixed. 
}
\lb{dclnu}
\end{figure*}
The two top panels in \fig{dclnu} show the relative 
change $\d C_l/C_l$ for CMB temperature and polarization spectra 
under $\Neff$ variation from $2.5$ to $3.5$.
The solid curves correspond to preserving the
parameters $\om_b$, $\Om=1$, $\Om_m$, $w\sd=-1$, and $h/\sqrt{1+\alnu}$,
as described above,  
as well as the primordial power spectrum and 
the primordial helium fraction~$Y\equiv \rho\sHe/\rho_b$. 
As seen from the plots, the model with a larger $\Neff$ 
has noticeably stronger damping on small scales.  
Indeed, for a fixed ionization fraction 
$x_e\equiv n_e/n_H\propto n_e/\om_H$, where $\om_H=(1-Y)\,\om_b$
specifies the primordial hydrogen density,
the ratio
\be
\fr{\t_c}{\t}=\fr1{\t a n_e\sigma}_{\!\rm Thompson}\!\!
    \propto ~ \fr{h}{x_e(1-Y)\,\omega_b\,\sigma}_{\!\rm Thompson}
\lb{t_c_var}
\ee
increases with $\Neff$ as $h\propto\sqrt{1+\alnu}$.

The full conformal equivalence of the photon subhorizon dynamics
can be straightforwardly implemented in numerical calculations.  
The dotted curves on the top panels of \fig{dclnu} show $\d C_l/C_l$
when the global cosmological parameters vary as before but
the numerically integrated equations in both models
use the same ${\t_c}/{\t}$ and the same 
integral visibility function~$g(z)$.
The remaining $C_l$ change for the modes $l\gg200$,
entering the horizon before the radiation-matter equality, is 
uniform power suppression and a constant phase shift, as predicted for 
the effects of neutrino perturbations.
The shift $l\to l+\d l$ in the  numerical calculations  
is only~$80\%$ of the expected $\d l=\D l_{\rm peak}\d\f/\pi\simeq 4.6$,
for the period of the acoustic oscillations 
$\D l_{\rm peak}\simeq 300$ from\ct{WMAPPeaks} and $\d\f$ 
evaluated from the radiation era result eq.\rf{Aph_res}.
Likewise, the power suppression is somewhat less than the
radiation era prediction $\d(1+\D_{\g})^2\simeq-4.8\%$,
with $\D_{\g}$ given by the leading term in eq.\rf{Aph_res}.
These discrepancies 
are caused by the residual effect of the large-scale
relativistic correction, \cf the first term in 
eq.\rf{d_g_rad_sol_k}, non-negligible matter
density during the horizon entry,
and $O(\Rnu^2)$ corrections in eq.\rf{Aph_res}.

As suggested by eq.\rf{t_c_var}, 
the ratio $\t_c/\t$ in the compared models 
could be matched by varying the helium 
abundance~$Y$.   Apparently, when all the helium has
recombined but the hydrogen remains fully ionized
($x_e=1$), the quantity\rf{t_c_var} remains constant
if $Y$ varies as $(1-Y)\propto h\propto\sqrt{1+\alnu}$.   
It is less obvious that the after such a 
rescaling, $\t_c/\t$ coincides in both models 
during the recombination, when $x_e$  violently changes
with time.  Despite the complicated nature of recombination,
the physical mechanism that is primarily responsible
for $x_e$ evolution around the peak of the photon visibility
function~$\dot g$ does lead to degenerate~$x_e(z)$
and so degenerate~$\t_c/\t$.

At the redshifts $1400\gtrsim z\gtrsim 800$, which
are of the most interest, the dominating process leading
to $H$ recombination to the ground state~$1s$
is the suppressed $2s\to 1s+\g+\g$ decay.
Faster transitions to $1s$ with the emission of a single photon
do not create more hydrogen in the ground state because
the emitted resonance photon soon ionizes or excites
another $H$ atom in the ground state.
Using the approximations of\ct{Peebles_rec}
(see\ct{rec_notes} for a recent review), one can find the 
rate of $x_e$ change by assuming that due to
the overcooling of the universe from the delay 
in recombination the fraction of all 
the excited $H$ atoms is negligible comparatively to 
the concentration of $H$ in the ground 
state $x_{1s}\equiv n_{1s}/n_H$ and to~$x_e$.  
Then $x_{1s}+x_e\simeq 1$, hence
$dx_{1s} \simeq - dx_e$.
The change in~$x_{1s}$ is mainly due to the spontaneous 
decay $2s\to1s+\g+\g$ with the lifetime $\L\simeq 8.2\,s^{-1}$:
${dx_{1s}}/{dt}\simeq \L\,x_{2s}$.
(The photo-excitation $1s\to2s$ is negligible when $z<1300$.)
$x_e$ decreases in $ep$ recombination and increases in hydrogen 
photo-ionization.
With realistic approximations,\ct{Peebles_rec}, one finds:
${dx_e}/{dt}= -\,\al x_e x_p n_H + \b x_{2s}$,
where $x_p\simeq x_e$, and
$\al$ and $\b$ at a given CMB temperature,
determined by the redshift, are 
independent of cosmological parameters.
The elimination of $x_{2s}$ from the equality
of ${dx_{1s}}/{dt}$ and $-{dx_e}/{dt}$ yields:
\be
\fr{dx_e}{dt}\simeq -\,\fr{\al\L}{\b+\L}\,x_e x_p n_H~.
\lb{rec_rate}
\ee
When both $d/dt=-(1+z)\,H(z)\,d/dz\propto h$ 
and $n_H\propto(1-Y)\,\om_b$ vary in
proportion to $\sqrt{1+\alnu}$, as discussed above,
eq.\rf{rec_rate} predicts that the function~$x_e(z)$ 
is unchanged, and so is unchanged the last
ratio in eq.\rf{t_c_var}.

Numerical calculations show that this conclusion holds very
well after the end of helium recombination at $z\simeq1500$.
For a variation of $\Neff$ from $2.5$ to $3.5$ and the corresponding
adjustment of~$Y$ by\footnote{
  Astrophysical constraints on the primordial helium abundance
  vary among groups and span a range 
  $Y=0.238\pm0.010$ $(2\sigma$),\cts{Y_OS_95,Y_OS_97,Y_ITL_97,Y_ITL_98,Y_PPL_02},
  a review in\ct{BBN_Steig}.
  Taking this prior, $Y$ needs to be considered as a free
  parameter in a CMB analysis only for the accuracies $|\D\Neff|\lesssim0.4$.
}
$-0.051$, the ionization fraction~$x_e$ changes 
at most by $0.5\%$, and even twice as little within the peak of 
the photon visibility function~$\dot g$.  (For comparison, 
without the adjustment of helium abundance, 
the change of $x_e$ reaches $6\%$ within the visibility peak.)  
At $2000\lesssim z\lesssim 5000$, when helium is singly ionized,
the decrease of the helium density decreases the total number of 
free electrons by about~$2\%$, and more at higher redshifts, 
when the second
helium electron unbounds.  But the related increase of the
Silk damping scale, eqs.\rfd{tc_def}{S_def}, is insignificant,
and the effect is quite negligible for~$dg/dz$.
The change $\d C_l/C_l$ for $TT$ and $EE$ spectra 
under the considered 
variation is seen from 
\begin{figure*}[tb]
\includegraphics[width=16cm]{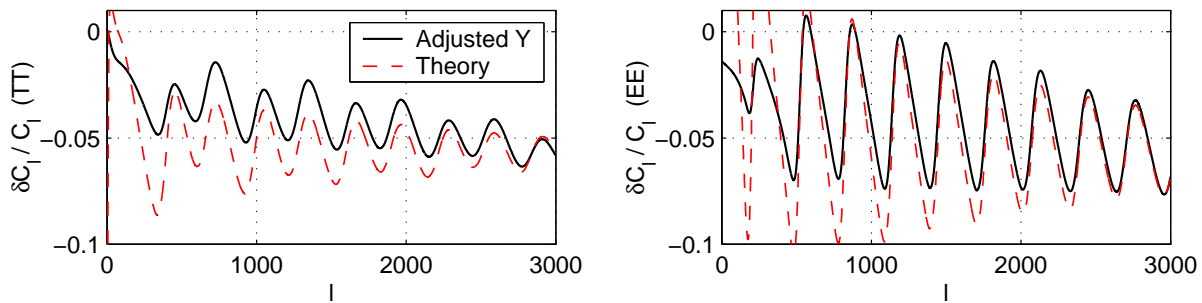}
\vspace{-2mm}
\caption{The relative change in $C_l^{TT}$ (left) and $C_l^{EE}$ (right)
for $N_{\nu}$ variation from 2.5 to 3.5 and the adjustment
of the helium abundance~$Y$ by~$-0.051$ (solid).
It is compared with the change from rescaling $C_l$'s
by $-4.8\%$ and shifting them to smaller~$l$'s
by~$\d l=3.7$ (dashed).
}
\lb{fig_ClHe}
\end{figure*}
\fig{fig_ClHe} (solid line).  
It is
compared with the theoretical reconstruction (dashed line) 
obtained by shifting $C_l$ by 
$\d l=0.8\D l_{\rm peak}\d\f/\pi\simeq 3.7$
and rescaling it by $\d(1+\D_{\g})^2=-4.8\%$.
The plots show that the $Y$~adjustment removes most
of the damping seen in \fig{dclnu}.

The approximate conformal degeneracy of CMB dynamics among the models
with {\it changed\/} dynamical time and length scales
should be distinguished from the well known degeneracy of primary CMB
anisotropies under any variation of $H(z)$ for $z< z\dec$
that preserves the angular diameter distance to the last scattering surface,\ct{LLbook}.  
The latter degeneracy is generally violated by the late ISW effect.  
The found one is respected by ISW but is somewhat violated by 
the gravity of matter perturbations, 
\cf footnote\rfp{CMB-CDM_par}, 
and, of course, by the gravity of neutrino 
perturbations.

The internal CMB dynamics is shown to be conformally invariant 
in an increment of~$\Neff$
if the helium abundance~$Y$ is {\it decreased\/}
as $\D Y/\D\Neff \simeq -0.05$.
In contrast, the standard Big Bang nucleosynthesis (BBN) then
predicts {\it increased\/} ${}^4$He production.\footnote{
  The sensitivity of the BBN yield of ${}^4$He and D/H 
  to variation of $\Neff$ and $\eta\equiv n_B/n\sg$ 
  about $\Neff=3$ and the WMAP~\cite{WMAPParams} 
  ``best fit'' value $\eta\approx 6.1\times 10^{-10}$
  is roughly
  \be
   \lf(\ba{c} \D Y \\ \D\ln(D/H) \ea\rt) \approx
   \lf(\ba{cc} 0.013 & 0.01 \\ 0.14 & -1.6~~ \ea\rt)
   \lf(\ba{c} \D\Neff \\ \D\ln\eta \ea\rt)~          
  \nn
  \ee
  (based on the numerical results from\cts{BargPLB03,Sarkar96,BBN_Walker91}.)
}
Indeed, for $\D\Neff>0$ the universe expands faster
and less neutrons decay by the time it cools sufficiently to 
allow their conversion into helium.  Of course, one should not presume
a priori that the ratio of neutrino and photon energy densities
is preserved since the nucleosynthesis until the matter era,
\ie\ that $\Neff$ in BBN and CMB physics are equal.  
But if they are not, the non-degeneracy of CMB dynamics
along the BBN-predicted curve $Y(N_{\nu})$ makes it easier 
to spot the discrepancy.

Neutrino perturbations can affect
the gravitational potentials even after the photon 
decoupling if radiation is still a significant component of
the total energy density. When the gravitational potentials are
time dependent, there is a contribution to the
CMB anisotropy\rf{los_zeta} that depends on the line of sight 
integral of $\dot\Phi+\dot\Psi$, the so-called 
integrated Sachs-Wolfe (ISW) effect. 
However, as discussed in Sec.~\ref{sec_superhor}, 
for superhorizon modes the derivative $\pd(\Phi+\Psi)/\pd\Rnu$
changes insignificantly after recombination in the 
standard models with matter radiation equality at $z\sim 3000$.
Thus the modification of the ISW effect by
neutrino anisotropic stress does not play a major role as 
a neutrino signature in CMB.
Specifically, 
on large scales ($l< l_{\rm 1st~peak}/3\simeq 70$)
the ISW effect becomes insensitive to perturbations
of relativistic neutrinos.

\subsection{Matter power spectrum} 
\lb{sbs_matter_P}

The growth of the acoustic perturbations 
in the tightly coupled photon-baryon fluid is halted by 
photon pressure, described in eq.\rf{ddd} by the term 
$- c_s^2\Nb^2d_{\g}$,
where $c_s\gtrsim 1/2$ before recombination.
Likewise, relativistic neutrinos and CMB photons after their decoupling
are stabilized against gravitational collapse by the
effective pressure, described by the same
term with $c_a=1/\sqrt3$, due to the  velocity dispersion.
In addition,  as noted in the introduction, the perturbation modes of 
free-streaming particles on subhorizon scales decay by ``directional'' damping.   
Assuming the dark energy too does not strongly cluster on 
small scales, only CDM and
after recombination baryons can cluster sufficiently
to generate non-negligible gravitational potential inside 
the subhorizon.
On subhorizon scales eq.\rf{Psi_eq} simplifies
to $\Nb^2\Psi=\g d$.  Then eqs.\rfd{ddd}{Phi_eq} 
applied to non-relativistic species~$n$ with negligible 
pressure ($c_{n}^2\simeq 0$) give
\be  ~~\qquad
\ddot d_n + \H\dot d_n 
          =  \sum_{m:\,c^2_m\simeq0}\g_m d_m~\rm\qquad(subhor)\,.\qquad
\lb{ddd_subhor_nr}
\ee

One can distinguish the following stages of the linear matter 
perturbation growth from their entry of the horizon in the radiation era 
to the present:
\begin{enumerate}
\item
Growth of CDM perturbations in the radiation-matter
universe until the recombination 
while baryons are tightly coupled to CMB.
\item
\lb{growth_stages}
Decoupling of the baryons from CMB and joining the matter 
gravitational collapse at $z_d\sim 1090$.
\item
Growth of the pressureless CDM-baryon matter perturbations
through the subsequent universe evolution, affected by
dark energy and, possibly, global curvature. 
\end{enumerate}

Eq.\rf{dc_trf_cor} for CDM density perturbations 
in the radiation era on subhorizon scales can be presented as 
\be
&&d^{(r,\,\l\ll\t)}_c(\t,k)=
\lb{d_c_b}
\\
&&\qquad=-6\zeta\i(1+\D_c)\lf\{A+\ln\lf[(1+\d c)\,\fr{k}{h}\rt]\rt\}
~,\quad
\nn
\ee
where $A$ is the same function of $\t h$ for all~$k$ and $\Rnu$:
\be~~
A(\t h)=\ln\lf(\fr{\t h}{\sqrt3}\rt)+\g-\fr12~~\qquad\rm(rad.~era)\,.\quad
\lb{b_rad}
\ee
At later times the matter density perturbation must be
of the form
\be
d_m(\t,k)= \!-6\zeta\i(1+\D_c)\lf\{\!A+B\ln\lf[\!(1+\d c)\,\fr{k}{h}\rt]\rt\}\!,~~
\lb{d_c0_lin}
\ee
where $A$ and $B$ depend on $\t h$ and on 
the cosmological parameters affecting the background 
evolution at low redshift.  
We argue that throughout linear evolution  $A$ and $B$ 
are independent of the mode wave-number~$k$ and of~$\Rnu$,
provided that the compared models agree in $\Omega_m$, in $\Omega_b$, 
and in the background
effective equation of state~$w\stot(z)\equiv p\stot/\rho\stot$.
This would immediately follow from the linearity of the evolution
with the initial conditions\rf{d_c_b} in the radiation era
if the equations of matter dynamics in terms of the
dimensionless variable~$\t H_0\propto\t h$ coincide.  
The latter is manifestly so by eqs.\rfd{ddd_subhor_nr}{g_cb}
for the stages $1$ and $3$ of the three stages of linear matter 
clustering identified above.  
As for the stage~$2$, the scale and $\Rnu$ invariances 
of the dynamics are violated
by the small oscillations of the baryon density and velocity at the 
decoupling due to the acoustic oscillations in CMB,\ct{HuSugSmall}.  
However, this effect becomes
negligible at sufficiently small scales because of the Silk damping.
Hence we assume that the baryons decouple with initially
negligible density and velocity perturbations.
Then all the stages of the matter evolution are scale- 
and $\Rnu$-invariant.  Then, from eq.\rf{d_c0_lin}, 
the matter density perturbation
modes~$d_m(z,k,\Rnu)$ can be obtained from the modes in the model 
with $\Rnu\to0$\footnote {
  In the models with dark matter, cosmological constant, 
  global curvature, 
  but negligible baryon density ($\rho_b/\rho_c\ll 1$),
  the linear matter density perturbation well after the radiation era
  on the small scales is
  \be
  d^{(\Rnu\to0)}_m\!= d\i\!\lf[\ln\lf(\fr{4k\t_e}{\sqrt3}\rt)+\g-\fr72\rt]
     \fr{15H(a)}{2a\eq H_0}\int_0^a \!da'\lf(a'\fr{H(a')}{H_0}\rt)^{\!-3}
  \!\!\!\!,
  \nn 
  \ee
  where $d\i=-3\zeta\i$.
  Analytical, albeit more complicated, expressions also 
  exist for
  non-negligible baryon density~\cite{HuSugSmall,WeinbSmall}. 
}
and the identical~$\Omega_m$, $\Omega_b$, and~$w\stot(z)$ as
\be
d_m\lf(z,\fr{k}{h},\Rnu\rt) \!=\! 
   (1+\D_c)~\!d_m^{(\Rnu\to0)}\!\!\lf(z,(1+\d c)\fr{k}{h}\rt)\!,~~
\lb{d_m_scaling}
\ee
where for the considered small scales, 
entering the horizon in the radiation era,
$\D_c$ and $\d c$ are given by eq.\rf{Ac_res}.

This effect of neutrino perturbations
on the matter density (dotted curve on the left bottom panel 
of \fig{dclnu}) is too small to provide by its own a useful information 
about $\Neff$ from the available cosmological probes.
A realistic data analysis constraining the abundance of
ultra-relativistic neutrinos should also include CMB data.  
However, the variations of CMB spectra with~$\Neff$, 
Sec.~\ref{sbs_Cl}, are less contaminated by neutrino unrelated
physics if $\omega_b=\Omega_bh^2$, rather than $\Omega_b$, is fixed.  
Likewise, the Big Bang Nucleosynthesis (BBN) 
constraints $\omega_b$ and not $\Omega_b$.
Thus it is more practical to consider the 
variation of matter density modes
with~$\Neff$ in the direction of the maximal CMB and BBN 
degeneracies -- under a fixed~$\omega_b$.

The magnitude of matter perturbations after the radiation 
era is sensitive to the parameter
\be
\b\equiv \fr{\rho_b}{\rho_c+\rho_b}= \fr{\omega_b}{\omega_m}~,
\ee
{\it e.g.}\ct{HuSugSmall}, which is currently estimated 
as~$0.17\pm 0.01$~\cite{WMAPParams}.
One reason for this sensitivity 
is the slower growth of CDM perturbations
prior to the recombination
in the model in which a larger fraction~$\b$ of 
the non-relativistic matter is withheld from gravitational 
collapse by the photon pressure.
The other reason is a greater dilution of the growing CDM perturbations
by almost unperturbed baryons when the latter finally 
decouple from photons.

It is apparently impossible to change 
the neutrino density preserving all of the parameters
$\omega_{m}/\omega_{r}$, $\omega_{b}$, and $\b$.  
If one varies $\Rnu$ while keeping $1+z\eq=\omega_{m}/\omega_{r}$ 
and~$\omega_b$ fixed, to minimize the changes in CMB spectra,
then the parameter~$\b$ will vary.  For example, 
$\b$ decreases by approximately $41\%$ 
for $\Neff$ change from  $0$ to~$3$, and by $14\%$ 
for $\Neff$ variation by $1$ around its standard value~$3.04$.
This is yet another source
of breaking the degeneracies between $\Neff$ and the 
other cosmological parameters.  
The significance of $\b$ variation for the growth
of matter perturbations is evident from comparing
the dotted (fixed~$\Om_b$ and $\Om_m$) and solid 
(fixed~$\om_b$ and $\Om_m$)
curves in \fig{dclnu}.

\section{Forecasts for Future Experiments}
\lb{sec_forecasts}

In this section we use numerical solutions from CMBFAST 
and apply them to predict the precision to which $N_{\nu}$ 
can be constrained from future experiments. We follow the 
basic approach of\ct{ZalScocHui_Lyal}: we evaluate
the standard error in a cosmological parameter~$s_i$ as
$\D s_i=(\bm{\al}^{-1})_{ii}^{1/2}$, where $\bm{\al}$ is the Fisher matrix: 
\begin{equation}
\alpha_{ij}=\sum_l \sum_{X,Y}\fr{\partial C^X_l}{\partial s_i}\,
{\rm Cov}^{-1}(\hat{C}^X_l,\hat{C}^Y_l)
\,\fr{\partial C^Y_l}{\partial s_j}~.
\end{equation}
Here, ${\rm Cov}^{-1}$ is the inverse of the covariance matrix,
$s_i$ are the unconstrained cosmological parameters,
and $X,Y$ stand for the observable power spectra. We limit the
analysis to  
CMB spectra $TT,EE,TE$, and
lensing convergence power spectrum $\kappa \kappa$ as measured 
from CMB itself.
For each $l$, one has to
invert the covariance matrix and sum over $X$ and~$Y$. The derivatives
were calculated by finite differences. The step was usually
taken to be about $5 \%$ of the value of each parameter and symmetric 
around the pivot point at the best fitted WMAP model,\ct{WMAPParams}.

An experiment is fully characterized by its sky coverage $f_{\rm sky}$, 
temperature, polarization, and convergence
instrument (or reconstruction) 
noise $w_{T}^{-1}$, $w_{P}^{-1}$, and $w_{\kappa}^{-1}$
and by the beam smearing window 
function $B_l^{-2}=e^{l(l+1) \theta_b^2/(8\ln 2)}$, with $\theta_b$ measuring the width of
the beam. For example, the covariance element for $TT$ is given by 
\be
{\rm Cov }\,(\hat{C}^{TT}_l,\hat{C}^{TT}_l)
      \!=\!{2\over (2l+1)f_{\rm sky}}(C^{TT}_l+
w_T^{-1}B_l^{-2})^2.~~\quad
\ee
The full covariance matrix for the CMB power spectra is given in\ct{MeasurPolar}.
For the lensing convergence noise spectrum we take the values from 
the maximum likelihood method developed in\ct{HS_reconst}, 
which gives the lowest reconstruction noise. 
In addition, we impose a maximum $l_{\rm max}=3000$ cutoff on all the 
spectra and we do not include the information from higher $l$. The 
justification for this is that scattering off moving electrons during 
and after reionization leads to an additional component in the CMB 
that cannot be separated from the CMB on the basis of frequency 
information. This component is rather uncertain since it receives 
contributions from perturbative and nonlinear structures, as well 
as from the patchiness in ionizing fraction. It is expected to be 
significantly less important for polarization, so our approach 
may be conservative for polarization sensitive experiments.

In our analysis we use parameters for several future experiments. 
While WMAP constraints on $N_{\nu}$ are not yet competitive
with the nucleosynthesis limits,
Planck satellite will improve the sensitivity of WMAP by an order of
magnitude. In parallel, there will be ground based experiments, 
such as SPT\footnote{http://astro.uchicago.edu/spt/} or the proposed ACT\footnote{http://www.hep.upenn.edu/$\sim$angelica/act/act.html} 
that will extend Planck to smaller angular scales. 
Whether or not they will be polarization sensitive remains to be 
determined, so we explore both possibilities. Finally, we 
also explore the prospects of an ambitious high resolution, low noise
satellite dedicated to polarization (CMBPOL), which will be able to measure
with high accuracy
not only CMB temperature and polarization but also matter power 
spectrum using the weak lensing induced signal in higher order
correlations. We also explore if adding information from 
more local probes, such as weak lensing, can further reduce the 
uncertainties.

We explore 11 parameters in our analysis. These are 
$\{\omega_m/\omega_r, \omega_b, 
\omega_{\nu\,\rm massive},N_{\nu},\Omega\sd,w\sd,
\tau_{\rm reion}, P, n_s, n_s',Y\}$. 
The first one is proportional to 
the ratio of matter density 
to radiation density (photons plus neutrinos), the second to 
proper baryon density, and the third to the proper density of massive 
neutrinos. $\Omega\sd$~is the dark energy density relative to
the critical density, 
$w\sd$~is its effective constant equation of state $p\sd/\rho\sd$,
$\tau_{\rm reion}$~is the reionization optical 
depth, $P$~is the amplitude of curvature power spectrum 
at $k=0.05/$Mpc,
and $n_s$ and $n_s'=dn_s/d\ln k$ 
are the scalar spectral index and its running at 
$k=0.05/$Mpc. 
We do the analysis for both 
fixed and unconstrained helium abundance~$Y$.

We ignore tensor perturbations since their
contribution is limited to 
large scales, where neutrinos do not play a major role. 
We also ignore $BB$ power spectrum, which is useful primarily as a 
tracer of tensors.  (On small scales it can be a  useful 
tracer of matter power spectrum and can provide additional constraints on
the convergence power spectrum as extracted from the nongaussianities in 
CMB; we ignore this additional information here.)  
We use the basis functions $\pd C_l^X/\pd s_i$ as defined by these parameters, but 
we also explore any possible numerical instabilities due to 
the variation in the
angular scale of the acoustic horizon, 
which is affected by variations in 
several of the parameters.

\fig{dclnu} shows the $N_{\nu}$ derivatives of 
$C^X_l$ for $X=TT$, $EE$, $\kappa \kappa$, 
and of the matter fluctuation $\delta_m$.
The derivatives of CMB spectra with lensing 
are plotted, 
but the unlensed ones are qualitatively similar
and the wiggles are not due to lensing, but due to the phase shift
(the lensed derivatives 
were used in the Fisher matrix analysis).
The solid curves on the $C_l$ plots
show the $N_{\nu}$ 
derivatives with the other parameters kept fixed. 
As discussed in previous sections, 
changing the effective number of neutrinos
changes not only the initial phase and 
amplitude of the acoustic oscillations
but also the angular scales of 
the acoustic horizon and oscillation damping.
To separate these effects, we also show with the dotted, blue line 
the change in the spectra
while keeping the visibility function and the photon free flight 
length in the units of $H_0^{-1}$ unchanged at every redshift.  

The modification of CMB spectra due 
to the change in the angular scale of acoustic oscillations
is described by multiplicative rescaling 
$l \rightarrow (1-0.002\Delta N_{\nu}) l$.
By itself, it is degenerate with other effects that change the 
angular acoustic scale, such as variation of~$w\sd$. 
The combined effect of the additive and multiplicative phase 
shifts is such that 
the phase shifts cancel exactly at $l \sim 1500$, but not at 
other values of $l$. For the temperature anisotropies, the phase shift is 
barely visible since temperature oscillations are weaker due to 
the competing effects from density and velocity terms. They are 
further suppressed by lensing. For polarization, which 
has more prominent acoustic oscillations,
the phase shift remains visible and can be clearly distinguished from 
the change in angular size of the acoustic horizon. 
This suggests that polarization information is crucial in extracting 
neutrino signature. Quantitative analysis in Table~\ref{table_forecasts} 
confirms that. 
Note that $\pd C_l/\pd N\snu$ approach zero at low~$l$. 
There is no significant neutrino dependent 
contribution coming from the integrated Sachs-Wolfe term
at the low~$l$, as discussed in Secs.~\ref{sec_superhor} and~\ref{sbs_Cl}.  
\begin{table*}[t]\small
\begin{center}
\begin{minipage}{5.2in}
{\sc Error Forecasts}\\
\end{minipage}
\end{center}
\begin{tabular}{|c|c|c|c|c|c|c|c|}
\hline
Experiment & $f_{\rm sky}$ & $\theta_b$ & $~w_T^{-1/2}$ & $~w_P^{-1/2}$
&  ~$\Delta N_{\nu}$~  & $\Delta N_{\nu}$  & ~$\Delta N_{\nu}$ (free $Y$)   \\
& & & [$\mu$ K'] & [$\mu$ K'] & TT & TT+TE+EE  & TT+TE+EE \\
\hline
Planck  & 0.8  &   7'  & 40 & 56 &  0.6 &  0.20 & 0.24 \\
ACT     & 0.01 &  1.7' &  3 &  4 &  1   &  0.47 & 0.9  \\
ACT + Planck  &  &    &     &    & 0.4  &  0.18 & 0.24  \\
CMBPOL &  0.8 &   4' &  1 & 1.4  & 0.12 &  0.05 & 0.09 \\
\hline
\end{tabular}\\[12pt]
\small
\caption{Standard deviations on $N_{\nu}$ as
expected from Planck, ACT, Planck+ACT and CMBPOL using 
temperature data only ($TT$) and added polarization ($TT+TE+EE$). 
The primordial helium abundance~$Y$ is considered 
a priori unconstrained for the last column results and fixed 
by independent measurements in the preceding two columns.
Adding weak lensing convergence as reconstructed from CMB
($TT+TE+EE+\kappa\kappa$) does not significantly improve the bounds, 
even assuming polarization information 
is available.   
}
\lb{table_forecasts}
\end{table*}
\begin{figure}[tb]
\includegraphics[width=9cm]{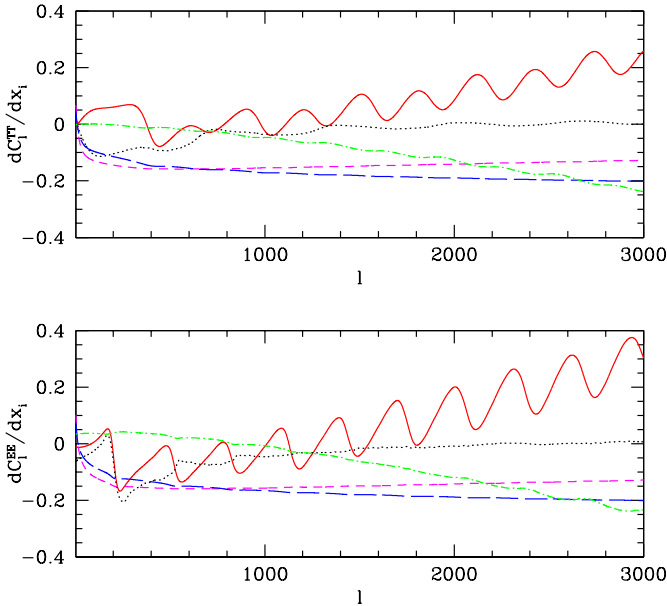}
\vspace{-8mm}
\caption{The derivatives of $C_l^{TT}$ (top) and $C_l^{EE}$ (bottom) 
with respect to $\om_b$ (solid, red), $\omega_m/\omega_r$ (dotted, black), 
$dn_s/d\ln k$ (short-dashed, cyan) and $n_s$ (long dashed, blue) and $Y$ (dotted-dashed, green).
All were shifted to match the angular scale of acoustic horizon.
This explains lack of oscillations when varying $\omega_m/\omega_r$ or~$Y$. Varying 
$\om_b$ changes the amplitude of the oscillations,
so adjusting the angular scale does not make their change vanish.
Note that the variation in~$Y$ appears as a change in damping length.
}  
\lb{dcldx}
\end{figure}
In \fig{dcldx} we show the derivatives of $C_l^{TT}$ and $C_l^{EE}$ 
with respect to some of the other parameters.
The displayed derivative are for
$\omega_m/\omega_r$, $\om_b$, $n_s$, $dn_s/d\ln k$, and~$Y$ and
they are taken by 
keeping the angular scale of the acoustic horizon constant.
Other derivatives are trivial and we do not show them here: the amplitude 
and reionization optical depth only rescale the amplitude, 
while dark energy density and its equation of state only change the 
angular scale of the acoustic horizon and make no effect after
this change is taken out.
Massive neutrinos have a minor effect on CMB
over the range of interest.
It is evident that none of these can mimic the additive phase shift 
generated by neutrinos. This signature thus uniquely identifies
the presence of neutrinos in the context of adiabatic initial 
conditions.

If matter power spectrum information is 
available, one can exploit the fact that while neutrino
perturbations {\it suppress\/} CMB anisotropies,
they {\it enhance\/} the matter power spectrum, Figs.~\ref{fig_comp}
and~\ref{dclnu}. 
For $\Delta N_{\nu}=1$ at a fixed $\om_b$, $k/h=1$\,Mpc${}^{-1}$,
and $z=0$ this gives $\Delta P_m/P_m\approx 0.12$,
of which 0.02 is due to neutrinos alone and the rest to 
the variation of $\omega_b/\omega_m$, as discussed in Sec.~\ref{sbs_matter_P}.
At the same time, the CMB power spectra are suppressed by $15\%$ 
at $l=3000$ if $Y$~is fixed. 
Whether or not this is a  useful method to break the degeneracies 
depends on the accuracy 
with which matter power spectrum can be extracted. We 
performed the analysis for the experiments in Table~\ref{table_forecasts} 
assuming weak lensing information as can be extracted
from the CMB itself. 
A Fisher matrix analysis with lensing reconstructed convergence 
power spectrum using  minimum variance maximum likelihood errors~\cite{HS_reconst} fails 
to improve significantly the accuracy on $N_{\nu}$, even with the 
polarization information, which allows for a better reconstruction of the 
convergence spectrum. This is in contrast to 
other parameters such as massive neutrinos, for which lensing 
information significantly improves the bounds~\cite{KaplKnoxSong_m_nu_CMB}.

Unless the astrophysical constraints on the primordial helium 
abundance~$Y$ will improve by an order of magnitude, 
$Y$ should be included in the list of unknown cosmological 
parameters for the accuracies of the order of $\D N_{\nu}\sim 10^{-1}$.
On the other hand, if we assume that the
neutrino fraction of the radiation is unchanged between 
BBN and recombination and there are no significant
$\nu/\bar\nu$ asymmetries
then BBN tightly limits the helium 
fraction, to $\D Y\sim 10^{-3}$ or better~\cite{HannestadConstr,BargPLB03}. 
For this reason, we performed the parameter prognosis 
with and without $Y$ parameter. 
As found in Sec.~\ref{sbs_Cl}, an adjustment of~$Y$ 
can compensate the changes 
in the acoustic and damping scales due to neutrino density
but it preserves the phase shift signature due to the gravity
of neutrino perturbations.  Correspondingly, the CMB 
limits on~$N_{\nu}$ somewhat broaden but remain tight
even with the unknown value of~$Y$.  Interestingly, the CMB 
data itself can be used to constraint the {\it primordial\/} 
helium abundance independently of the astrophysical measurements 
and the related systematic uncertainties.   For the considered
CMBPOL experiment with the polarization data included we find
$\sigma(Y)\simeq 0.005$ from the CMB alone.

\section{Conclusions}
\lb{sec_concl}

In this paper we study analytically the evolution of cosmological perturbations 
in the presence of ultra-relativistic neutrinos.  
While dynamical equations for cosmological perturbations
have been known for a while~\cite{Lifshitz46,PeeblesYu,BondSzalay,XiangKiang,MaBert}, 
their analytical solutions exist only in a handful of cases 
and are restricted to the fluid description. The best known examples,
\eg\ct{KS84}, 
are the solutions for CDM and photon-baryon plasma 
in the matter and radiation eras or in the subhorizon limit, 
and for superhorizon metric perturbations. 
In contrast, neutrinos cannot be modeled by a fluid and 
their phase space distribution should be considered.

Most of the recent publications abandoned the analytic approaches
and relied on numerical results from Boltzmann integrator codes.
While, in principle, there is nothing wrong with this, 
analytic solutions often lead to deeper understanding of the problem
that can reveal the new directions of exploration. They sharpen 
the focus on the features that are unique and cannot be mimicked by 
the variation of other parameters. 
Care must be exercised when performing 
numerical analysis and parameter forecasting for future experiments. 
The computational errors must be well controlled, 
otherwise they can lead to artificial breaking of degeneracies. 
Besides, the parameter space of forecasting is often 
small and with the addition of new parameters new degeneracies may 
open up. For example, while so far only simple parameterization of 
the primordial power spectrum have been explored, one could 
consider its more general parameterization, including
the running of the slope, its running, etc.\ 
as free parameters.
In this case analytic solutions can provide a better understanding 
of whether the limits on a given parameter are robust against 
adding new parameters.

We obtain analytic perturbative in $\rho\snu/\rho_r$
solutions for cosmological perturbations in the presence of 
ultra-relativistic neutrinos.   Much of the success is due
the following result, equally useful for fluid models.
We find that a simple redefinition of the independent 
dynamical variables that is consistent with their 
classical interpretation and preserves them on small scales,
eliminates all the time derivatives of the non-dynamical 
metric perturbations from the evolution equations 
in the Newtonian gauge.  The resulting description 
of cosmological perturbations acquires the advantages of 
an initial value Cauchy's problem while remains formulated
in the Newtonian gauge, which is fully fixed 
and is especially suitable for describing the physics of CMB.
Moreover, it turns out that even in the solvable 
fluid models the solutions for matter or radiation density 
perturbations, \cf eqs.\rfd{d_g_rad_sol_k}{dc_trf0}, 
appear far simpler in the redefined variables.  
In addition, these variables
are generally constant on superhorizon scales.

While most of the previous literature has focused on 
Fourier space analysis, we also consider perturbation Green's 
functions in real space~\cite{BB_PRL,BB_GRF}.
The latter become indispensable for the analytical study of
neutrino perturbations.  They also allow one to prove quickly that 
without neutrinos the cosine form of the acoustic density oscillations
in the radiation era {\it cannot} be 
modified by the gravitational feedback processes.

We use the zero, in the powers of $\rho\snu/\rho_r$, 
order solutions for neutrino perturbations to derive the 
analytic expressions for the CMB and matter density fluctuations
in the linear order.
We show that these first order solutions 
are for the most part sufficient for a quantitative interpretation 
of numerical solutions. Finally, we use the full numerical solutions 
from CMBFAST to derive parameter forecasts for various planned 
experiments. 
The presented methods can be straightforwardly extended to other 
applications such as tensor modes and massive neutrinos. We plan to address 
some of these in future work.

The distinctive cosmological features of ultra-relativistic 
neutrinos are due to their free streaming at the speed of light. 
The free streaming creates neutrino anisotropic stress, 
perturbing Newtonian metric even 
for superhorizon modes.
In real space, it also leads to perturbing the photon-baryon 
plasma beyond the acoustic horizon of a primordial perturbation.
In Fourier space this manifests itself as the phase shift in the
acoustic oscillations that is generated at horizon crossing. This phase 
shift is unique in the sense that for adiabatic perturbations 
no non-relativistic or fluid matter can generate it. 
The effect changes the phase additively and corresponds to 
$\Delta l \sim -4$ for $\Delta N_{\nu}=1$.  In contrast, 
any change in the angular scale of acoustic horizon 
acts as multiplicative rescaling $l \rightarrow \alpha l$. 
The two shifts are only degenerate at a single~$l$ and can be 
distinguished in general, \fig{dclnu}.
The effect is more visible in polarization, which has sharper 
acoustic peaks relative to temperature anisotropy, where the
density and velocity contributions to the $C_l$ oscillations 
partially cancel.
As a result, the precision of determining the
effective number of neutrino species 
can be improved dramatically if polarization information is included. 

Phase shift is not the only signature of neutrinos in CMB. 
The free streaming neutrinos also suppress the oscillation 
amplitude of the CMB modes entering the horizon in the radiation era. 
A change $\Delta N_{\nu}=1$ leads to $\Delta C_l/C_l\approx-0.04$.
Since the CMB modes entering in the matter era and not experiencing
the suppression
are limited to large scales, where 
sampling variance is large, this effect by itself cannot be 
extracted with high precision. However, neutrino perturbations 
amplify the CDM modes entering the horizon in the radiation era. 
The effect is further enhanced by the fact that 
while CMB physics is more sensitive to the ratios 
$\om_b/\om\sg$ and $\omega_m/\omega_r$, 
specifying the acoustic dynamics and the background evolution,
matter fluctuations are also sensitive to the ratio $\omega_b/\omega_m$, 
which, if fixing the natural CMB variables, cannot be held fixed
under varying~$\Neff$.
This changes the present matter fluctuation spectrum on small scales by
$\Delta P_m/P_m\approx 0.12$ for $\Delta N_{\nu}=1$, 
of which 0.04 is due to neutrinos and the rest to $\om_b/\om\sg$ 
variation.

It is unclear how accurately can this effect be extracted from local 
probes of large scale structure, such as galaxy clustering and 
weak lensing, since nonlinear evolution will complicate or,
in the case of galaxy clustering, prevent its determination on small 
scales. Weak lensing of CMB traces matter fluctuations on 
larger scales and higher redshifts than any other method.
It may be the optimal tool to use here since nonlinear reconstruction 
methods using the nongaussian information, especially from polarization 
data~\cite{HuOk_massfromCMB,HS_reconst},
can achieve high signal to noise on the projected convergence power spectrum. 
However, from the Fisher matrix analysis we find that lensing of CMB
 cannot improve the limits from primary CMB and its 
polarization significantly.

Finally, $\Neff$ variation changes the relativistic 
energy density and thus changes the relation between 
the expansion factor and time. This leads to a change in 
the proper size of the acoustic horizon and so in its 
angular size, which determines the 
positions of acoustic peaks. The angular size of the horizon, 
however, is degenerate with other parameters, such as those
changing the angular diameter distance to recombination. 
The change of the expansion time scales also modifies
the recombination process, the visibility function, and 
the angular damping scale.  The effect on the CMB power spectrum
can be significant, reaching $15\%$ power suppression 
at $l=3000$ for $\Delta N_{\nu}=1$,
\fig{dclnu}.  However, this can be mimicked by different
primordial helium abundance: a change of $\D\Neff=0.1$
is compensated by 
$\D Y\simeq -5\times10^{-3}$.
If CMB data is used to constrain $n_B/n\sg$ at BBN 
then the standard BBN
limits on~$Y$ are already at the level of 
$\D Y\lesssim 10^{-3}$~\cite{HannestadConstr,BargPLB03}, suggesting that 
$Y$ can be assumed fixed.  These limits are
not applicable in the models where 
the photon entropy changes between BBN and CMB decoupling
or in non-standard BBN models with $\nu/\bar\nu$ asymmetries 
or particle decays.

In summary, the effects of ultra-relativistic neutrinos on CMB and 
matter power spectrum are generally small.  This is why only weak 
limits on the neutrino background density
have been placed from the available observations.
On the other hand, neutrinos give rise 
to unique effects which exist on small scales and are thus 
less limited by sampling variance. As a consequence, future 
CMB experiments should be able to improve the limits significantly.  
While Planck will be able to 
determine $N\snu$ with a standard deviation 0.24, or 0.20 if $Y$ is constrained, 
a dedicated CMB polarization 
experiment should improve this bound even further, 
reaching accuracy levels of 0.09 without $Y$ constraint,
or 0.05 if $Y$ is constrained. 
This will allow one to test the details of neutrino decoupling 
and the scenarios giving 
rise to a nonstandard number of neutrino species.

\begin{acknowledgments}

SB is grateful to E.~Bertschinger, J.~R.~Bond, and A.~Loeb
for their encouragement in studying neutrino perturbations,
and to C.~Lunardini for a discussion of BBN physics.
We both thank E.~Bertschinger and V.~Mukhanov for
useful discussions, A.~Makarov and P.~McDonald for 
valuable help with numerical calculations,
and S.~Weinberg for suggestions allowing
to extend the superhorizon conservation laws of 
Sec.~\ref{sbs_init} to the inflationary epoch.
SB has been supported by Princeton University 
Dicke Fellowship;
US by Packard Foundation, Sloan Foundation,
NASA NAG5-1993 and NSF CAREER-0132953.

\end{acknowledgments}
\appendix

\section{Cosmological Dynamics}
\lb{apx_cosmo}

Given mutually non-interacting, except than gravitationally, 
groups of cosmological species~$\{a\}$, one can define\footnote{
  The definition is 
  $T^{\mu\nu}_a(x)\equiv ({2}/{\sqrt{-g}})\,{\d S_a}/{\d g_{\mu\nu}(x)}$,
  where $S_a$ are the terms of the action that describe the species~$a$.
}  
for each group 
an energy-momentum tensor $T^{\mu\nu}_{a}$ that 
satisfies the local conservation law $T^{\mu\nu}_{a}{}_{;\nu}=0$.
The total energy-momentum tensor
\be
T^{\mu\nu} = \sum_a T^{\mu\nu}_a~
\lb{T_tot}
\ee 
sources space-time curvature perturbations, 
as described by the Einstein equations 
$R_{\mu\nu}-\fr12g_{\mu\nu}R=8\pi G T_{\mu\nu}$.

\subsection{Background}

We study cosmological perturbations relative to
a spatially homogeneous and isotropic model with the 
metric 
\be
ds^2=a^2(\t)\lf(-d\tau^2+\g_{ij}dx^idx^j\rt)~.
\ee
The spatial part of the background metric may be written as
\be
\g_{ij}dx^idx^j=\fr{dr^2}{1-Kr^2}+r^2d\Omega^2~.
\ee
For most applications of this paper, except 
for the late time evolution of the matter perturbations,
the background curvature~$K$ can be neglected.  
In this case, we take $\g_{ij}=\d_{ij}$.

The background expansion rate with respect to 
the conformal time~$\t$ is denoted by $\H\equiv{\dot a}/{a} = aH$.
By the Friedmann equation, it equals
\be
\H^2&=& \fr{8\pi Ga^2}{3}\,\rho-K~,
\lb{Fr_eq}
\ee
where $\rho=-T^0{}_0$.

In the unperturbed universe, 
\be
T_a^0{}_0= -\rho_a~,\quad 
T_a^0{}_i=0~,\quad
T_a^i{}_j=\d^i_j\, p_a~.
\ee
Assuming that the species pressure
$p_a$ is uniquely specified locally by the species energy 
density $\rho_a$, we introduce
\be
w_a&\equiv& \fr{p_a}{\rho_a}~,   \lb{w_a_def}\\
c_a^2&\equiv& \fr{\,d p_a}{\,d\rho_a}=w_a+\fr{d w_a}{d\ln\rho_a}~.
\lb{c_a_def}
\ee
For the similar quantities applied to all the cosmological 
species together, with $\rho\equiv\sum_a \rho_a$ and $p\equiv\sum_a p_a$, 
we have
\be
w&\equiv& \fr{p}{\rho}~,  
\qquad \fr1{1+w} = \sum_a \fr{x_a}{1+w_a}~,
\lb{w_tot}
\\
c_w^2&\equiv&  \fr{\dot p}{\dot \rho}=
\lf(\fr{\d p}{\d\rho}\rt)_{\rm adiab}
=\sum_a x_a c_a^2~.
\lb{c_tot}
\ee
Here
\be
x_a\equiv\frac{\rho_a+p_a}{\rho+p}~,
\lb{x_def}
\ee
are species enthalpy abundances, satisfying
$\sum_a x_a = 1$.

It will prove useful to introduce ``reduced''  
enthalpy background densities 
\be
\g_a &\equiv& 4\pi Ga^2(\rho_a+p_a)~,
\lb{ga_def}
\\ 
\g &\equiv& 4\pi Ga^2(\rho+p) = \sum_a \g_a~.
\lb{g_def}
\ee
From definition\rf{x_def} and Friedman equation\rf{Fr_eq},
\be
\g_a = x_a\g~, \quad \g=3(1+w)(\H^2+K)/2~.
\lb{g_Fr}
\ee

Finally, we give the rate of change of some of the above
quantities with respect to the conformal time~$\t$.
Energy conservation requires
\be
\dot\rho_a = -3\H(\rho_a+p_a)~,
\lb{cons_back_a}
\ee
This and the Friedmann equations give
\be
\dot\H= \fr1a\,\fr{d^2a}{dt^2}= -\,\fr{4\pi Ga^2}{3}\,(\rho+3p)\,=~\, 
\lb{dH}
\\
=-\,\fr{1+3w}{2}\lf(\H^2+K\rt)~.
\nn
\ee
By eqs.\rfd{c_tot}{cons_back_a},
\be
\dot w = 3\H(1+w)(w-c_w^2)~.
\ee
By eqs.\rfd{c_a_def}{c_tot} and\rf{cons_back_a}, 
\be
\dot\g_a = -\H(1+3c_a^2)\,\g_a\,,\quad\dot\g = -\H(1+3c_w^2)\,\g\,.\qquad
\lb{gammaH}
\ee
From the last two equations we also see that
\be
\dot x_a = \lf(\fr{\g_a}{\g}\rt)\odot = -3\H(c_a^2-c_w^2)x_a~.
\lb{dot_x_a}
\ee

\subsection{Matter perturbations}

We parameterize $T_a^{\mu}{}_{\nu}$ perturbations 
by the {\it particle number} overdensity\footnote{
  Many authors use $\d_a$ for the {\it energy\/} overdensity
  $\d\rho_a/\rho_a = (1+w_a)\d_a$, in our notations.
}~$\d_a=\d n_a/n_a$, 
the peculiar velocity vector $v^i_a$,
and the anisotropic stress perturbation~$\Pi_a^{ij}$:
\be  
 \ba{rcl}
T_a^0{}_0&=&-\,(\rho_a+\d\rho_a)~,\qquad  
               \d\rho_a \equiv (\rho_a+p_a)\,\d_a~,
\vsp\\
T_a^0{}_i&=& (\rho_a+p_a)\,v_{i\,a}~,
\vsp\\
T_a^i{}_j&=& \d^i_j(p_a+\d p_a)+(\rho_a+p_a)\,\Pi_a^i{}_j~,
      \quad  \Pi_a^i{}_i \equiv 0~.
\vsp
  \ea
\lb{Tmn} 
\ee
In scalar modes the 3-vectors~$v_{i\,a}$ and
3-tensors~$\Pi_a^i{}_j$ are derivatives of 
scalar velocity potentials~$u_a$,
\be
v_{i\,a} = -\Nbi u_a~,
\lb{u_def}
\ee
and anisotropic stress potentials~$\pi_a$,
\be
\Pi_a^i{}_j = \fr32\lf(\Nb^i\Nbj-\fr13\,\d^i_j\Nb^2\rt)\pi_a~.
\lb{pi_def}
\ee
With the normalization\rf{pi_def}, $\Nbi\Nbj\Pi_a^{ij}=\Nb^4\pi_a$.
The potentials~$\pi_a$ 
are related to some of the alternative variables
used to describe anisotropic stress 
as $\sigma=-\Nb^2\pi$ for the variable of\ct{MaBert} 
and $\Pi=-(3/2)(1+\rho/p)\Nb^2\pi$ for\ct{KS84}.

For the perturbations in the total energy-momentum tensor~$T^{\mu}{}_{\nu}$,
parameterized analogously to eqs.\rf{Tmn},
\be
\d= \sum_a x_a\d_a\,,\quad 
u= \sum_a x_a u_a\,,\quad  \pi=\sum_a x_a \pi_a\,, \qquad
\lb{u_tot}
\ee
where $x_a$ were defined in eq.\rf{x_def}.

\subsection{Metric and Gauges}
\lb{sbs_gauges}

Under a reparameterization of spacetime coordinates 
(gauge transformation)
\be
\~\t= \t + \d\t(\t,\r)~,\qquad
\~x^i = x^i + \d x^i(\t,\r)~,
\lb{gauge_var}
\ee
the perturbation variables transform as
\be
\~\d_a = \d_a + 3\H\d\t\,,\quad
\~u_a = u_a - \d\t\,,\quad   \~\pi_a =  \pi_a\,. \quad
\lb{gauge_tr}
\ee
Note that the following quantities are gauge invariant:  
$\d_a + 3\H u_a$, or $\d_a-\d_b$, or $u_a - u_b$.

The perturbed metric of space-time can be parameterized as
\be
  \ba{rl}
ds^2=a^2(\t)&\!\!\lf[(-1-2A)d\tau^2 - 2B_i\,d\t dx^i \,+ \rt. \vsp
\\ 
     &\lf.+\,(1-2H_L)d\r^2 - 2\bar H_{ij}dx^idx^j\rt], \vsp
  \ea
\lb{ds_gen}
\ee
where $\bar H^i_i\equiv0$.   For scalar modes,
the 3-vector~$B_i$ and the 3-tensor~$\bar H_{ij}$
can be written as spatial derivatives of
scalar functions:
\be
B_i= \Nbi\,b~,\qquad 
\bar H^i_j= \lf(\Nb^i\Nbj-\fr13\,\d^i_j\Nb^2\rt)\chi~.
\lb{ds_scalar}
\ee

The metric perturbations transform 
under the gauge transformation\rf{gauge_var}
as
\be
\~A&=&A-\d\dot\t - \H\d\t~, \Vsp 
\lb{g_transA}\\ 
\~B_i &=& B_i - \Nbi\d\t + \d\dot x_i~, 
\lb{g_transB}\\
\~H_L&=&H_L+\H\d\t+\fr13\Nbi\d x^i~,\Vsp 
\lb{g_transHL}\\ 
\~{\bar H}{}^i_j &=&\bar H^i_j
                + \fr12\lf({\Nb^i\d x_j + \Nbj\d x^i}\rt)
                - \fr13\,{\d^i_j\Nb_{\!k}\d x^k}\,.\qquad
\lb{g_transHT}
\ee
For scalar perturbations with
\be
\d x_i \equiv \Nbi\,\d\lambda~
\nn
\ee
the potentials $b$ and $\chi$ in eq.\rf{ds_scalar} transform as
\be
\~b = b-\d\t+\d\dot\lambda~, \qquad  
\~\chi = \chi + \d\lambda~.\qquad
\lb{g_trans_scal}
\ee

In the {\it conformal Newtonian\/} (longitudinal)
gauge the gauge conditions on scalar perturbations 
are $b^{(N)}\equiv 0$ and $\chi^{(N)}\equiv 0$.
For brevity, we refer to this gauge as the ``Newtonian''.
Defining $\Phi\equiv A^{(N)}$ and $\Psi\equiv H_L^{(N)}$
we arrive at the metric of eq.\rf{Newt_gauge_def}.  

  In the {\it synchronous} gauge one sets 
$A^{(s)}\equiv 0$ and $B_i^{(s)}\equiv 0$. 
The observers who are at rest in the synchronous gauge
are free falling in the gravitational field and their
locally measured proper time sets the coordinate time.
By eqs.\rf{g_trans_scal}, the gauge transformation from 
the synchronous to the Newtonian gauge
is $\d\t=-\dot\chi^{(s)}$, $\d\lambda=-\chi^{(s)}$.
Hence, from eqs.\rfd{g_transA}{g_transHL},
the Newtonian potentials  are related to the 
scalar metric perturbations in the synchronous gauge as
\be
\Phi=\ddot\chi^{(s)}+\H\dot\chi^{(s)},\,~
\Psi=H_L^{(s)}-\H\dot\chi^{(s)}-\fr13\,\Nb^2\chi^{(s)}\!.~\quad
\lb{Synch2Newt}
\ee

For the reverse transformation from the Newtonian
to the synchronous gauge, by
eqs.\rfd{g_transA}{g_trans_scal}, $\d\t(x)$ 
and $\d\l(x)$ are any functions satisfying
\be
\d\dot\t+\H\d\t=\Phi~,\qquad \d\dot\l=\d\t~.
\lb{gtr_N2s}
\ee
The initial values $\d\t(\t\i,\r)$ and $\d\l(\t\i,\r)$
can be chosen arbitrarily, corresponding to the residual
gauge freedom of the synchronous gauge.
The metric perturbations in the synchronous gauge are obtained as
\be
H_L^{(s)}=\Psi+\H\d\t+\fr13\Nb^2\d\l~,\qquad
\chi^{(s)}=\d\l~.
\lb{Newt2synch}
\ee

In the {\it spatially flat\/} gauge the scalar perturbations of
the spatial part of the metric are absent: $H_L^{(f)}\equiv 0$
and $\chi^{(f)}\equiv 0$.  By eqs.\rfd{g_transHL}{g_trans_scal},
this gauge has no~residual gauge freedom.
It is obtained from the Newtonian gauge with $\d\t=-\Psi/\H$, $\d\l=0$.
In terms of the Newtonian potentials, the scalar metric perturbations in 
the spatially flat gauge are
\be
A^{(f)}=\Phi+\Psi+\lf(\fr{\Psi}{\H}\rt)\odot~,\qquad
b^{(f)}=\fr{\Psi}{\H}~.
\ee

  The {\it comoving\/} gauge is defined as the gauge in which the scalar 
components of the total 
$T^0_i$ vanishes: $u^{(c)}\equiv 0$.  By eq\rf{gauge_tr},
the transformation from Newtonian to the comoving
gauge is achieved with $\d\t=u^{(N)}$.  For the second gauge condition, 
fixing~$\d x^i$ in eq.\rf{gauge_var}, it is convenient to 
choose $\chi^{(c)}\equiv 0$.  Then $\d x^i=0$.  
By eq.\rf{g_transHL},
the spatial curvature potential in the comoving gauge is
related to the Newtonian gauge variables as
\be
\R \equiv H_L^{(c)} = \Psi + \H u^{(N)}~.
\lb{comov_curv_def}
\ee
For the remaining scalar metric perturbations 
eqs.\rfd{g_transA}{g_trans_scal} give
\be
A^{(c)}=\Phi-\dot u^{(N)} - \H u^{(N)}~,\qquad
b^{(c)}=-u^{(N)}~.\
\ee
By the following eq.\rf{dot_psi} and eq.\rf{g_Fr},
the ``comoving curvature''~$\R$ can be easily transformed
into its conventional form
\be
\R =  \Psi + \fr2{3(1+w)}\lf(\Phi+\fr{\dot\Psi}{\H}\,\rt)~.
\lb{comov_curv_st}
\ee

The {\it uniform density\/} gauge corresponds to the condition
$\d^{(u)}\equiv 0$.
Hence, eq\rf{gauge_tr}, it is obtained from the Newtonian gauge
with $\d\t=-\d^{(N)}/(3\H)$.  Taking $\chi^{(u)}\equiv 0$
to be the second gauge condition,
one finds that the 3-curvature potential~$H_L$ in 
the uniform density gauge is
\be
\zeta\equiv  H_L^{(u)} = \Psi - \fr13\,\d^{(N)}~.
\lb{zeta_def}
\ee
By Einstein equations\rfs{nb2_psi}{dot_psi},
the curvature perturbations in the comoving,  uniform density,
and Newtonian gauges are related as 
\be
\R = \zeta + \fr{\Nb^2\Psi}{3\g}~.
\lb{curv_related}
\ee

\subsection{Conservation and Einstein Equations}

\subsubsection*{\it Newtonian gauge:}

The dynamics of the species density and velocity 
perturbations follows from the conservation 
law $T^{\mu\nu}_{a}{}_{;\nu}=0$ as
\be
 \ba{rcl}
\dot\d_a&=& \Nb^2u_a+3\dot\Psi~,\Vsp\\
\dot u_a &=& c_a^2\d_a-\chi_a u_a
               +\Nb^2\pi_a+\Phi~
 \ea
\lb{dot_du}
\ee
(for the scalar mode), where
$\chi_a \equiv \H\lf(1-3c_a^2\rt)$
is the Hubble drag rate for the species~$a$.
The evolution of the anisotropic stress potential~$\pi_a$
is determined by the internal dynamics of the species.

The linearized Einstein equations in Newtonian 
gauge,\ct{MaBert,Bert_LesHouches}, 
are easily reduced to
\be
\Nb^2\Psi-3\H\lf(\dot\Psi+\H\Phi\rt)&=& \g\d~,\vsp
\lb{nb2_psi}\\
\dot\Psi+\H\Phi &=& \g u~,\vsp
\lb{dot_psi}\\
\ddot\Psi+\H\lf(2\dot\Psi+\dot\Phi\rt)-3w\H^2\Phi
  &=&  \g\fr{\d p}{\d\rho}\,\d  + \g\Nb^2\pi,\qquad \quad \vsp
\lb{ddot_psi}\\
\Psi-\Phi &=& 3\g\pi~,\vphantom{\fr11}
\lb{psi-phi}
\ee
where $\g$ is introduced by eqs.\rfd{g_def}{g_Fr}
and the background is assumed spatially flat.
If for all the species $\d p_a = c_a^2 \d \rho_a$ then
$({\d p}/{\d\rho})\,\d=\sum_a x_a c_a^2\d_a$.
By equations\rfs{nb2_psi}{dot_psi},
\be
\Nb^2\Psi&=& \g\d^{(c)}~,
\lb{Psi_Pois}
\ee
where~$\d^{(c)}=\sum_a x_a\d^{(c)}_a$ 
is the averaged particle number 
density perturbation in the {\it comoving\/} gauge,
\be
\d_a^{(c)} = \d_a + 3\H u~.
\lb{d_comov}
\ee

\subsubsection*{{\it Synchronous gauge:}}

In Appendix~\ref{apx_local} we refer to the evolution equations
in the synchronous gauge.  In this gauge the
energy and momentum conservation equations give
\be
 \ba{rcl}
\dot\d_a^{(s)}&=& \Nb^2u_a^{(s)}+3\dot H_L^{(s)}~,\Vsp\\
\dot u_a^{(s)} &=& c_a^2\d_a^{(s)}-\chi_a u_a^{(s)}
               +\Nb^2\pi_a~.
 \ea
\ee
The corresponding linearized Einstein equations, \eg\ct{MaBert},
in spatially flat background are  
\be
\Nb^2\lf(H_L^{(s)}-\fr13\Nb^2\chi^{(s)}\rt)-3\H\dot H_L^{(s)}&=& 
                     \g\d^{(s)}~,\vsp
\lb{nb2_psi_syn}\\
\dot H_L^{(s)}-\fr13\Nb^2\dot\chi^{(s)} &=& \g u^{(s)}~,\vsp
\lb{dot_psi_syn}\\
\ddot H_L^{(s)} +\H\dot H_L^{(s)}
  =  \g\lf(\fr{\d p}{\d\rho} + \fr13\,\rt)\d^{(s)},  
\hspace{-1.8cm}  &&
\lb{ddot_psi_syn}\\
\ddot\chi^{(s)}+2\H \dot\chi^{(s)}-\lf(H_L^{(s)}-\fr13\Nb^2\chi^{(s)}\rt)  
               &=& -\,3\g\pi~.\qquad\quad
\lb{psi-phi_syn}
\ee

\subsection{Dynamics in phase space}

Six variables specify the coordinates of a particle
in phase space at a given time.  For them, 
following\cts{BondSzalay,MaBert},
we take the comoving  
coordinates of the particle $r^i$ and the comoving momenta
\be
q_i\equiv a p_i
\lb{q_def}
\ee
where $p_i$ are the proper momenta measured by a comoving observer,
who is at rest with respect to the coordinate frame.  
For a particle with a mass~$m_a$ the momenta that are canonically 
conjugate to the variables~$r^i$ are
$P_i = m_a{dx_i}/{\sqrt{-ds^2}} = (1-\psi)q_i$.

The particle density in phase space is specified by the canonical
phase space distribution~$f_a(r^i,P_j,\t)$:
\be
dN_a=f_a(r^i,P_j,\t)\, d^3r^i d^3P_j~
\lb{f_def}
\ee
for every species of particles and their states of polarization~$a$.
The energy-momentum tensor of the species~$a$ is given in 
the Newtonian gauge by the following simple expression up to the first order 
of cosmological perturbation
theory~\cite{MaBert,BondSzalay}:
\be
T^{\mu}_a{}_{\nu}=\int d^3p_i\,\fr{p^{\mu}p_{\nu}}{p^0}\,f_a~,
\lb{Tmunu_expl}
\ee
with $p^0\equiv -p_0 \equiv \sqrt{(q/a)^2+m_a^2}$ and
$p^i\equiv p_i = q_i/a$. 
Below we drop the species label 
$a$ when referring to any sort of particles in general.

The evolution of the phase space distributions
obeys the Boltzmann equation:
\be
\dot f+\dot r^i\,\fr{\pd f}{\pd r^i}
+\dot q\,\fr{\pd f}{\pd q}+\dot n_i\fr{\pd f}{\pd n_i}
=\lf(\fr{\pd f}{\pd\t}\rt)_C,
\lb{Bolz_gen}
\ee
where $f$ is considered as a function of the coordinates 
$r^i$, $q\equiv |q_i|$, $n_i\equiv {q_i}/{q}$, and~$\t$.
The right hand side of eq.\rf{Bolz_gen} describes the change of
the phase space density due to particle collisions.
Linearizing the Boltzmann equation relative to an unperturbed
background phase space distribution,
$f= \bar f+\d f$,
in the Newtonian gauge one finds,\ct{MaBert}: 
\be
\lf(\d f\rt)\dot{\vphantom{t}}
+\fr{q_i}{\eps}\,\Nbi\lf(\d f\rt)
+q\,\fr{\pd\!\bar f}{\pd q}\lf(\dot\Psi-\fr{\eps}{q}\,n_i\Nbi\Phi\rt)=~\qquad
\lb{Boltz_lin}
\\
=\lf(\dot f\rt)_C-\lf(\dot{\bar f}\rt)_C~,
\nn
\ee
where $\eps\equiv \sqrt{q^2+a^2m^2}$.

\section{Locality of Adiabatic Green's Functions}
\lb{apx_local}

In this Appendix for any ``reasonable'' cosmological system 
we construct the perturbation Green's functions in that all the scalar 
gravitational and matter distribution potentials are identically zero 
in the Newtonian or synchronous gauges
beyond the Green's function particle horizon $|x|>\t$.   
We argue that the perturbations formed
with these Green's functions by their convolution with any smooth 
kernel are adiabatic, as defined in footnote\rfp{def_adiab}.
A Green's function Fourier component with any wave number~$k$ 
presents a non-zero non-decaying adiabatic mode.

We assume that all the scalar perturbations of matter distributions, 
classical fields, or the metric tensor in our system can be 
parameterized by a sum of $3$-scalar functions (matter potentials)
acted by polynomials of~$\Nbi$. 
This is easily achieved for classical fields that are scalars,
vectors, or tensors of a higher integer rank.  
Complications would arise for spinor and other non-integer spin
fields.  However, being fermions, such fields can not support 
coherent semiclassical excitations.
Without sacrifice of generality, we require that the 
polynomials of~$\Nbi$ are homogeneous $l=0,\,1,$ etc.\ 
degree polynomials
that transform under spatial rotations as irreducible tensors of 
the rank~$l$.
We treat all the dynamical degrees of freedom, whether they are
dynamical ``coordinates'' or ``velocities'', on equal footing, 
implying the Hamiltonian formulation of the system
classical dynamics in the coordinate time of the chosen gauge.

We set the Green's function initial conditions at 
a spatial slice of an infinitesimal thickness at 
a coordinate time $\t\i>0$, which is eventually sent to zero.
Within this slice we impose the Newtonian gauge conditions
$b\equiv 0$ and $\chi\equiv 0$ on the scalar metric  
potentials\rf{ds_scalar}. 
We require that, in the Newtonian gauge,
all the matter distribution or field perturbation
{\it potentials\/} initially vanish 
for $|x|>\t\i$. (The Green's functions considered
in this paper are homogeneous in the normal to~$x$ 
spatial directions $y$ and $z$, Sec.~\ref{sec_grf}.) 
Inside the interval $|x|\le\t\i$, to be specific, 
we set all the $l\ge 1$ matter distribution potentials
at~$\t\i$ to zero and adjust the initial conditions
for classical fields, if any are present, so that
$u(\t\i)=0$.  (Then the metric on the initial spatial slice 
satisfies the comoving gauge condition as well.  
The spatial slices in the Newtonian and comoving gauges will, 
in general, differ for $\t>\t\i$.)  At the time~$\t\i$ by eq.\rf{Psi_eq}
\be
\Nb^2\Psi-3\g\Psi=\g d~.
\lb{Phi_eq_tin}
\ee

Let us demand that $\Psi(\t\i,x)=(1/\t\i)\,\bar\Psi_0(x/\t\i)$ 
where $\bar\Psi_0(\chi)$ is an arbitrarily chosen {\it even\/} function 
that {\it vanishes\/} for $|\chi|>1$ 
and that has a non-zero integral over $d\chi$ from
$-1$ to $1$.  Then the coordinate ``particle number'' density
perturbation~$d(\t\i,x)$ should be given by the right hand side 
of eq.\rf{Phi_eq_tin}.
The corresponding proper number density perturbation~$\d$ at~$\t\i$ is
\be
\d=d+3\Psi=\fr1{\g}\,\Nb^2\Psi~.
\ee
We set all the species initial densities and the 
initial conditions for classical fields so that
$\d_a(\t\i,x)=(1/\g)\Nb^2\Psi$ and all the other matter potentials
are {\it even\/} and {\it vanish\/} for $|x|>\t\i$.
For interacting species, the separation of the total energy-momentum 
tensor~$T^{\mu\nu}$ into $T^{\mu\nu}_a$ and so the definition of
the species density perturbations~$\d_a$ may be ambiguous.  In this case, 
we take any of the possible definitions.  In the limit~$\t\i\to0$
all of them lead to the physically identical adiabatic
Green's functions.  If any of the matter potentials or fields remain 
unfixed, we take them unperturbed at~$\t\i$ for all~$x$.

The motivation for these initial conditions is to construct 
an initially localized almost ``pure'' curvature perturbation.  
For any~$\t\i>0$, the curvature perturbation must be generated 
by some non-zero matter disturbance.  The above procedure attempts to 
restrict the required matter inhomogeneity so that in
the limit $\t\i/\l\to 0$ only curvature appears to be 
perturbed.  Indeed, after the convolution of the constructed Green's 
function with a primordial fluctuation field~$A(x)$ that 
is smooth on the comoving scales below some~$\l>0$, 
the number density perturbation of any species at~$\t\i$ 
by eq.\rf{Phi_eq_tin} tends to
\be
\lim_{\t\i\to0}\langle\d(x)\rangle 
&=& \lim_{\t\i\to0} \int dx' A(x')\,\fr{\Nb^2\Psi(x-x')}{\g}=\qquad~~
\\ 
&=&\lf(\int_{-1}^{1}d\chi\,\bar\Psi_0(\chi)\rt)
   \lim_{\t\i\to0}\fr{\Nb^2\! A(x)}{\g}~.
\nn
\ee
The last limit vanishes as $\t\i^2/\l^2$.  Since the other initial
matter distributions are chosen unperturbed, 
they remain unperturbed after the smoothing.  
However, the $\t\i\to0$ limits of 
the smoothed comoving curvature perturbation~$\R=\Psi+\H u$
as well as the variable $\zeta=-d/3$ at~$\t\i$ 
are non-zero:
\be
\lim_{\t\i\to0}\langle\R(x)\rangle \!= \!\lim_{\t\i\to0}\langle\zeta(x)\rangle
\!=\!\! \lf(\int_{-1}^{1}d\chi\,\bar\Psi_0(\chi)\rt) \!A(x)\,.~~
\lb{conv_curv_init}
\ee
Given the locality of the Green's function, which is proved next, 
and the initially vanishing smoothed matter perturbations, 
we argue at the end of Sec.~\ref{sbs_init} that the smoothed 
perturbations $\zeta$ and $\R$ remain constant while $\t\ll\l$.
The Fourier transformation is a convolution
with the kernel $A(x)=\exp(-ikx)$.  
Therefore, it gives non-decaying non-zero  
curvature perturbation modes\rf{conv_curv_init}, as long as 
$\int_{-1}^{1}d\chi\,\bar\Psi_0$ is chosen different from zero.

The smoothed density perturbation $\langle\d(x)\rangle$ will remain as small
as $O(\t^2/\l^2)$ for $\t\i<\t\ll\l$ in the comoving gauge, where it is 
always related to $\Psi$ by eq.\rf{Psi_Pois}.
In other gauges, including the Newtonian, the 
proper density perturbation $\langle\d(x)\rangle$ 
and other proper matter perturbations do not need to vanish 
for~$\t\i<\t\ll\l$.  For example, the species proper density
and velocity perturbations are given by eq.\rf{gauge_tr}, where
$\d\t$ is the time lapse between the comoving and the considered gauge.

Now we prove that in a synchronous gauge and the set 
initial conditions at~$\t\i$ all the matter and gravitational
Green's functions remain zero for $\t>\t\i$ 
beyond the particle horizon $|x|>\t$.  
The coordinate transformations $\d\t$ and $\d x_i=\Nbi\d\l$ 
from the Newtonian to the synchronous gauge can be chosen
so that $\d\t(\t\i)=0$ and $\d\l(\t\i)=0$ for all~$x$.
Then, from eqs.\rf{gtr_N2s},
$\d\dot\t(\t\i)=\Phi(\t\i)$ and $\d\dot\l(\t\i)=0$.
Eq.\rf{Newt2synch} gives that the synchronous gauge potentials 
$H_L^{(s)}$ and $\chi^{(s)}$ and their first time derivatives vanish
at $\t\i$ for $|x|>\t\i$. Same is true for the transformed
initial perturbations of all the matter distributions and their 
rates of change.

Causality requires that in the synchronous gauge the matter 
and the metric remain unperturbed beyond
the particle horizon of the original perturbation.  
This means that $H_L^{(s)}$, $\Nb^2\chi^{(s)}$,
\cf eq.\rf{ds_scalar}, $\Nb u^{(s)}$, $\Nb^2\pi$, etc.\ are 
zero for~$|x|>\t$.  We argue that the potentials
$\chi^{(s)}$, $u^{(s)}$, $\pi$, etc.\ themselves are also zero
for~$|x|>\t$.
As an illustration, let us consider the Euler's equation
for  $\dot v_x\equiv -\Nbs{x}\dot u$
that in the synchronous gauge reads
\be
\Nbs{x}\,\dot u^{(s)} = \Nbs{x}\lf[c_w^2\d^{(s)}-\H(1-3c_w^2)u^{(s)}
    +\Nb^2\pi\rt]\,.
\lb{pot_evol_example}
\ee
The gradient in this and the similar equations can be dropped 
after we {\it define} the expressions under the gradients to 
be equal at a certain~$x_0$, which may depend on~$\t$.  
Consistent definition of the potentials is obtained if $x_0$ 
is chosen outside of the perturbation horizon; suppose, 
$x_0(\t)<-\t$.
Since a sufficiently high spatial derivative of every 
synchronous gauge potential gives a perturbation of some
matter distribution or a metric tensor component and thus, by above, 
vanishes for $x<-\t$, the potentials themselves can be chosen
to vanish in that interval for all~$\t$. 
We prove that the potentials then also vanish for $x>\t$.

Let $\{f_l\}$ be multipole potentials for some matter distribution
perturbation~$F(\r,\n)$, which may depend on other, 
not displayed, phase space or internal coordinates,
\be
F(\r,\n)\equiv \sum_{l=0}^{\infty}(-1)^l(2l+1)\,P_l
                \lf(\fr{n_i\Nbi}{\Nb}\rt)\Nb^l f_l~,
\lb{mult_expan_ap}
\ee
where $i$ runs over the spatial coordinates~$x$, $y$, and~$z$.
We suppose that the distribution perturbation dynamics can be 
described as
\be
\dot F(\r,\n) 
   =\lf({\mbox{local in space functional}\atop 
     \mbox{of perturbation fields}}\rt).
\lb{Fevol_abstract}
\ee
The functional on the right hand side is linear in the linear order
of perturbation theory.  
Multiplying both sides of eq.\rf{Fevol_abstract} by a
spherical function~$Y_{lm}(\n)$, integrating over the solid angle 
$d^2\Omega_{\n}$, and applying the identity
\be
\int d^2\Omega_{\n}~Y_{lm}(\n)~P_l\lf(\fr{n_i\Nbi}{\Nb}\rt)=
\fr{4\pi}{2l+1}\,Y_{lm}\!\lf(\fr{\bm{\Nb}}{\Nb}\rt)\,,
\ee
on the left hand side of eq.\rf{Fevol_abstract}
we find $\dot f_l$ times a homogeneous 
$n$-degree polynomial $Q_{lm}(\Nbs{x},\Nbs{y},\Nbs{z})$.
The polynomials $\{Q_{lm}\}_{m=-l}^l$
transform under spatial rotations as~$Y_{lm}$.
Since all the perturbation potentials are defined to be 
3-scalars,  $\Nbi$ are the only quantities that may appear
on the right hand side and that transform non-trivially
under spatial rotations.
Therefore, after the convolution with~$Y_{lm}(\n)$,
every term on the right hand side of the linear
equation\rf{Fevol_abstract} must contain the same polynomial
$Q_{lm}(\Nbs{x},\Nbs{y},\Nbs{z})$ as that appears on the left hand side.
Putting $m$ to $0$ and applying the resulting evolution equation
to the $y$ and $z$ independent Green's functions,
we find $(\Nbs{x})^l\dot f_l$ on the left and at least~$l$ derivatives
$\Nbs{x}$ in every term on the right hand side.
After $l$ integrations over $dx$ with zero initial values
at $x<-\t$ we find an equation of the form
\be
\dot f_l(\r) = \lf({\mbox{local in space functional}\atop 
     \mbox{of scalar potentials}}\rt).
\lb{fevol_abstract}
\ee

The corresponding equations for the studied in the main text 
system of photon-baryon and CDM fluids and ultra-relativistic neutrinos
are eqs.~(\ref{dot_d}--\ref{dot_u},\,\ref{dot_dl}).
The scalar potential evolution can be generalized
to describe photon-baryon Thompson scattering and photon polarization
as shown in the Appendix of\ct{BB_GRF}.
The evolution of the synchronous metric potentials
$H_L^{(s)}$, $\chi^{(s)}$, and $\dot\chi^{(s)}$ 
can as well be presented in the form\rf{fevol_abstract}
using eqs.\rft{nb2_psi_syn}{dot_psi_syn}{psi-phi_syn}.

Thus the dynamics of all the potentials 
in the synchronous gauge can be reduced
to an initial value Cauchy's problem.  
The initial conditions at $\t=\t\i$ were chosen even.
Assuming the dynamical equations are invariant under $x\to -x$
inversion, the resulting solution of the Cauchy's problem
will remain even for all~$\t$.
Hence, the potential values for $x>\t$ are equal to
those at $x<-\,\t$, \ie, vanish in the synchronous gauge.

The reverse transformation to the Newtonian gauge
is achieved with $\d\t=-\dot\chi^{(s)}$ and 
$\d\l=-\chi^{(s)}$, see Sec.~\ref{sbs_gauges}.
These functions have been proved
to vanish outside of the horizon.
Therefore, all the matter multipole potentials in the Newtonian gauge 
and the gravitational potentials $\Phi$ and $\Psi$,
related to the synchronous metric perturbations by eqs.\rf{Synch2Newt},
as well vanish for $|x|>\t$.

\begin{widetext}
\vspace{1mm}

\section{$O(\Rnu)$ order calculations in the radiation era}
\lb{apx_ORnu}

\subsubsection*{Gravitational potentials}

In Secs.~\ref{sbs_grf_pot}--\ref{sec_nuphot} 
we derived integro-differential 
equations satisfied by the potentials 
$\Phi_{\pm}\equiv (\Psi\pm\Phi)/2$
during radiation domination.  We can rewrite
these equations, eqs.~(\ref{eqcls}\,,\ref{D_l_sol_real}) 
and~(\ref{Php_sol},\,\ref{Fm_def},\,\ref{p_g2Phi}--\ref{p_g_sol})
as
\be
\bar\Phi_-'(\chf)
    &=&\Rnu\!\lf[\fr32\,\zeta\i\,\chf\lf(1-\chf^2\rt)\theta(1-|\chf|)
   -6\!\int_{-1}^{\chf}\!\!d\chf'\!
       \int_{-1}^{1}\!\! d\chf''~\fr{\bar\Phi_+(\chf'')\, \chf' P_2(\chf')
        +\bar\Phi_+(\chf')\, \chf''P_2(\chf'')}
        {\chf''-\chf'}
         \rt]
\lb{dPhm_sol}
\ee
and
\be
\bar\Phi_+(\chf)=\lf({\chf^2-\frac13}\rt)\lf[
       p_{\Phi}\,\theta\lf(\fr1{\sqrt3}-|\chf|\rt) - F_-(\chf)\rt],
\lb{Php_sol_ap}
\ee
with
\be
F_-(\chf)&=& \fp\int_{-1}^{\chf}d\chf'\,
        \frac{\chf'^2-1}{\lf(\chf'^2-\frac13\rt)^2}\,\bar\Phi_-'(\chf')~,
\lb{Fm_def_ap}\\
p_{\Phi}&=&-\,\fr{\sqrt3(1-\Rnu)}{1-2\Rnu}\lf[\fr32\,\zeta\i
       -\int_{-1}^1 d\chf\,F_-(\chf)\rt]~.
\lb{p_Ph_ap}
\ee
In this Appendix we use them to calculate in $O(\Rnu)$ order 
the prefactors at the 
singular terms in the photon and CDM density Green's functions.
Up to small $O(\Rnu^2)$ corrections, this yields
the density modes on subhorizon scales.

The quantity $\bar\Phi_-'$ in $O(\Rnu)$ order is obtained by
substituting  the zeroth order solution for $\bar\Phi_+$, eq.\rf{Php_0},
into the right hand side of eq.\rf{dPhm_sol}.
The substitution gives
\be
\bar\Phi_-'&=&\Rnu\zeta\i\lf\{\lf[I_1(\chf)-I_0\sign\!\chf\rt]\theta(1-|\chf|)
       - \lf[I_2(\chf)-I_0\sign\!\chf\rt]\theta\!\lf(\fr1{\rtt}-|\chf|\rt)\rt\} 
      +O(\Rnu^2)~ \nn
\lb{Phi_m_1}
\ee
where
\be
&& I_1(\chf) \,\equiv\, 
     9\chf\lf(\fr12\chf^4-\fr7{18}\chf^2+\fr1{9}\rt)
      - \fr{(3\chf^2-1)^3}{4\rtt}
          \ln\!\lf|\fr{\chf+\fr1{\rtt}}{\chf-\fr1{\rtt}}\rt|
~,\\
&& I_2(\chf) \,\equiv\, \fr{3\rtt}4\lf(1-\chf^2\rt)\lf[
         2\chf\lf(3\chf^2+1\rt)-(3\chf^4+1)\ln\!\lf|\fr{1+\chf}{1-\chf}\rt|
         \rt]~,
\nn\\
&& \qquad\Vsp 
   I_0\,\equiv\,I_1(1)\,=\,I_2\!\lf(\fr{1}{\rtt}\rt)=2-\fr2{\rtt}\,\ln(2+\rtt)~.\Vsp \nn 
\ee
\end{widetext}

\subsubsection*{Photon singularities}

The magnitude of the photon density acoustic spike~$p_{\g}$,
eq.\rf{p_g_sol}, can be calculated analytically using 
$F_-$ definition\rf{Fm_def_ap} and the
above expression for $\bar\Phi_-'$.
For this we employed the symbolic 
calculation program ``Mathematica''
({\tt http://www.wolfram.com}\/).  
The resulting expression turns out rather lengthy but is 
easily evaluated~to
\be
p_{\g}\simeq \fr32\,\zeta\i\lf[1-0.2683\Rnu+O(\Rnu^2)\rt]\,.
\lb{pg1}
\ee
The residue~$r_{\g}$, as defined by eq.\rf{p_g_sing_gen},
follows from eqs.\rfd{dgam_gen}{Php_sol} as
\be
r_{\g}\!= \!\bar\Phi_+\!\!\lf(\fr1{\sqrt3}\rt)\!
  = \!\zeta\i\Rnu\lf(\fr{I_0}{\sqrt3} - \fr19\rt)\!
  \simeq \!0.1656\,\zeta\i\Rnu.~
\lb{rg1}
\ee
Substitution of the found values in
eq.\rf{p_g_lg_gen} leads to the results\rfs{dg_trf_cor}{Aph_res}.


\subsubsection*{CDM singularities}

Now we proceed to the calculation of the neutrino corrections
$\D_c$ and $\d c$ in the CDM density perturbation\rf{dc_trf_cor}.
The equation for the radiation era CDM Green's function\rf{dc_real_eq}
gives
\be
(\chi\bar d_c)'=\fr1{\chi}\,\bar\Phi'+p_{c1}\dd(\chi)~.
\ee
Integrating this expression, fixing~$p_{c1}$
by the requirement that $\chi\bar d_c$ is an odd function 
that vanishes at $|\chi|>1$, and dividing the result by~$\chi$, 
we obtain
\be
\bar d_c =
\lf[\fr{G(\chi)}{\chi}-\fr{G(1)}{|\chi|}\rt]\theta(1-|\chi|)
              + p_c\dd(\chi)~,\qquad
\lb{dc_gen_sol}
\ee
where
\be
G(\chi)\equiv \int_{0}^{\chi}\fr{d\~\chi~\bar\Phi'(\~\chi)}{\~\chi}~.
\lb{G_def}
\ee
The initial condition $\int_{-1}^1 d\chi~\bar d_c=-3\zeta\i$
fixes the new constant~$p_c$ as
\be
p_c= -3\zeta\i-\int_{-1}^1d\chi
     \lf[\fr{G(\chi)}{\chi}-\fr{G(1)}{|\chi|}\rt]~.
\nn
\ee
The~$d\chi$ integral of $G(1)/|\chi|$, which would be
infinite in Riemann sense, is equal to $0$ in the sense 
of generalized function integration, see Table\rfp{FPtable}.
Hence,
\be
\fp p_c=-3\zeta\i-\int_{-1}^1\fr{d\chi\,G(\chi)}{\chi}~.
\lb{p_c}
\ee

The CDM density perturbation modes in the radiation era are given 
by the Fourier components of the Green's function\rf{dc_gen_sol}. 
In the radiation era for adiabatic initial conditions 
the potential $\bar\Phi(\chi)$, generated by the photon and neutrino 
perturbations, is regular at $\chi=0$ and even.  Hence, 
the function~$G(\chi)$ in eq.\rf{G_def} is regular at 
$\chi=0$ and odd, and so is $G(\chi)/\chi$ regular at the origin. 
The subhorizon $(k\t\gg1)$ values of the CDM Fourier modes
are fully specified by the singular terms in eq.\rf{dc_gen_sol}, 
which are proportional to $1/|\chi|$ and 
$\dd(\chi)$.  From Table~\ref{FTtable} on p.~\pageref{FTtable}
we find that
\be
d_c(\t,k)=A_c\lf(\ln\fs+c\rt)+O(\fs^{-1})~,
\lb{dc_trf_genlim1}
\ee
with $\fs=k\t/\sqrt3$ and
\be
\qquad
A_c=2G(1)~,\quad
c=\g+\fr{\ln3}2 + \fr{\fp p_c}{2G(1)}~.~~
\lb{dc_trf_genlim2}
\ee

In the $\Rnu\to0$ limit, with the potentials\rfs{Phm_0}{Php_0},
\be
G^{(\Rnu\to0)}(\chi)=
    -3\zeta\i\lf\{\ba{cl}
  \chi\sqrt3~, & |\chi|\le \fr1{\sqrt3}~,\vsp\\
  {\sign\chi}~, & |\chi|\ge \fr1{\sqrt3}~.\vsp
  \ea \right.
\ee
Then $G^{(\Rnu\to0)}(1)=-3\zeta\i$ and
\be
&&\bar d_c^{(\Rnu\to0)}=
\lb{d_c_rad_sol_x_ap}
\\
&&~~=-3\zeta\i\!\lf(\sqrt3-\fr{1}{|\chi|}\rt)
                  \theta\!\lf(\fr1{\sqrt3}-|\chi|\rt)
           +p_c\dd(\chi)~,
\nn
\ee
where by eq.\rf{p_c}
\be
\qquad \fp p_c^{(\Rnu\to0)} = 3\zeta\i\lf(1+\ln3\rt)~.
\lb{Fppc0}
\ee
Fourier transforming eq.\rfs{d_c_rad_sol_x_ap}{Fppc0},
where the singular function $1/|\chi|$ can be transformed 
using Table~\ref{FTtable}, we obtain the CDM density perturbation 
mode\rf{dc_trf0}.
Of course, the radiation era, $\Rnu\to0$ Fourier mode
could be obtained directly in $k$-space 
by integrating the evolution equation\rf{dc_evol_rad}.
Now we consider how the CDM density perturbation changes 
when neutrinos are added.

Although the analytical calculation of the integrals in 
eqs.\rfd{G_def}{p_c} in the $O(\Rnu)$ order
may be possible,
it does not appear easy.  On the other hand the numerical
evaluation of the absolutely convergent 
integrals in eqs.\rfd{G_def}{p_c},
given the potentials\rfd{Phi_m_1}{Php_sol},
is straightforward and yields
\be
G(1)&\simeq&  -3\zeta\i\lf[1+0.2297\Rnu+O(\Rnu^2)\rt] \,, \\
\fp p_c&\simeq& 3\zeta\i\lf[1+\ln3+1.746\Rnu+O(\Rnu^2)\rt]\,.\qquad
\ee
Then for $A_c$ and $c$, eqs.\rfs{dc_trf_genlim1}{dc_trf_genlim2},
\be
 \ba{rcl}
A_c&\simeq& -6\zeta\i\lf[1+0.2297\Rnu+O(\Rnu^2)\rt]\,,\vsp \\
c&\simeq& \g-\fr12-0.6323\Rnu+O(\Rnu^2)~.\vsp
 \ea
\lb{Ac_fin}
\ee
The corresponding values for $\D_c =  A_c/A_c^{(\Rnu\to0)}-1$
and $\d c = c - c^{(\Rnu\to0)}$ are used in 
Sec.~\ref{subsec_CDMrad}, eq.\rf{Ac_res}.


\bibliography{nubib}

\end{document}